\documentclass[acmsmall, screen]{acmart}
\usepackage{caption}
\usepackage{subcaption}
\usepackage{array}
\usepackage{multirow}
\usepackage{rotating}
\usepackage{float}
\usepackage{MnSymbol}
\usepackage[flushleft]{threeparttable}
\usepackage{csquotes}

\usepackage{booktabs}
\usepackage{hyperref}
\usepackage{xcolor}
\definecolor{light-gray}{gray}{0.95}

\newif\ifshowchanges
\showchangestrue
\showchangesfalse 

\ifshowchanges
    \usepackage[commandnameprefix=ifneeded]{changes}
\else 
    \newcommand{\added}[1]{#1}
    \newcommand{\deleted}[1]{}
    \newcommand{\replaced}[2]{#1}
\fi
\AtBeginDocument{%
  \providecommand\BibTeX{{%
    \normalfont B\kern-0.5em{\scshape i\kern-0.25em b}\kern-0.8em\TeX}}}
    
\usepackage{booktabs}
\usepackage{graphicx}
\usepackage{longtable}
\usepackage{multirow}
\usepackage[many]{tcolorbox}

\definecolor{main}{HTML}{757575}    
\definecolor{sub}{HTML}{EEEEEE}     

\tcbset{
    sharp corners,
    colback = white,
    before skip = 0.2cm,    
    after skip = 0.5cm      
}                           

\newtcolorbox{boxC}{
    colback = sub, 
    colframe = main, 
    boxrule = 0pt, 
    toprule = 6pt 
}

\begin{document}
\title[Exploring Multidimensional Checkworthiness]{Exploring Multidimensional Checkworthiness: Designing AI-assisted Claim Prioritization for Human Fact-checkers} 

\author{Houjiang Liu}
\email{liu.ho@utexas.edu}
\orcid{0000-0003-0983-6202}
\affiliation{%
  \institution{School of Information, University of Texas at Austin}
  \city{Austin}
  \state{Texas}
  \country{USA}
}

\author{Jacek Gwizdka}
\email{jacekg@utexas.edu}
\orcid{0000-0003-2273-3996}
\authornote{Both authors contributed equally.}
\affiliation{%
  \institution{School of Information, University of Texas at Austin}
  \city{Austin}
  \state{Texas}
  \country{USA, }
}
\affiliation{%
  \institution{Institute of Applied Computer Science, Łódź University of Technology}
  \city{Łódź}
  \country{Poland}
}

\author{Matthew Lease}
\authornotemark[1]
\email{ml@utexas.edu}
\orcid{0000-0002-0056-2834}
\affiliation{%
  \institution{School of Information, University of Texas at Austin}
  \city{Austin}
  \state{Texas}
  \country{USA}
}


\begin{abstract}
Given the 
volume of potentially false claims 
online, claim prioritization is essential in allocating limited human resources available for fact-checking. In this study, we perceive claim prioritization as an information retrieval (IR) task: just as multidimensional IR relevance, with many factors influencing which search results a user deems relevant, checkworthiness is also multi-faceted, subjective, and even personal, with many factors influencing how fact-checkers triage and select which claims to check. Our study investigates both the multidimensional nature of checkworthiness and effective tool support to assist fact-checkers in claim prioritization. Methodologically, we pursue \textit{Research through Design} combined with mixed-method evaluation. 

Specifically, we develop an AI-assisted claim prioritization prototype as a probe to explore how fact-checkers use multidimensional checkworthy factors to prioritize claims, simultaneously probing fact-checker needs and exploring the design space to meet those needs. With 16 professional fact-checkers participating in our study, we uncover a hierarchical prioritization strategy fact-checkers implicitly use, revealing an underexplored aspect of their workflow, with actionable design recommendations for improving claim triage across multidimensional checkworthiness and tailoring this process with LLM integration. 


\end{abstract}



\maketitle

\section{Introduction} \label{sec:intro}

The scale of potentially false claims circulating online far exceeds limited human resources for manual fact-checking. While Natural Language Processing (NLP) research has sought to fully or partially automate fact-checking \citep{Guo2022-km, Nakov2021-sq, Graves2018-ne, Zhou2020-mf}, even state-of-the-art NLP still cannot match human capabilities in many areas. NLP technology continues to rapidly advance \citep{Min2023-yy}, but experts argue that the complexity involved in fact-checking requires subjective judgment and expertise \citep{Arnold2020-ly, Nakov2021-sq}, continuing to necessitate human work for the foreseeable future \citep{Das2023-fg}.  

Given the need for human fact-checking, claim prioritization is key to efficiently allocating human resources \citep{Westlund_undated-po, Micallef2022-vp, Sehat2023-xa}. 
Prioritization seeks to triage and select the most consequential claims by considering \textit{checkworthiness} factors aligned with the goals and news values of fact-checking organizations \citep{Vinhas2023-hx, Belair-Gagnon2023-fr}. While many NLP methods have been developed to identify and monitor misinformation, NLP research has typically sought to automate rather than develop mixed-initiative \cite {Horvitz1999-wp} tools to assist human fact-checkers in claim prioritization. Given the complexity and uncertainty underlying the assessment of checkworthiness \citep{Micallef2022-vp, Procter2023-lu, Liu2023-ur}, better tooling could significantly help.

In this study, we perceive claim prioritization as an information retrieval (IR) task in which a fact-checker has an \textit{information need} \cite{Wilson1981-hc} and seeks relevant information to address that need. Just as relevance is multidimensional, with many factors influencing which search results a user will deem relevant to their personal information need \citep{Xu2006-sj, Zhang2014-yk}, checkworthiness is also multi-faceted, subjective, and even personal, with a diverse set of factors influencing how fact-checkers triage and select which claims to check \citep{alam-etal-2021-fighting-covid, Neumann2022-ip}. Prior work has identified a variety of dimensions to checkworthiness \citep{Micallef2022-vp, Procter2023-lu, alam-etal-2021-fighting-covid}, such as whether the claim is checkable at all \citep{Graves2017-rb, Hassan2017-fn}, the potential harm it might cause if left unchecked \citep{Sehat2023-xa}, how difficult it might be to check \citep{Singh2021-du}, etc. However, it remains unclear today how fact-checkers perceive the relative importance of these different factors, as well as how they dynamically apply them in an IR context for claim prioritization. As reported by fact-checkers in prior studies \citep{Sehat2023-xa, Liu2023-ur, Procter2023-lu}, this process is a less organized, complex, and highly context-dependent task, making it difficult to develop an optimal design solution.

To explore this design challenge, we adopt a \textit{Research through Design} (RtD) approach \citep{Zimmerman2007-gh}. Unlike traditional empirical methods, such as interviews or focus groups, RtD uses design interventions as a methodological tool to uncover knowledge that informs both the understanding of user practice and the creation of innovative solutions \citep{Zimmerman2010-fb}. By designing an intervention and observing how people react to it, we gain new insights into user practices and needs \citep{Zimmerman2007-gh, Zimmerman2014-cs}. To this end, we developed an AI-assisted claim-prioritization prototype that provides customizable filters to help fact-checkers search and filter claims over multiple dimensions of checkworthiness. We use this prototype as a probe to both explore fact-checker work practices and to better understand their needs for claim prioritization.

Our prototype provides two key capabilities beyond a basic search box for entering query terms. First, we provide automated models that predict four checkworthiness dimensions (``Verifiable'', ``Likely harmful'', ``Likely false'', and ``Interest to the public''), coupled with simple UI slider widgets that support dynamically varying the relative weight assigned to each dimension of checkworthiness, individually or in combination, to customize claim ranking in real-time. The second key capability enables fact-checkers to develop additional zero-code, custom search filters using large language model (LLM) technology. This allows additional dimensions of checkworthiness to be introduced and influence claim ranking beyond the four dimensions supported natively. More generally, the flexibility and power of the customized LLM search filter can help fact-checkers to overcome the limitations of traditional keyword search. Just as LLMs enable non-programmers to quickly formulate new AI tasks without model training and data acquisition, our goal was to enable fact-checkers to specify custom search criteria for claim prioritization in natural language.

Our user study with 16 professional fact-checkers employed mixed-method evaluations to collect and analyze participant experiences and reflections. Guided by an RtD process, the prototype enabled us to probe and observe how fact-checkers flexibly triage claims across multidimensional checkworthiness. We investigate: 1) how participants assessed the relative importance of different checkworthy dimensions and developed priorities in claim selection; 2) how they created customized LLM-based search filters and the corresponding benefits and limitations; and 3) their overall user experiences with our prototype. 

\textbf{Research Contributions:} In this paper, we examine how fact-checkers dynamically prioritize claims by considering various checkworthiness factors and uncover specific user needs for tools to support this process. Specifically, we developed an interactive claim prioritization prototype as a design probe to investigate these dynamics in-depth. We uncovered a hierarchical prioritization strategy that they implicitly use, shedding light on an underexplored aspect of their workflow. We also synthesized actionable design recommendations learned from fact-checkers, suggesting mechanisms to better triage claims across multidimensional checkworthiness and to tailor this process by integrating LLMs. These insights deepen our understanding of fact-checker work practices and the supporting tools they require, while also offering broader design implications for improving relevance judgment and triage in other user activities.

\section{Related Work} \label{sec:lit review}

\subsection{Journalistic Fact-checking and Digital Tools Used for Claim Prioritization} \label{subsec:fact-checking}

Our information ecosystem has become severely contaminated with the rise of different false content circulating through online media, including text, images, and videos  \citep{Wardle2017-fq, Jack2017-cn}. This pollution is leading to various social problems, including public health crises \citep{Swire-Thompson2020-vi}, political polarization \citep{Hameleers2020-sp}, and increased tensions among different social groups. In a survey representing a diverse demographic of 1207 Americans, it was found that 49\% have encountered online misinformation \citep{Saltz2021-yg}. Journalistic fact-checking plays a crucial role in addressing this emerging issue \citep{Graves2017-rb}. Fact-checking not only assists individuals in assessing information accuracy but also raises public awareness of pre-bunking misinformation \citep{Comlekci2022-ch}. 

%
\citet{Graves2017-rb} describes a five-stage process of traditional fact-checking, including claim selection, contacting the claimant, tracing the claim, consulting experts, and making the verification process public. Currently, the scope of fact-checking has expanded beyond checking political claims to investigating more general false information spread throughout social media platforms (e.g., rumors, hoaxes, and conspiracy theories), and fact-checking practices evolved \citep{Graves2020-gw}. One of the important changes, as described by \citet{Westlund_undated-po}, includes the extensive usage of technological tools to counter the massive scale of online misinformation \citep{Das2023-fg, Beers2020-cw}. 



Fact-checkers use many tools to search for, monitor, filter, and collect potential claims to check. As reported in prior work \citep{Micallef2022-vp, Beers2020-cw, Westlund_undated-po, Procter2023-lu, Das2023-fg}, off-the-shelf tools include Google search, search provided by the social media platforms, third-party monitoring software\footnote{TweetDeck has now become X Pro: \url{https://pro.twitter.com/}, CrowdTangle (suspended): \url{https://www.crowdtangle.com/}.}, for example, TweetDeck and CrowdTangle, and other open-source intelligent tools. More advanced tools were also built to better identify checkable claims that contain factual statements. For example, \citet{Majithia2019-eu} build ClaimPortal, a tool incorporating Claimbuster \citep{Hassan2017-fn}, to identify checkable claims from Tweets and perform traffic analytics. The UK fact-checking organization, Full Fact\footnote{Full Fact AI: \url{https://fullfact.org/ai/about/}.}, built a claim monitoring tool that helps identify different types of claims from different media and news outlets, including statistical, opinionated, and predicted claims. Meta also provided fact-checking tools\footnote{Meta’s fact-checking partnership (suspended at the time of publication): \url{https://transparency.meta.com/features/how-fact-checking-works}.} that enabled fact-checking organizations they partnered with to monitor, search, and check claims on their platforms, including Facebook and Instagram. 

%

From previous research and industry reports, most tools mentioned above do not directly assist fact-checkers in claim prioritization. This might stem from several causes, including the lack of transparency, personalization, or other unmet user needs. For example, as described in \citet{Arnold2020-ly}'s report, fact-checkers complained that the ranking provided in the current Meta monitoring tool was not transparent. Although the tool ranks claims based on different sources, such as feedback from social media users flagging potential false or harmful claims and the popularity indicated by social media metrics, fact-checkers did not understand how these factors are combined in a ranking list. This lack of clarity made it difficult for them to trust the tool and agree that checkworthiness rankings align with their goals. 

In \citet{Liu2023-ur}'s co-design study with fact-checkers, participants expressed a need for more personalized claim filtering and selection. Because the assessment of checkworthiness is multi-faceted, they preferred tools that help them triage claims across multiple dimensions of checkworthiness. Similar findings arose in other recent work \citep{Procter2023-lu, Sehat2023-xa}. Given this gap in tooling support, both researchers and tool developers would benefit from a deeper understanding of how fact-checkers prioritize claims, as well as exploring new tools to assist with claim prioritization.


\subsection{Relevance Judgment and Claim Checkworthiness Assessment} \label{subsec:checkworthiness}

Since fact-checkers primarily prioritize claims within the context of information search, we perceive their process of searching for and selecting claims as analogous to a traditional Information Retrieval (IR) task, where a user has an \textit{information need} \cite{Wilson1981-hc} and seeks relevant information to address that need. Just as relevance is multidimensional, with many factors influencing which search results are deemed relevant \citep{Xu2006-sj}, checkworthiness is also multi-faceted \added{and }affected by diverse factors \citep{alam-etal-2021-fighting-covid}. 
Given this parallel, we review key literature from both IR and fact-checking to ground our understanding of claim prioritization and to inform our tool design.

In IR studies, evaluating relevance through multiple dimensions provides a more accurate assessment of search results than considering only unidimensional topical relevance. \citet{Jiang2017-fi} integrates four factors---``Novelty,'' ``Understandability,'' ``Reliability,'' and ``Effort''---with user experience measures to evaluate an IR system. This multidimensional approach to judging relevance showed a stronger statistical relationship with user experience than just topical relevance. It is also recognized that individuals perceive the importance of these multidimensional factors differently, which can influence their final judgment of search relevance. \citet{Zhang2014-yk} use structural equation modeling to examine the relationship between five-dimensional factors and the overall relevance of search results. Findings from their model indicated a considerable difference in the weight that individuals assigned to different relevance factors. Additional research shows that user perceptions of multidimensional relevance are further influenced by user domain expertise \citep{Tamine2017-bx} and biases \citep{Krieg2022-wl}. This suggests that relevance judgment is highly subjective and individualized.

The notion of checkworthiness also varies across individuals, organizations, and time, just as relevance is multidimensional, subjective, and personal. Fact-checkers consider different factors in assessing claim checkworthiness.  
For example, \citet{Procter2023-lu} describes three checkworthy factors regarding claim prioritization, including ``Spread,'' ``Severity,'' and ``Amplification.'' Additionally, \citet{Sehat2023-xa} highlights three other factors, including "Urgency of Claims," "Resource Allocation and Claim Scope," and "Interests of Different Stakeholders." 
\citet{Singh2021-du} also describe ``Claim Difficulty'' based on the claim ambiguity, the poor ranking and unreliable sources in evidence retrieval, and the difficulty of inferring veracity from the evidence. To better ground our understanding of multidimensional checkworthiness that influences claim prioritization, we present a summary of these factors with definitions in Table \ref{tab:multi-dimensional checkworthiness}.

\begin{table}[ht]
\begin{tabular}{@{}>{\raggedright}p{4cm}>{\raggedright\arraybackslash}p{9.5cm}@{}} \toprule
Factors & Definitions \\ \midrule
Already checked \citep{Shaar2020-eb} & A fact-check of this claim or similar claims have already been conducted. \vspace{0.1cm} \\
Amplification \citep{Micallef2022-vp, Procter2023-lu} & Publishing a fact-check of this claim is likely to cause a risk of raising the profile and thus increasing public awareness of this claim. \vspace{0.1cm} \\
Checkable \citep{Micallef2022-vp} (or verifiable) & A checkable claim is a factual statement (e.g., numbers, geographical references) that can be checked. \vspace{0.1cm} \\
Difficulty \citep{Singh2021-du} & The claim is difficult to fully verify because of its term ambiguity, unreliable evidence, and other limitations. \vspace{0.1cm} \\
Harmful \citep{Sehat2023-xa} (or severity \citep{Procter2023-lu}) & The claim either directly or indirectly causes harm to people and society. \vspace{0.1cm} \\
Likely false \citep{Graves2016-tc} & The claim is likely to contain false information. \vspace{0.1cm}\\
Public interest \citep{alam-etal-2021-fighting-covid} & The claim includes topics such as healthcare, political news, and current events, which tend to be of higher interest to the general public. \vspace{0.1cm}\\
Spread \citep{Procter2023-lu} (or virality) & The claim is widely spread across different social media platforms and different languages or countries. \vspace{0.1cm}\\
Susceptibility \citep{Babaei2022-so} & The likelihood of people believing in this claim. \vspace{0.1cm} \\
Urgency \citep{Sehat2023-xa} & Immediate action is needed to fact-check this claim due to the negative impacts it might cause or has already caused. \\ \bottomrule
\end{tabular}
\vspace{0.2cm}
\caption{Dimensions of checkworthiness that have been directly mentioned by fact-checkers or identified in prior studies.} 
\label{tab:multi-dimensional checkworthiness}
\end{table}

NLP research has also annotated claims for different checkworthy dimensions in order to build and test predictive models \citep{Hassan2017-fn, Konstantinovskiy2021-ga,alam-etal-2021-fighting-covid}, with annotation guidelines using linguistic features to infer various aspects of checkworthiness \citep{Allein2020-jy}. A well-known NLP competition, CLEF \textit{CheckThat!} \citep{Barron-Cedeno2024-gi, Barron-Cedeno2023-ec}, emphasizes the multidimensional nature of evaluating claim checkworthiness and helps capture some of them from human annotators. Despite the prominence of this work, there is little insight into the provenance of annotated dimensions: how the particular dimensions were selected or their relative importance, what other dimensions might exist, how fact-checkers used these dimensions in practice, etc. 



Informed by decades of study of relevance in IR \cite{saracevic1975relevance, saracevic2007relevanceII, saracevic2007relevanceIII}, we can infer that fact-checkers perceive the importance of various dimensions of checkworthiness differently, which influences how they identify and select claims during searches. Therefore, several open questions arise that exemplify the gaps in existing claim prioritization: How do fact-checkers evaluate the relative importance of various dimensions? Are different dimensions of checkworthiness considered serially or in parallel by fact-checkers to conduct claim selection effectively? How do fact-checker self-reported perspectives on these dimensions align with their actual behavior during claim search and selection? When different dimensions compete in claim ranking, how do fact-checkers navigate these trade-offs dynamically? In Section \ref{subsubsec:data collection and analysis}, we structure these questions around the primary research goal and explain how we address them. As we seek to address some of these questions with an RtD approach, we synthesize important literature on RtD in the next section and describe our motivation for using RtD.

\subsection{Research through Design (RtD) in Misinformation Research} \label{subsec:rtd}

Researchers in Computer-Supported Cooperative Work (CSCW) and Human-computer Interaction (HCI) have extensively used RtD. Originating from traditional arts and design practice, RtD as described by \citet{Frayling1993-sp} documents how artifacts are created and communicated through art, craft, and design activities. In HCI, RtD involves using designed artifacts as tools to uncover new knowledge and insights that inform the understanding of user practice and the creation of innovative solutions \citep{Zimmerman2014-cs}. Unlike traditional empirical methods, such as interviews or focus groups, RtD allows researchers to tackle emergent, context-dependent, complex, or multi-faceted questions, providing insights that are not just theoretical but grounded in the practices of making and doing \citep{Gaver2012-yw}. Additionally, RtD emphasizes the quality of the design process and its implications, rather than merely measuring tool usability \citep{Prochner2022-lg}. 

Scholars in CSCW and HCI have widely adopted RtD to investigate the phenomenon of misinformation and develop workable solutions. As \citet{Venkatagiri2023-wu} write: 
\begin{quote}
    ``Misinformation on social media is a wicked problem because: 1) it is a symptom of another problem (e.g., political polarization or psychological biases), 2) it can be interpreted and solved in many different ways (e.g., social, psychological, or technological), and 3) solving it is identical to completely understanding it, and there are no clear criteria for sufficient understanding.''
\end{quote}
RtD is particularly useful for studying misinformation and developing solutions because its iterative design process allows researchers to develop a deeper understanding of the problem as it evolves while simultaneously modifying design interventions. RtD has been applied to develop various interventions, digitally or socially, to address different societal problems brought about by misinformation. For example, \citet{Venkatagiri2023-wu} developed a new platform that fosters competition and collaboration among crowd workers to identify and debunk misinformation. \citet{Zade2023-vu} employed an RtD process to design contextual cues to inform credibility assessment on social media. Similarly, \citet{Lovlie2023-df} designed a tool to help readers better understand evidence and uncertainty in science journalism. Additionally, throughout years of misinformation research, \citet{Arif2018-hn} implemented community-engaged programs to enhance people's digital literacy regarding online misinformation in their everyday environments \citep{Wilner2023-wo}.

As discussed at the end of Section \ref{subsec:checkworthiness}, our study investigates how fact-checkers perceive the relative importance of checkworthiness dimensions and apply this in claim prioritization during search to better reveal potential user needs. This knowledge is situational and contextualized within fact-checking, especially in fact-checker behavior of searching, filtering, and selecting claims. RtD is a well-suited approach for addressing this challenge. It emphasizes using design as a method to uncover nuanced, situational knowledge rather than focusing on developing definitive solutions, particularly when user practices are not yet fully understood \citep{Gaver2012-yw, Zimmerman2010-fb}. In our context, this involves exploring how fact-checkers prioritize various dimensions of checkworthiness, how they apply this situational understanding to the process of searching for claims, and what AI solutions we can design to facilitate this process more effectively. 

To better articulate how we adopt RtD, we clarify how RtD helps us conceptually define research goals and methodologically support the design and study process in Section \ref{sec:design process}.


\section{Research through Design} \label{sec:design process}

In \citet{Zimmerman2010-fb}'s examinations of RtD practices across different design projects, they found that ``All of the projects employed an RtD approach, creating artifacts that included products, prototypes, and models that illustrated future visions, uses of new materials, and potential ideas.'' \citet{Bowers2012-iq} also emphasize that the artifacts created via RtD aim to ``provide the design research community with information about how to design.'' Thus, RtD differs from traditional empirical studies, such as interviews or focus groups, where these methods typically do not involve user interactions with artifacts. To understand user practices, these methods primarily rely on verbal accounts and thematic analysis of participant narratives, rather than gaining insights through direct observations of participant actions and processes. RtD also contrasts with tool evaluations, where the study focuses on assessing a tool usability rather than better uncovering user practices and generating design knowledge to inform future solutions \citep{Frankel2010-sw, Pierce2014-am}.

Given these methodological differences, we first describe how RtD helps conceptualize our research goals (Section \ref{subsec:frame}). We then detail our design process of creating the artifact used for RtD (Section \ref{subsec:dicover and define} and \ref{subsec:prototype and iterate}) and the evaluative approach (Section \ref{subsec:evaluate and deploy}) to achieve our research goals.

\subsection{Frame Research Goals} \label{subsec:frame}

After interviewing fact-checkers around the globe, \citet{Sehat2023-xa} reported ``no established systematic approach towards claim prioritization'' today. Additionally, fact-checker decisions to focus on specific claims typically depend on case-by-case situations and are heavily influenced by the local news context \citep{Liu2023-ur, Procter2023-lu, Micallef2022-vp}. Thus, both theoretical and empirical understanding highlight the complexity and uncertainty involved when fact-checkers use tools to search and select claims to check. This poses a further design consideration: new tools developed for claim prioritization should support such complexity and uncertainty with the necessary flexibility to adapt to dynamically changing priorities. While prior work has explored aspects of this challenge, it often stops short of offering clear solutions in tool support. This gap presents an opportunity to leverage RtD to better inform and develop actionable design outcomes.

In particular, we believe using artifact-driven actions and reflections is more effective in first untangling the complexity and uncertainty of fact-checker tasks in claim prioritization. As noted by \citet{Bowers2012-iq} and \citet{Pierce2014-am}, the value of using RtD lies in creating exploratory artifacts to uncover design knowledge. In our context, these artifacts act as probes, enabling fact-checkers to demonstrate how they dynamically triage claims across various checkworthiness factors. By observing the strategies they develop and adapt over time to improve the efficiency of searching and selecting claims, we could gain valuable insights into user behaviors and patterns, which subsequently guide the design of future tools. 



\textbf{We define our two research goals based on RtD as follows}:
\begin{itemize}
  \item [\bf RG1] \textbf{A practice-based examination of fact-checker practice and needs for claim prioritization.} We aim to build a prototype as a probe to elicit fact-checker insights into how they flexibly triage claims among multidimensional checkworthiness by searching, filtering, and selecting claims in real time. Additionally, as RtD contributes to the creation of innovative solutions \citep{Gaver2012-yw, Zimmerman2014-cs}, we also synthesize insights, including user behaviors and patterns, into more sophisticated needs, as well as concrete design suggestions for claim prioritization based on what we observed when fact-checkers use the prototype.
  
  \item [\bf RG2] \textbf{An evaluation of fact-checker use experiences for the claim prioritization prototype.} We aim to employ a lab-based RtD approach that uses both quantitative and qualitative data to inform new design knowledge. As described by \citet{Zimmerman2014-cs}, a lab-based RtD approach helps explore ``semi-articulated hypotheses for better forms of user interaction.'' Therefore, another goal of this study is to present a comprehensive evaluation of fact-checker use experiences of the AI-assisted claim prioritization prototype we developed. While this prototype might not represent the final form of user interactions in claim prioritization, presenting its evaluative results helps inform the future design and development of more advanced tools.
\end{itemize}

\begin{figure}[ht]
    \centering
    \includegraphics[width=1\textwidth]{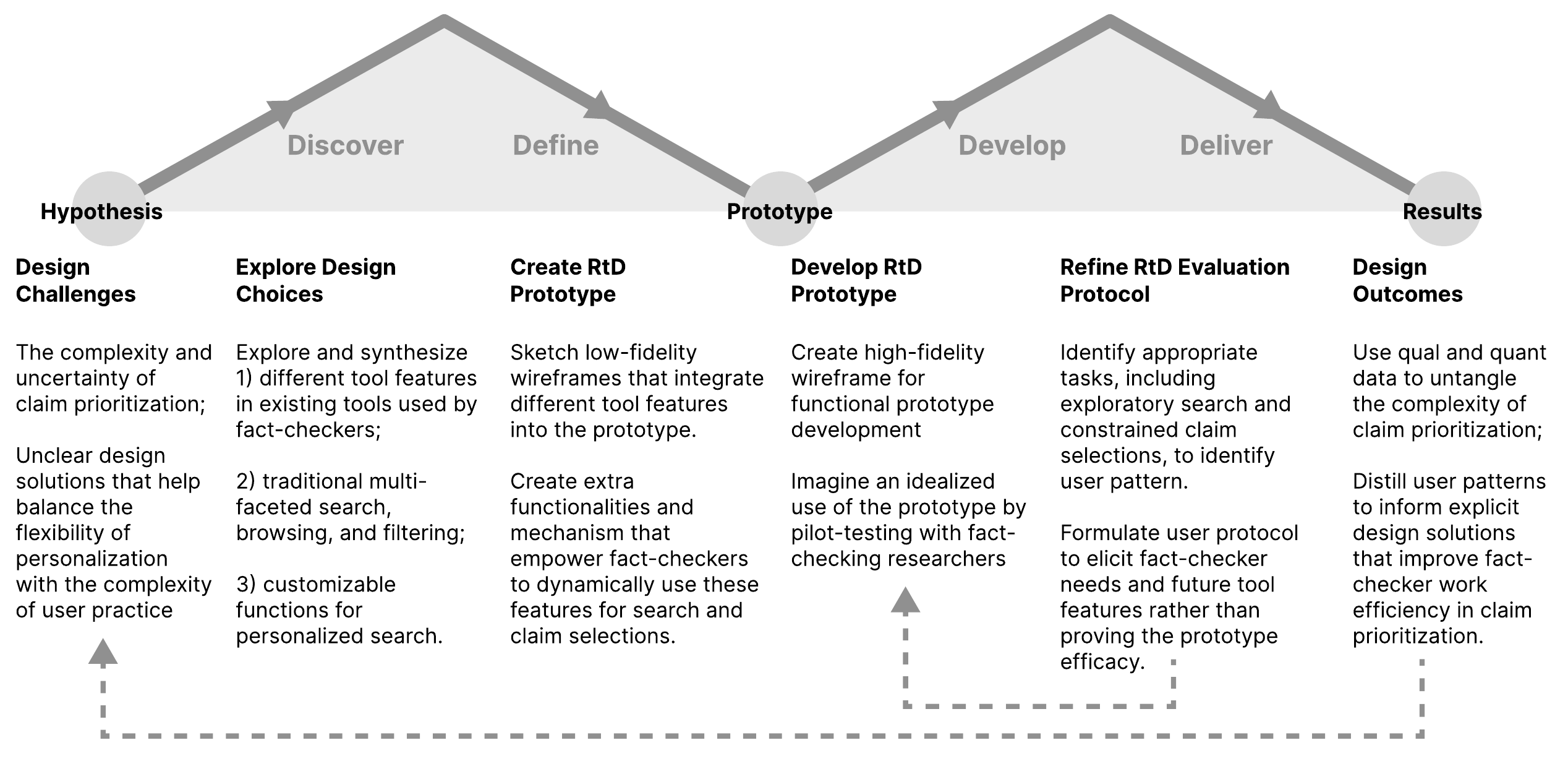}
    \caption{Our lab-based RtD method integrated into a classic double diamond process \cite{Design_Council2015-wu}. We document our design challenges as the research goals, described in Section \ref{subsec:frame}. Steps of \textit{Discover} and \textit{Define}, i.e., how we explore and finalize design choices, are documented in Section \ref{subsec:dicover and define}. The step of \textit{Develop}, i.e., how we create and deploy the prototype, is documented in Section \ref{subsec:prototype and iterate}. The step of \textit{Deliver}, i.e., how we finally evaluate the design to gather use-inspired insights, is documented in Section \ref{subsec:evaluate and deploy}. The iterative refinement of RtD evaluation protocol is documented in Appendix \ref{sec:pilot_findings}.}
    \Description{A visual representation of lab-based RtD method}
    \label{fig:design process}
\end{figure}

As described by \citet{Zimmerman2014-cs}, a lab-based approach ``blends design methods to envision the unimagined and both analytic and experimental methods to evaluate the novel design offerings.'' We thus follow the classic double-diamond design framework \cite{Design_Council2015-wu} to scaffold our design process (Figure \ref{fig:design process}). This involves steps of \textit{Discover} and \textit{Define} (i.e., how we explore and finalize design choices, documented in Section \ref{subsec:dicover and define}) and \textit{Develop} (i.e., how we prototype and deploy the design concept technically, documented in Section \ref{subsec:prototype and iterate}) and \textit{Deliver} (i.e., how we finally evaluate the design to gather use-inspired insights, documented in Section \ref{subsec:evaluate and deploy}). Unlike the traditional double-diamond design, where the final design concept is ideally perceived as a tentative yet optimal solution to a user problem, we formulate a design probe in the define stage. In the delivery stage, this probe then serves to collect analytical data, aiming to better serve the discover stage. Findings from this loop help guide the development of future optimal solutions.

\subsection{Discover and Define Design Probe} \label{subsec:dicover and define}

As described in Section \ref{subsec:fact-checking}, claim prioritization mainly happens when fact-checkers use search-related tools. Furthermore, Section \ref{subsec:checkworthiness} discussed how assessing claim checkworthiness parallels multidimensional relevance judgment of search. This suggests that, in claim prioritization, fact-checkers rely on information seeking and retrieval to address their information needs when identifying important claims. In order to meet different user information needs, scholars in information seeking and retrieval have been exploring different search features, such as metadata \citep{Yee2003-cr} and graphical facets \citep{Guo2023-iv}, and different types of search result presentations \citep{Capra2007-ph}, including standard website, hierarchical text-based faceted UI, and dynamic query faceted UI. These insights require us to examine design work from this area as the initial phase of design exploration.

To synthesize design work from both academic research and industry practices for our tool-building, our literature search focused on three main aspects: 1) existing tools used by fact-checkers to identify and monitor claims; 2) traditional multi-faceted search, browsing, and filtering that support general information seeking; and 3) customizable functions that provide users with a more personalized search experience. We conducted an academic literature search on Google Scholar using keywords such as ``multi-faceted search,'' ``personalized search,'' and ``browsing and filtering.'' Most of the academic research we identified originated from conferences related to \textit{SIGCHI}, \textit{SIGIR}, and \textit{CHIIR}. To explore tools currently used by fact-checkers, we examined a range of existing resources and tools related to claim prioritization. These included a collection of tools curated by nonprofit research organizations, such as RAND Corporation\footnote{RAND Corporation: \url{https://www.rand.org/research/projects/truth-decay/fighting-disinformation/search.html}} and Credibility Coalition\footnote{Credibility Coalition: \url{https://credibilitycoalition.org/}}.

We summarize our search results in Table \ref{tab:tool features from prior work}. These include standard keyword and semantic search, multi-faceted filters, personalized weighting, and user-generated facets. These features illustrate the different levels of customization that assist users in their search and browsing activities. We make a simplifying assumption by restricting our scope to textual claims. While future work should investigate the multi-modal setting as well, we show that even this simplified setting is sufficient to reveal broadly useful insights into fact-checker needs and corresponding design implications. 

\begin{table}[ht]
\begin{tabular}{@{}>{\raggedright}p{1.2cm}>{\raggedright}p{2cm}>{\raggedright}p{5cm}>{\raggedright\arraybackslash}p{4.5cm}@{}}
    \toprule
    Categories & Features & Description & Tool implementations \\
    \midrule
    Search & Keyword search & Users enter keywords to look for exact matches of documents where the keyword appears. & \textit{Academic:} \citep{Yee2003-cr, Jasim2022-rm, Majithia2019-eu, Guo2023-iv} \linebreak \textit{Industrial:} Google fact-check tools, Trendolizer, Meta fact-checking tool, Full Fact Alpha \\
    \noalign{\vskip 2pt}
    \cline{2-4}
    \noalign{\vskip 2pt}
            & Semantic search & Users enter a search query to retrieve documents that provide contextual meanings similar to the query. & \textit{Academic:} \citep{Jasim2022-rm, Majithia2019-eu} \linebreak \textit{Industrial:} Google fact-check tools, Trendolizer, Meta fact-checking tool, Full Fact Alpha  \\
    \noalign{\vskip 2pt}
    \cline{2-4}
    \noalign{\vskip 2pt}
            & Image reverse search  & Users use an image as a search query to find similar images from the database.  &  \textit{Academic:} \citep{Cai2019-fz} \linebreak \textit{Industrial:} Google fact-check tools \\
    \midrule                    
    Filtering & Multi-faceted filters  & Users select various criteria from different categories to dynamically filter documents. Only documents relevant to the criteria are updated. &  \textit{Academic:} \citep{Yee2003-cr, Kern2018-vy, Cai2019-fz, Jasim2022-rm, Majithia2019-eu, Guo2023-iv} \linebreak \textit{Industrial:} Trendolizer, Meta fact-checking tool, Full Fact Alpha  \\
    \noalign{\vskip 2pt}
    \cline{2-4}
    \noalign{\vskip 2pt}
            & Personalized weighting & Users adjust the importance of certain criteria or elements to tailor the retrieved results based on individual preferences. &  \textit{Academic:} \citep{Kern2018-vy, Cai2019-fz} \\
    \noalign{\vskip 2pt}
    \cline{2-4}
    \noalign{\vskip 2pt}
            & User-generated facets  & Users create and define the criteria or categories used for organizing and filtering research results. &  \textit{Academic:} \citep{Kern2018-vy, Papenmeier2023-cx} \\
    \bottomrule
\end{tabular}
\vspace{0.2cm}
\caption{Tool features implemented in academic or industry practice for search and filter}
\label{tab:tool features from prior work}
\end{table}


We engaged in an iterative process of creating low-fidelity wireframes (see Appendix \ref{appendix:low-fidenlity wireframe}). This process helped design a claim prioritization tool that integrated different design features (as described in Table \ref{tab:tool features from prior work}) in a meaningful way to meet our research objective. 

For example, at this design stage, we first decided to develop a multi-dimensional ranking system in order to enhance algorithmic transparency, enabling fact-checkers to explore how different facets of checkworthiness could influence a unified, overall checkworthiness ranking.
We also know from prior work in other expert search domains (e.g., legal \citep{Paul2006-mg, Oard2013-qu} and medical search \citep{Wallace2013-ho}) that experts highly value transparent, controllable ranking functions. Our tool seeks to provide such transparent ranking by 1) selecting established dimensions of checkworthiness, 2) making those dimensions explicit, visible, and actionable in the user interface, and 3) providing real-time updates to search results in response to user weight changes. Our work thus builds on both prior technical work and a prior understanding of user needs in expert search domains, extending these capabilities to support fact-checking tasks today.

In addition, because fact-checkers (individually or organizationally) may value different aspects of checkworthiness that were not captured in our standard facets, we further integrated LLM customization as another feature, allowing fact-checkers to extend our tool by incorporating additional checkworthiness. Relatedly, we also expected that algorithmic transparency would be promoted by keeping the search bar (topical relevance) separate from checkworthiness, motivating separation between the search bar vs.\ the standard and LLM-customized checkworthiness facets.

More generally, we recognized that separating these functionalities aligns more effectively with our research goals. As RtD prioritizes the discovery of design knowledge over the pursuit of optimal solutions, particularly in contexts where user practices are ambiguous \citep{Zimmerman2014-cs, Gaver2012-yw, Frankel2010-sw} (in our case, i.e., claim prioritization), we viewed these three features -- topical search bar, standard checkworthiness filters, and LLM-customized filters -- as distinct yet valuable tools. They enabled participants to articulate the differences and uncover use-inspired insights, guiding decisions about whether to merge these features together or maintain them as separate components for designing future tools.

The final design specifications of the RtD prototype were: 

\begin{enumerate}
  \item The prioritization tool needs to offer standard features comparable to those found in commonly used tools for fact-checkers, such as keyword and semantic search. 
  \item Given that the assessment of claim checkworthiness is similar to a multidimensional relevance judgment, fact-checkers should be able to filter claims based on multidimensional factors. 
  \item Considering the subjectivity involved in claim prioritization, fact-checkers should have the flexibility to determine the relative importance of multidimensional factors in different situations. To enable this, personalized weighting on multi-faceted filters is important.
  \item As fact-checkers may have additional checkworthy factors that are important to them or their organizations, the tool should offer a customizable function enabling fact-checkers to explore additional checkworthy dimensions beyond those natively supported in the tool. 
\end{enumerate} 


\subsection{Prototype and Deploy the Design} \label{subsec:prototype and iterate}

After completing a high-fidelity wireframe based on the above design specifications, we implemented design features into a functional software prototype. We then conducted pilot user testing to identify potential usability issues that prevent our design from meeting the research goal and to establish specific tasks and protocols for the formal evaluation studies to be conducted with professional fact-checkers. In this section, we describe the user interface (Section \ref{subsubsec:UI}) and the technical implementation (Section \ref{subsubsec:implementation}). See Appendix \ref{sec:pilot_findings} for a discussion of preliminary findings from pilot tests. 

\subsubsection{User interface (UI)} \label{subsubsec:UI}

\begin{figure}[!t]
    \centering
    \includegraphics[width=0.9\textwidth]{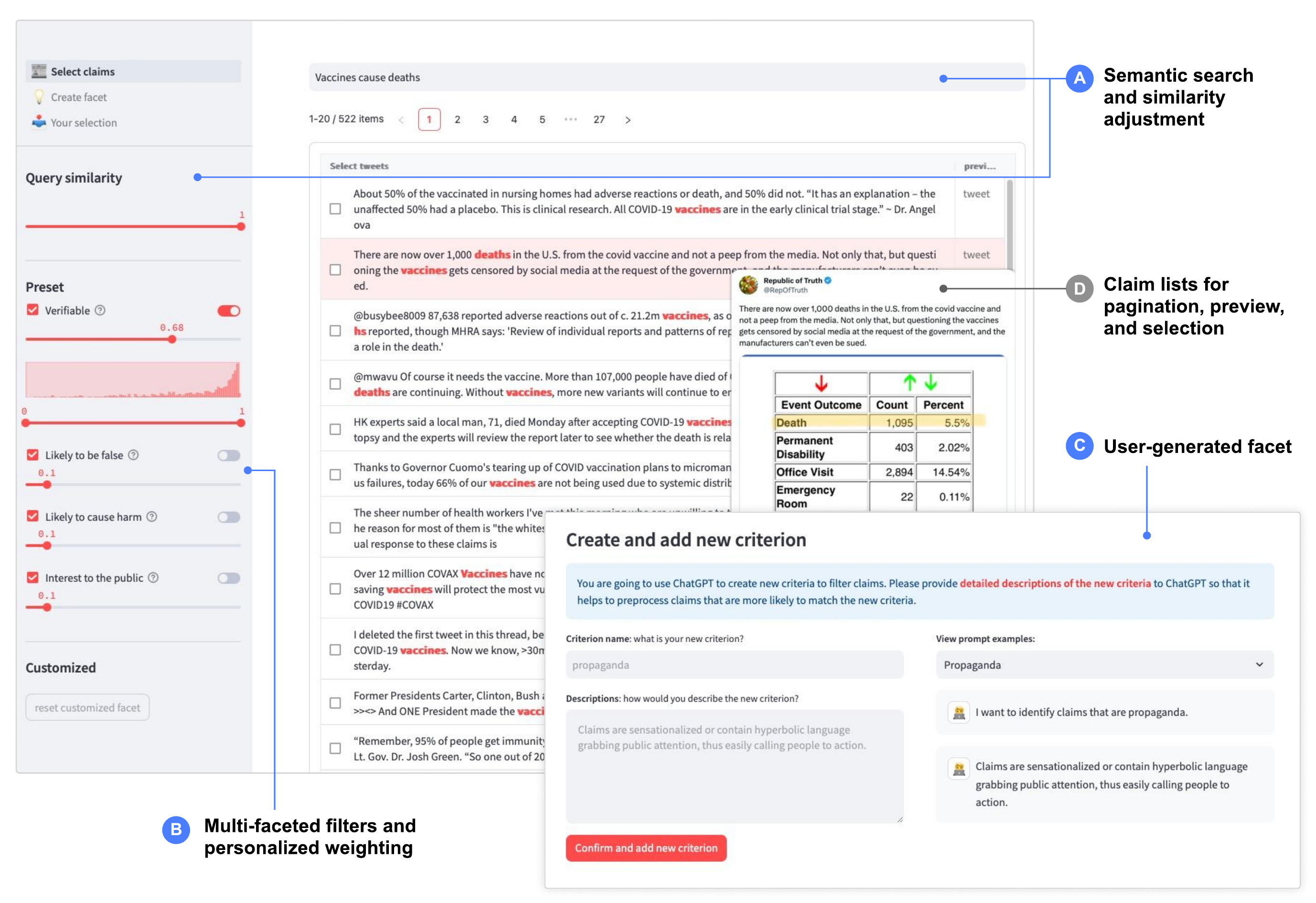}
    \caption{Screenshots of our claim prioritization mockup. The mockup includes three main functions to search, filter, and select claims over multidimensional checkworthiness: (A) semantic search to retrieve the most relevant claims based on query similarity; (B) Multi-faceted filters and personalized weighting to filter or rank claims that meet certain preset checkworthy dimensions; 
    and (C) User-generated facets to create new, customized dimensions using LLMs. Fact-checkers preview and select claims in view (D).}
    \Description{Screenshots of our claim prioritization mockup}
    \label{fig:claim prioritization UI}
\end{figure}

In the initial phase of pilot tests, we carried out heuristic evaluations \citep{Hvannberg2006-nb} with five graduate students with years of experience in misinformation research and fact-checking. We asked them to experiment with the prototype to identify potential usability issues on whether existing design features align with our design specifications. We present the final UI features in Figure \ref{fig:claim prioritization UI}.
Collaborating with our pilot study participants, we imagined an idealized scenario for how professional fact-checkers would use this UI in a real-world setting.

\paragraph{\textbf{Imagining an Idealized Use of the UI}} John, a professional fact-checker, is searching for potential claims to check about COVID-19. He starts by using the search function (A) to look for claims that might say, ``COVID vaccine causes deaths.'' John wants to focus on claims that express the same meaning, similar to his query. He assigns a higher weight to the similarity slider. This action reorganizes the results, positioning claims similar to his query at the top. To further refine his search, John filters claims only ``verifiable.'' He checks the ``verifiable'' criterion at (B) and assigns a higher weight to this criterion than other criteria. This helps him bring more verifiable claims to the top. As he browses the results, he hovers over the text at (D) and previews its content to see additional social media metrics, such as the number of reposts, quotes, and likes it has received. However, he doesn't find any particularly interesting claims. He then creates his own customized criteria, named ``political propaganda'' at (C). He provides a detailed description of what a propaganda claim might look like. This could include emotionally charged language, oversimplification of complex issues, or a clear bias towards a particular political viewpoint. After creating this new criterion, he assigns a higher weight to it, just like he did previously. John retrieves a different set of claims with this new filter to meet his ``political propaganda'' criterion. He is more satisfied with these new claims as they align with his current investigation focus. 


\subsubsection{Technical implementation} \label{subsubsec:implementation}

We used the COVID-19 claim dataset developed by \citet{alam-etal-2021-fighting-covid} to build a domain-specific claim prioritization tool. The dataset contains 4,542 COVID-19-related tweets, annotated for seven dimensions of checkworthiness as a multi-label classification task. Each dimension includes binary labels on whether the tweet satisfies that criterion. To further simplify our study, we selected four of the seven dimensions: 

\begin{enumerate}
  \item [] \textbf{Verifiable}: ``A verifiable factual claim is a sentence claiming that something is true, and this can be verified using factual, verifiable information such as statistics, specific examples, or personal testimony.''
  \item [] \textbf{Likely false}: ``The stated claim may contain false information. False information appears on social media platforms, blogs, and news articles to deliberately misinform or deceive the readers.''
  \item [] \textbf{Likely harmful}: ``The stated claim aims to and can negatively affect the society as a whole, specific person(s), company(s), product(s), or spread rumors about them.''
  \item [] \textbf{Interest to the public}: ``In general, topics such as healthcare, political news and findings, and current events tend to be of higher interest to the general public.''
\end{enumerate} 

Regarding the other three dimensions annotated in \citet{alam-etal-2021-fighting-covid}'s dataset, we used the overall ``Needs Verification'' for our unidimensional baseline. We describe our baseline condition used for experimental study in Section \ref{subsubsec:baseline condition}. The two checkworthy dimensions not used were another variant on likely to cause harm and whether the claim merited government attention. Although the dimensions we selected are highlighted as commonly known and important checkworthy factors, as included in Table \ref{tab:multi-dimensional checkworthiness} and also reported by fact-checkers in \citet{Liu2023-ur} and \citet{Procter2023-lu}'s studies. Compared to a full list of Table \ref{tab:multi-dimensional checkworthiness}, other factors are omitted due to the lack of available datasets. Additionally, as part of our research limitations, we discuss this in Section \ref{subsec:limitation}.

We employed different NLP models to achieve each design specification (outlined in Section \ref{subsec:dicover and define}). First, we used SentenceBERT \citep{Reimers2019-pc} as an embedding model to perform query semantic search. This involved transforming each sentence in the dataset and the user query into the same embedding space. This transformation enables us to retrieve claims similar in meaning to user queries using cosine similarity. 
Second, to create multi-faceted filters, we split the dataset into training and testing sets with a 2:1 ratio and built classification models based on the binary labels for each dimension. 

Each classifier represents one dimension and was implemented using Scikit-learn \citep{Pedregosa2011-tl} using   
logistic regression, using random undersampling to address imbalanced labels methods\footnote{\url{https://imbalanced-learn.org/stable/references/generated/imblearn.under_sampling.RandomUnderSampler.html}}. All use the same textual features, combining n-grams and Word2Vec embeddings\footnote{\url{https://radimrehurek.com/gensim/auto_examples/tutorials/run_word2vec.html}}. Classifiers had an average accuracy rate of approximately 70-75\%. Initially, we implemented a ``hard'' filter effect in which 
claims predicted as negative by the trained classifiers were completely filtered out. However, to support personalized ranking, we changed this to a ``soft'' filter, ranking all claims by classifier probability for each dimension rather than filtering any claims out.

As a baseline UI for unidimensional claim ranking, we also trained another logistic regression classifier for a different annotation. After 
\citet{alam-etal-2021-fighting-covid} first asked annotators to label different checkworthy dimensions, annotators were finally asked whether or not they thought the claim should be fact-checked.  
These annotations provide a unidimensional criterion for our baseline condition. The same training process yielded an accuracy of 71\% for predicting these labels. 

To allow users to create customized checkworthy dimensions as new facet filters, we employed LLMs as a flexible classifier. One of the exciting capabilities of LLMs is their ability to provide zero-code solutions, where users express what they want in natural language. This LLM classifier identifies whether claims meet the new dimension based on the written prompt (see Appendix \ref{appendix:Prompt} for prompt template and Table \ref{tab:examples of customized dimensions} for prompt examples created by our participants). Participants can use LLMs to create multiple checkworthiness dimensions. In both experimental and baseline conditions, they can use LLMs with the prototype to create unlimited checkworthiness dimensions through prompts, with each prompt defining one faceted filter.

The personalized ranking function multiplies each aforementioned model's predicted probability score by a user-customized weight in the range [0,1] and then aggregates the scores (see \(S_{l}\) as a linear weighting function in Equation \ref{eq:x-1}). Based on the user assessment of the relative importance of various dimensions, they can directly influence the ranking results by assigning weights to the customized variable for each model, including semantic search, trained classifiers, and LLM classifiers. To further enhance the sensitivity to weight changes and make the system more responsive to user adjustments, we squared the output for each weighted score (see \(S_{s}\) in Equation \ref{eq:x-2}). 

With \(S_{l}\), changes in the score primarily reflect variations in \added{AI-predicted }probability, offering limited leverage over user weights. In contrast, \(S_{s}\) ensures that the rate of change in the ranking score increases alongside the user weight. Additionally, to satisfy that \(S_{s}\) changes more rapidly than \(S_{l}\) (i.e., \(\frac{\partial S_{s}}{\partial W_{i}} > \frac{\partial S_{l}}{\partial W_{i}}\)) and given that both \(P_{i}\) and \(W_{i}\) range from 0 to 1, \(W_{i}\) must fall within the range of \(\left(\frac{1}{2P_{i}}, 1\right]\), for \(1\ge P_{i}\geq\frac{1}{2}\). This probability range corresponds to scenarios where the model predicts a positive match for the checkworthy factor. Consequently, the squared weighting function increases sensitivity to user-assigned weights for positive predicted claims\added{ (i.e., AI predicts that claims satisfy this criterion)}. However, we acknowledge that this approach also risks disproportionately diminishing the influence of user weights for negative predicted claims, underscoring the need for further refinement.

\begin{equation}\label{eq:x-1}
    S_{l}=W_{1}P_{1}+W_{2}P_{2}+...+W_{i}P_{i}
    \quad 
    \frac{\partial S_{l}}{\partial W_{i}} = P_{i} \end{equation}
\begin{equation}\label{eq:x-2}
    S_{s}=W_{1}^{2}P_{1}^{2}+W_{2}^{2}P_{2}^{2}+...+W_{i}^{2}P_{i}^{2}  
        \quad
    \frac{\partial S_{s}}{\partial W_{i}} = 2W_{i}P_{i}^{2}
\end{equation}

We implemented our models on the Streamlit server and built the front-end\footnote{Streamlit: \url{https://streamlit.io/}, AG-Grid: \url{https://www.ag-grid.com/}} using Python, AG-Grid JS, and CSS. We used OpenAI's \texttt{gpt-3.5-turbo} as our LLM to create customized classifiers\footnote{While more advanced GPT models now exist, the log probability needed for building the ranking function was only available from \texttt{gpt-3.5-turbo} at the time of our implementation.}. Codes can be accessed here \footnote{Codes: \url{https://github.com/JialingJia/claim_prioritization}}.


\subsection{Evaluate via a Mixed-Method Approach} \label{subsec:evaluate and deploy}

Given our design prototype and study protocol, we proceeded to conduct a mixed-method evaluation with 16 professional fact-checkers. In this section, we describe our formal evaluation protocol, including the study procedure and tasks (Section \ref{subsubsec:study procedure}), participant recruitment (Section \ref{subsubsec:recruitment}), \added{baseline condition (Section \ref{subsubsec:baseline condition}), }and data collection and analysis (Section \ref{subsubsec:data collection and analysis}). 

\subsubsection{Study procedure and tasks} \label{subsubsec:study procedure}

\begin{figure}[ht]
    \centering
    \includegraphics[width=1\textwidth]{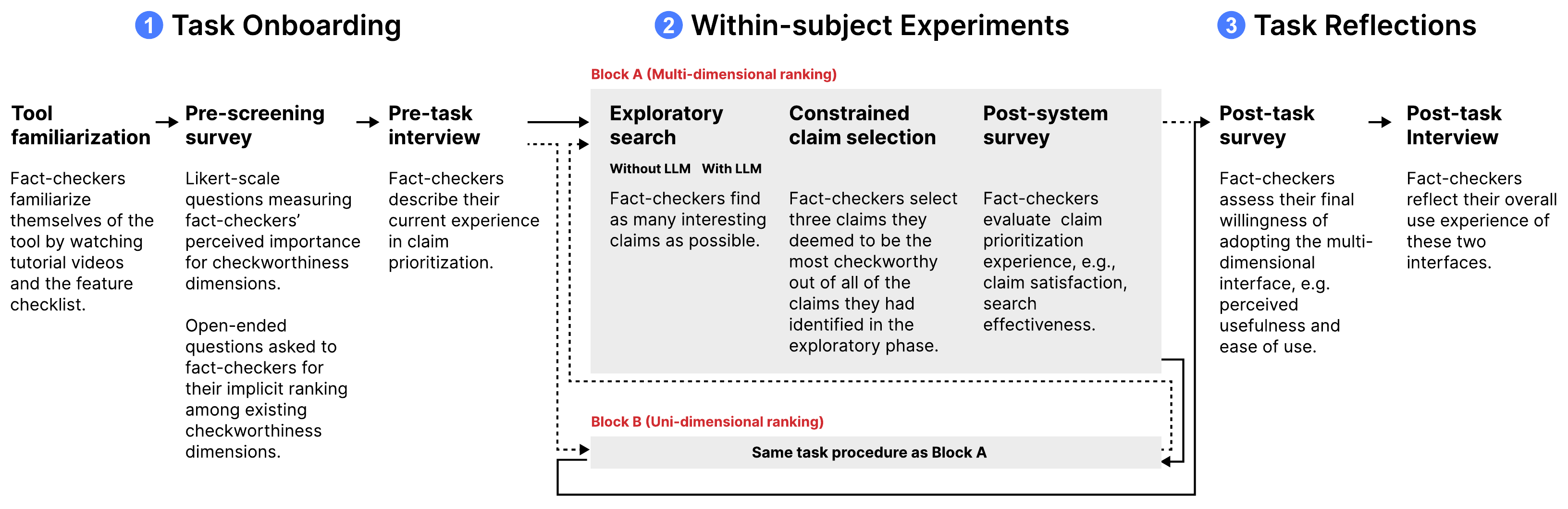}
    \caption{Flowchart of the within-subjects experimental procedure}
    \Description{Flowchart of the within-subjects experimental procedure}
    \label{fig:experimental procedure}
\end{figure}

Our evaluation protocol includes three phases, as described in Figure \ref{fig:experimental procedure}, including 1) task onboarding, 2) a within-subjects experiment, and 3) task reflection. 

During the onboarding phase, we first prepared a tutorial video and a checklist to help fact-checkers understand how to use each feature provided in the tool. 
%
Before the experiment phase, participants were asked to fill out a pre-screening survey to measure the perceived importance of the four checkworthiness \replaced{dimensions }{dimentions }(described in Section \ref{subsubsec:implementation}) with a pre-task interview to talk about their current claim prioritization experience.

Next, we scheduled an online meeting with each participant remotely to conduct the experimental study. The within-subjects experiment required participants to perform claim prioritization using two different interfaces: baseline (unidimensional claim) vs.\ treatment (multidimensional checkworthy claim ranking). 
%
Participants were randomly assigned to two groups (i.e., Block A or B), each using the two interfaces in a different order. 
Recall that our claims are drawn from \citet{alam-etal-2021-fighting-covid}'s COVID-19 dataset (Section~\ref{subsubsec:implementation}). For participant use, the dataset's test split was further bifurcated into two portions: one used during the familiarization phase and the other used in our actual experiment. We also used different sets of claims in each interface. While this creates a potential difference due to topical effects between baseline vs.\ treatment conditions, we believe this was justified by the benefit of preventing any learning or familiarization effects that could occur if we had used the same claims in both conditions. 

For the experiment, participants were asked to complete a series of claim prioritization tasks using both interfaces. The tasks involved two stages. In the initial exploratory search, participants identified as many interesting claims as possible, first without customized LLM filters and then after creating them. Then, in a constrained selection phase, participants selected three claims they deemed to be the most checkworthy out of all of the claims they had identified in the exploratory phase. They were then asked to complete a post-system survey to evaluate their claim prioritization satisfaction and search experience.

Finally, during the task reflection, participants were first required to assess their subjective use experience via a post-task survey. Unlike the post-system one, this survey aims to assess fact-checker final willingness for tool adoption in claim prioritization. They were then asked to recall their claim selection experience by looking at their final selected claims and describing reasons for selecting them. Additionally, we conducted a semi-constructed interview, asking them to reflect on their overall use experience by comparing the baseline and experimental interfaces. A more detailed task description of our claim prioritization tasks is presented in Appendix \ref{appendix:task description}. More details about our pre-/post- interview questions, subjective evaluation metrics in the post-system/task survey, and behavioral measurement across the three-stage experimental procedure are documented in Appendix \ref{appendix:measurement}.


\subsubsection{Baseline condition} \label{subsubsec:baseline condition}
In both the experimental and baseline conditions, participants could access features such as query search, user-generated facets powered by the LLM, and personalized weighting options. The key distinction, however, lies in the ranking method. The baseline condition employs a uni-dimensional checkworthiness ranking system trained on the overall ``Needs Verification'' label from \citet{alam-etal-2021-fighting-covid}'s dataset. This single checkworthiness facet, labeled ``checkworthy,'' is displayed in the left panel of the interface. Participants could adjust its weight using a slider, similar to how they could prioritize other shared factors like query similarity (i.e., topical relevance to the search query) and LLM-generated facets.

\begin{table}
\resizebox{\textwidth}{!}{%
\begin{tabular}{@{}p{0.4cm}p{1cm}p{2.3cm}p{2cm}p{1cm}p{2cm}p{3cm}@{}}
\toprule
ID & Gender & Region & Position & Years & Organizational Context & Language \linebreak Fact-checked \\ \midrule
  1 & Male & United States & Fact-checker & 4 & Media & English \& Spanish\\
  2 & Female & South Africa & Fact-checker & 2.5 & Independent & English \& Afrikaans\\
  3 & Male & United Kingdom & Fact-checker & 5.5 & Media & English\\ 
  4 & Female & Australia & Fact-checker  & 2.5 & Media & English\\ 
  5 & Male & United States & Fact-checker & 16 & Independent & English\\ 
  6 & Female & United States & Researcher & 17 & Independent & English\\ 
  7 & Male & South Africa & Fact-checker & 4 & Independent & English \& Afrikaans\\
  8 & Male & India & Fact-checker & 2 & Independent & English \& Hindi\\ 
  9 & Male & Nepal & Fact-checker & 2.5 & Independent & English \& Nepali\\
  10 & Female & India & Fact-checker & 5 & Independent & English \& Hindi\\
  11 &  Male  & India  & Fact-checker & 1.5 & Independent & English \& Hindi\\ 
  12 & Female & United States & Fact-checker & 2.5 & Media & English\\ 
  13 & Female & United States & Fact-checker & 2 & Independent & English\\ 
  14 & Female & United States & Fact-checker & 4 & Media & English\\ 
  15 & Male & United States & Researcher & 2 & Media & English\\
  16 & Male & Nepal & Fact-checker & 4.5 & Independent & English \& Nepali\\ 
  \bottomrule
\end{tabular}
}
\vspace{0.2cm}
\caption{Participant Demographics}
\label{tab:participants}
\end{table}

\subsubsection{Participant recruitment} \label{subsubsec:recruitment}

We screened global fact-checking websites listed in the Duke Reporters' Lab\footnote{Duke Reporters' Lab: \url{https://reporterslab.org/locations/}} and sent recruitment emails to various organizations. To ensure relevant expertise in using claim-monitoring tools, we specifically targeted organizations that mentioned their partnerships with Meta or other technical fact-checking entities, such as Meedan or Full Fact. Fact-checkers in these organizations are more likely to have experience using claim-monitoring tools provided by their partnered tech companies. We recruited 16 participants, including 14 full-time professional journalists who conduct fact-checks and 2 researchers affiliated with the organization with previous experience as fact-checkers (Table \ref{tab:participants}). All participants took part in our study remotely via Zoom. Each session lasted approximately 1.5 hours. Participants received Amazon Gift Cards or other redeemable options available in their respective countries as compensation.

We intentionally sampled participants who had experience checking social media claims during the recruitment process. We believe these participants help provide valuable insights as they are more familiar with different search-related tools to look for claims. From the pre-screening survey, all of our participants have used social media monitoring tools such as CrowdTangle, TweetDeck, Trendolizer, Newswhip, or others. Eleven of them also used tools specifically developed for fact-checking, such as the Meta fact-checking tool, Full Fact Alpha, Meedan Check, or others. Additionally, eight participants reported they always found claims using either social media monitoring or fact-checking tools. Another eight participants said they frequently found claims from these tools. Regarding how many claims they finally checked were from these tools, two participants mentioned almost all claims they checked come from these tools; eight participants reported a very large portion (60\% to 90\%); 
and six participants stated a fair amount (40\% to 60\%). 

\subsubsection{Data collection and analysis} \label{subsubsec:data collection and analysis}

Data were collected throughout the three-phase evaluation protocol from multiple sources: the pre-screening survey, user interaction logs, post-task / post-system questionnaires, and recordings from retrospective think-aloud and semi-constructed interviews. We detail our measurement and data collection procedure in Appendix \ref{appendix:measurement}. This section describes how we conducted mixed-method analyses to achieve our two research goals (RG1 and RG2) and associated RQ defined in Section \ref{subsec:frame}. 

To guide our data analysis, we define two research questions for each goal:
\begin{itemize}
    \item [\bf RG1] \textbf{A practice-based examination of fact-checker practice and needs for claim prioritization}
    \begin{itemize}
        \item [] \textbf{RQ1.1} How did participants assess the relative importance of the four checkworthy dimensions?
        \item [] \textbf{RQ1.2} How did participants apply different priorities among the four dimensions in claim selection?
    \end{itemize}
    \item  [\bf RG2] \textbf{An evaluation of fact-checker use experiences for the claim prioritization prototype}
    \begin{itemize}
        \item [] \textbf{RQ2.1} How did participants create customized LLM filters, and what were the benefits and limitations?
        \item [] \textbf{RQ2.2} What were overall user experiences with our prototype (e.g., usage behaviors, efficacy of claim selection, and subjective feedback)?
    \end{itemize}
\end{itemize}

\textbf{RQ1.1:} To investigate the relative importance of the four checkworthy dimensions, we compared what participants said in the pre-screening survey (i.e., self-assessment data) vs.\ their actions using the prototype (i.e., user interaction logs). The self-assessment included three 5-point Likert scale questions to evaluate each dimension <X>: 
\begin{itemize}
    \item \textbf{Perceived importance}: ``<X> is an important factor resulting in the final fact-checked claim.''
    \item \textbf{Ease of finding}: ``It is easy for me to identify <X> claims.''
    \item \textbf{Criterion accuracy}: ``Claims that I finally checked are usually <X> as they first appeared'', i.e., how accurately could participants predict whether a claim would ultimately satisfy dimension <X> prior to conducting the fact-check?
\end{itemize}
Complementing \textit{Perceived importance}, we further asked participants to implicitly rank the relative importance of the four dimensions. Appendix \ref{sec:implicit_ranking} compares \textit{Perceived importance} ratings vs.\ this implicit ranking of dimension importance. 

Regarding the other two rating questions, while \textit{Ease of finding} asks how easy it is to identify claims satisfying a given checkworthy dimension, \textit{Criterion accuracy} asks how accurately fact-checkers can predict whether a claim would satisfy the dimension prior to the fact-check. In other words, while both ask fact-checkers about assessing checkworthy dimensions, \textit{Ease of finding} gets at initial impressions, whereas \textit{Criterion accuracy} focuses on ultimate determinations. 

Quantitative, observational statistics drawn from the user interaction logs include the following:
\begin{itemize}
    \item \textbf{Weight at selection}: UI slider weights assigned to each checkworthy dimension at the time of claim selection.
    \item \textbf{Overall weight}: slider weights assigned to each dimension at all time points sampled throughout the task. 
    \item \textbf{Use frequency}: The number of times participants adjusted each of the checkworthy dimension sliders.
\end{itemize}
Whereas the unidimensional baseline UI had only a single slider (beyond the query similarity slider), the multidimensional UI had a slider for each of the four checkworthy dimensions. 
\textbf{We present findings for RQ1.1 in Section \ref{subsec:finding-multi-dimensional}}.

\textbf{RQ1.2:} We adopted a similar mixed-method evaluation approach to understand how participants apply different priorities to triage claims. First, we asked participants to describe how they would filter and select claims before using the multidimensional interface. We then mapped out different usage behaviors onto a step-series diagram. These usage behaviors include searching, changing checkworthy sliders, making claim selections, etc. Additionally, we asked them to think aloud their actions retrospectively after using the tool. By analyzing participant qualitative reflections and cross-referencing these reflections with the step-series diagrams, we analyzed if any systematic processes or behavior patterns participants developed to conduct claim selection efficiently. \textbf{Findings for RQ1.2 are in Section \ref{subsec:finding-hierachy}.}

\textbf{RQ2.1:} To understand how participants create customized filters, we performed content analysis over LLM prompts written and thematic analysis of qualitative reflections. We first extracted prompts written by participants from the user interaction logs, which covered both conditions when they used the unidimensional interface and the multidimensional one. During the post-task interview, we asked participants about the benefits and limitations of using LLMs to create customized filters. By comparing the written prompts with their qualitative reflections, we identified if any written patterns exist and how they relate to the user intents of claim triage. \textbf{We present findings for RQ2.1 in Section \ref{subsec:finding-prompting}}.

\textbf{RQ2.2:} To assess overall user experience, we 
combine quantitative data analysis on observational data from user interaction logs with a post-task questionnaire. {\bf Findings for RQ2.2 are presented in Section \ref{subsec:finding-user experience}}.

Interaction logs were used to identify usage behaviors and effectiveness of claim selection and recorded for both unidimensional and multidimensional interfaces.  We compare the following metrics:
\begin{itemize}
    \item \textbf{\# Queries}: The number of queries submitted by the participant.
    \item \textbf{\# Checkworthy slider changes}: The number of times the checkworthy slider(s) were changed.
    \item \textbf{\# Query similarity slider changes}: The number of times the query similarity slider was changed.
    \item \textbf{\# Selected claims}: The number of interesting claims identified in the initial exploratory stage (with or without using customized filters).
    \item \textbf{\# Final claims found checkworthy}: Out of the three final claims selected, the number of these that were initially found with or without customized filters. 
    \item \textbf{Conversion rate}: the ratio \textit{\# Final claims found checkworthy / \# Selected claims} 
    
\end{itemize}

The post-task questionnaires were designed based on \citet{Marchionini2006-wk}'s three types of search activities for exploratory search: how participants \textit{Learn}, \textit{Lookup}, and \textit{Investigate} claims. Also, the post-system questionnaire was conducted to collect participant \textit{Perceived usefulness} and \textit{Ease of use} of the tool. We then conducted a thematic analysis of the post-task interview recordings. By cross-validating these quantitative data with participant qualitative reflections, we reported whether participants preferred using a unidimensional or multidimensional interface. 

\section{Findings} \label{sec:findings}
We now address the four research questions from the previous Section (\ref{subsubsec:data collection and analysis}), using measures and statistics defined there. First, how did participants assess the relative importance of the four checkworthy dimensions? (RQ1.1, Section \ref{subsec:finding-multi-dimensional}). Second, how did participants apply different priorities among the four dimensions in claim selection? (RQ1.2, Section \ref{subsec:finding-hierachy}). Third, how did participants create customized LLM filters, and what were the benefits and limitations? (RQ2.1, Section \ref{subsec:finding-prompting}). Finally, what were overall user experiences with our prototype (RQ2.2, Section \ref{subsec:finding-user experience}).  

\subsection{Fact-checker Perceptions and Priorities of Multidimensional Checkworthiness} \label{subsec:finding-multi-dimensional}

We assess how fact-checkers evaluate the four-dimensional checkworthiness from descriptive statistics and significance testing (RQ1.1). We organize the quantitative results based on the different measurements (Section \ref{subsubsec:data collection and analysis}) and then comparatively analyze different results. We also discuss reasons fact-checkers report for their different priorities. 

Given the relatively small scale of data in our user study, we primarily conducted non-parametric tests. The Friedman test was used to evaluate differences in participant quantitative data across four-dimensional checkworthiness. Subsequent pairwise Wilcoxon signed-rank tests were conducted if the Friedman test first found that at least two checkworthy dimensions showed significant differences vs.\ one another (see post-hoc results in Appendix \ref{appendix:post-hoc test}).

\subsubsection{Perceived importance} As shown in Table \ref{tab:dimension-statistics}, ``Likely harmful'' had mean average rating of (\(M = 4.81\)). The score decreased from ``Likely false'' (\(M_{false} = 4.63\)) to ``Interest to the public,'' (\(M_{public-interest} = 4.50\)) and ``Verifiable'' (\(M_{verifiable} = 4.44\)). The median scores were the same for each dimension (\(Median = 5\)). ``Verifiable'' received the lowest average rating but had the largest standard deviation (\(SD = 1.21\)), indicating the greatest variation in opinions among our participants. No significant differences were found across these four dimensions (\(X^2 = 1.824\), \(p > 0.05\)). 

\begin{table}[ht]
\resizebox{\textwidth}{!}{%
\begin{tabular}{@{}>{\raggedright}p{3.4cm}llllllll>{\raggedleft\arraybackslash}p{1.8cm}@{}} 
\toprule
\multirow{3}{*}{Measures} & \multicolumn{8}{c}{\textbf{Four dimensions of checkworthiness}} &\multirow{3}{*}{Friedman \linebreak \(X^2\)}\\
& \multicolumn{2}{c}{Verifiable} & \multicolumn{2}{c}{Likely false} & \multicolumn{2}{c}{Likely harmful} & \multicolumn{2}{c}{Interest to public} &  \\ 
& \textit{M(SD)} & \textit{Median} & \textit{M(SD)} & \textit{Median} & \textit{M(SD)} & \textit{Median} & \textit{M(SD)} & \textit{Median} &
\\\midrule
& \multicolumn{8}{c}{\textbf{Self-assessment Ratings}} & \\
Perceived importance &  4.44(1.21) & 5 & 4.63(0.62) & 5 & 4.81(0.40) & 5 & 4.50 (0.82) & 5 &  1.82(0.60) \\
Ease of finding &  4.13(0.96) & 4 & 3.69(1.08) & 4 & 4.31(0.87) & 4.5 & 4.25(1.00) & 4.5 &  \textbf{11.73(0.00)} \\
Criterion accuracy &  4.13(1.06) & 4 & 3.87(0.74) & 4 & 4.20(0.77) & 4 & 4.07(0.70) & 4 &  1.87(0.59) \\ \midrule
& \multicolumn{8}{c}{\textbf{Observational Data}} & \\
Weight at selection &  0.79(0.27) & 0.93 & 0.75(0.28) & 0.82 & 0.64(0.36) & 0.69 & 0.45(0.41) & 0.29 &  2.37(0.30) \\
Overall weight &  0.77(0.26) & 0.83 & 0.73(0.28) & 0.81 & 0.62(0.31) & 0.69 & 0.39(0.35) & 0.26 &  \textbf{0.98(0.04)} \\
Use frequency &  2.69(1.96) & 2.50 & 2.25(1.24) & 2.00 & 3.06(3.49) & 1.00 & 2.13(2.75) & 1.50 &  6.15(0.61) \\ \bottomrule
\end{tabular}
}
\vspace{0.2cm}
\caption{Mean (M), standard deviation (SD), and median statistics over participant data for different measures (rows, Section \ref{subsubsec:data collection and analysis}) and checkworthy dimensions (columns). Self-assessment data (top 3 rows) come from 5-point Likert scale answers, while observational data (bottom 3 rows) is drawn from interaction logs for the different UI sliders. Bold results indicate at least two checkworthy dimensions showed statistically significant differences (Friedman test at \textit{p} < 0.05). We observe significant differences between checkworthy dimensions for ``Ease of finding'' and ``Overall weight'' measures. Post-hoc test results are presented in Appendix \ref{appendix:post-hoc test}.}
\label{tab:dimension-statistics}
\end{table}

\subsubsection{Ease of finding} \deleted{Recall that \textit{Ease of finding} gets at initial impressions in assessing a checkworthy dimension whereas \textit{Criterion accuracy} focuses on final assessments post-check (Section \ref{subsubsec:data collection and analysis}). }Participants generally agreed that ``Verifiable'' and ``Likely false'' claims were more difficult to initially identify (\(M_{verifiable} = 4.13\), \(Mdn_{verifiable} = 4\), \(M_{false} = 3.69\), \(Mdn_{false} = 4\)). Significant differences were found in the Friedman test (\(X^2 = 11.735\), \(p < 0.05\)). From the post-hoc test (see Table \ref{tab:post-hoc ease of findings} in Appendix \ref{appendix:post-hoc test}), ``Likely false'' was rated significantly lower than ``Likely harmful'' and ``Interest to the public'' but had no significant difference from ``Verifiable.'' \deleted{Regarding \textit{Criterion accuracy}, ``Likely false'' also received the lowest rating (\(M = 3.89\)), but no significant difference was observed (\(X^2 = 1.878\), \(p > 0.05\)). These results suggest that among the four dimensions, ``Likely false'' claims may be the most challenging to find, not only initially, but that initial impressions of likely false claims may also prove incorrect after conducting the fact-check.}

\subsubsection{Criterion accuracy} \added{Recall that \textit{Ease of finding} gets at initial impressions in assessing a checkworthy dimension whereas \textit{Criterion accuracy} focuses on final assessments post-check (Section \ref{subsubsec:data collection and analysis}). In this final assessment, ``Likely false'' also received the lowest rating (\(M = 3.89\)), but no significant difference was observed (\(X^2 = 1.878\), \(p > 0.05\)). These results suggest that among the four dimensions, ``Likely false'' claims may be the most challenging to find, not only initially, but that initial impressions of likely false claims may also prove incorrect after conducting the fact-check.}

\subsubsection{Weight at selection} Slider weight for ``Verifiable'' had the highest mean weight at claim selection (\(M = 0.79\)). Lower mean weights at claim selection were seen for ``Likely false'' (\(M_{false} = 0.75\)), ``Likely harmful'' (\(M_{harmful} = 0.64\)), and ``Interest to the public'' (\(M_{public-interest} = 0.45\)). No significant differences were found (\(X^2 = 2.375\), \(p > 0.05\)). 

\subsubsection{Overall weight} For slider weights across all times during the task, a similar pattern with significant differences (\(X^2 = 6.156\), \(p < 0.05\)) is observed. Further tests (see Table \ref{tab:post-hoc overall weights} in Appendix \ref{appendix:post-hoc test}) show that the weight of ``Verifiable'' was only significantly higher than ``Interest to the public,'' which was significantly lower than the other dimensions. 

\subsubsection{Use frequency} Regarding the number of times participants adjusted each dimension's slider weight, participants used ``Likely harmful'' more frequently (\(M = 3.06\)) than other dimensions. Frequency of usage decreased with ``Verifiable'' (\(M_{verifiable} = 2.69\)), ``Likely false'' (\(M_{false} = 2.25\)), and ``Interest to the public'' (\(M_{public-interest} = 2.13\)). However, no significant differences were found across dimensions for \textit{Use frequency} (\(X^2 = 0.980\), \(p > 0.05\)).

\subsubsection{Comparison and conclusions} Self-assessments did not show a high variation across participant perceptions of the importance of the four dimensions. However, user interaction logs indicated that ``Verifiable'' and ``Likely harmful'' were considered relatively more important as these two dimensions either received the highest average weight or were used most frequently. In contrast, ``Interest to the public'' was not deemed important or used often, as evidenced by comparing participant self-assessment with their actual behaviors. These results confirm that our participants have different priorities over the checkworthy dimensions. 


\subsubsection{Why do fact-checkers have different priorities?}
One reason is that fact-checking organizations have different priorities. For example, some participants (P4, 5, and 13) mentioned that their fact-checking organizations primarily check political claims. Another important reason reported is the changing news environment: as news events develop, priorities over checkworthiness dimensions also evolve. Such dynamism is further discussed in Section \ref{subsubsec:hierarchy}. 

As the name of the profession indicates, fact-checkers check facts, so whether a claim is ``Verifiable'' is clearly at the heart of fact-checking. For example, in their training sessions, the first lesson fact-checkers often reported learning is to identify ``Verifiable'' facts, such as numerical assertions. However, some participants (P9, 10, 16) viewed ``Verifiable'' as the least important criterion (see Table \ref{tab:dimension-responses} in Appendix \ref{sec:implicit_ranking}) among the four checkworthy dimensions, and more participants consider ``Likely harmful'' as the most or equally important. One might assume fact-checkers prioritizing potential harm ahead of verifiability do so simply as a logistical matter, e.g., it being faster or easier to first consider one before the other. However, we were surprised that three of our participants from India (P8, 10, 11) seemed to articulate the fact-checking enterprise to have a broader social responsibility beyond fact-checking, including preventing or mitigating harms unrelated to factuality. Given an opinionated claim that was not verified, one might try to balance it against other relevant opinions to help stave off civil unrest and violence. We discuss this emphasis on harm over verifiability further in both the next Section \ref{subsec:finding-hierachy} and in later discussion in Section \ref{subsec:tensions}.





\subsection{Fact-checker Hierarchical Approach for Claim Prioritization} \label{subsec:finding-hierachy}

How did participants apply different priorities among the four dimensions in claim selection (RQ1.2)?
\citet{Sehat2023-xa} reported the absence of any systematic approach to claim prioritization based on their interviews with fact-checkers, particularly in prioritizing harmful claims. In contrast,   
our participants appeared to develop an inherent hierarchical approach to filtering and selecting claims according to the relative importance of different dimensions. This difference likely stems from our study's inclusion of observation beyond participant self-reporting. In particular, our understanding arose from two distinct sources of evidence: 1) qualitative responses describing how participants believed they would filter and select claims using the four checkworthy dimensions, and 2) observational data showing how participants actually selected claims through an iterative process that involved applying different checkworthy filters. 

\begin{figure}[ht]
    \centering
    \includegraphics[width=0.95\textwidth]{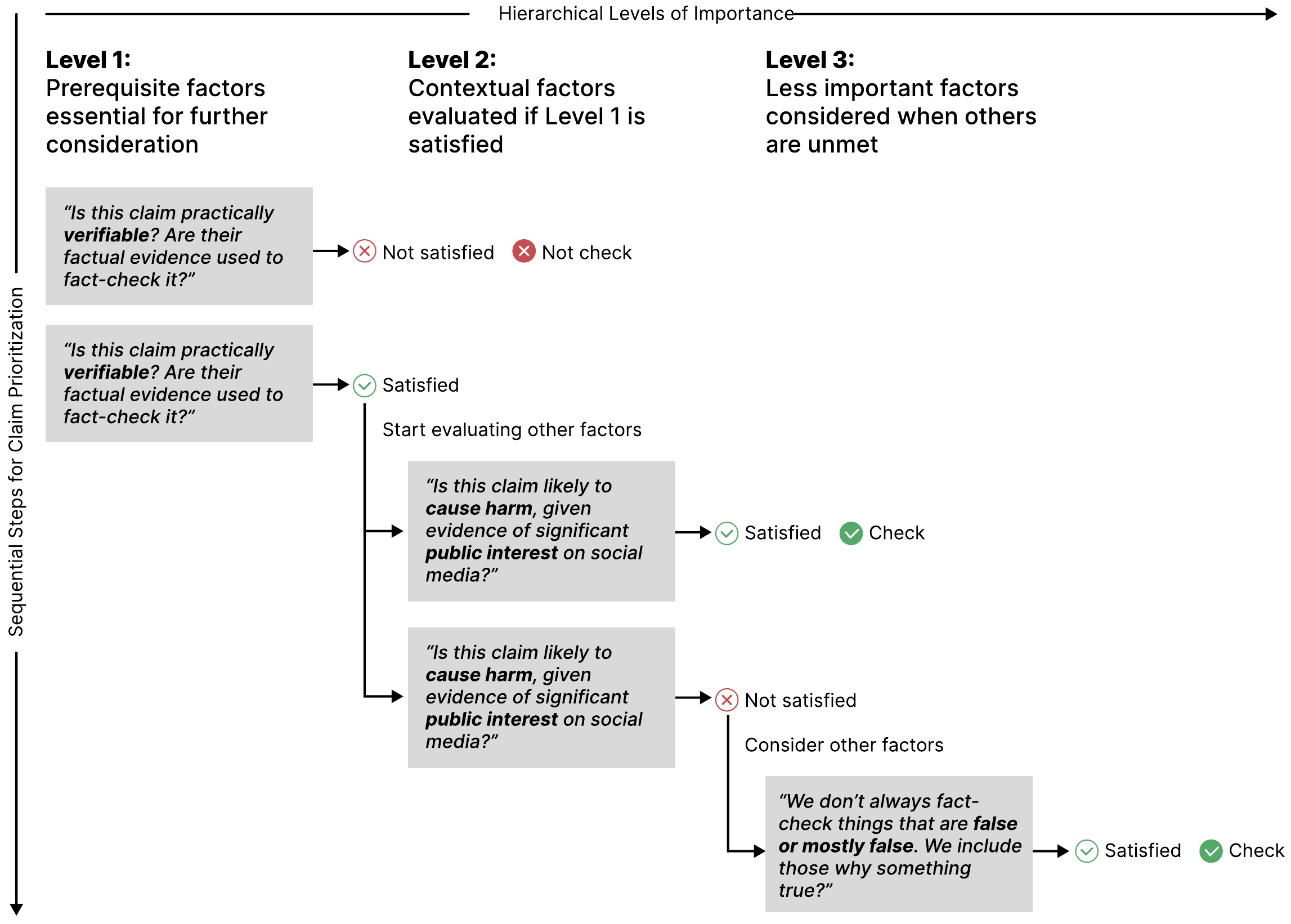}
    \caption{Claim prioritization strategy. Participants follow a sequential process to evaluate various factors for fact-checking. Level one includes prerequisite factors that must be satisfied before considering other factors. Levels two and three involve additional factors, but they differ in their importance. If the more important factors are not met, fact-checkers then evaluate less important factors.}
    \Description{A visual representation of a three-level hierarchical prioritization strategy.}
    \label{fig:hierarchial procedure}
\end{figure}


Our analysis suggests a three-level hierarchy (Figure \ref{fig:hierarchial procedure}). First, claim assessment begins with filtering based on \textit{prerequisite} dimensions: mandatory criteria that must be met. 
Next, participants assessed other criteria we refer to as \textit{important but more contextual}. Finally, fact-checkers may also consider other, \textit{less important} criteria as time allows. 
The most important dimensions (and time) are thus prioritized while remaining open to other dimensions when possible. To support this finding, we present qualitative responses and quantitative data detailing this hierarchical process.

\subsubsection{Prerequisite dimensions.} Some participants (P2, 3, 4, 5, 13, 14) reported that they would first find a set of claims that are ``Verifiable'' because this dimension is the \textit{prerequisite} criterion for fact-checking: \begin{displayquote}
    ``\textit{Verifiability is non-negotiable, so that comes first, public interest and harm are looked at together, and weighed up against each other, as they interact with one another.}'' (P2) ``\textit{Of course, when you're doing an organic search, you are looking for something that's verifiable. But then, when you find something that's verifiable, I guess those other factors come through.}'' (P3)
\end{displayquote}

As noted in the previous Section \ref{subsec:finding-multi-dimensional}, we were surprised to learn that three participants from India (P8, 10, 11) viewed ``Likely harmful'' as the prerequisite condition. Potential regional differences are discussed in Section \ref{subsec:tensions}. 
\begin{displayquote}
    ``\textit{So when you're checking a claim, when something is really harmful, even if it is not a direct fact check, we will try to balance it with many relevant opinions. That opinion can respond and further trigger a domino effect of misinformation}'' (P8) ``\textit{If a claim is harmful, I think that should be given more attention than the verifiable one. As fact-checkers, we can move to a verifiable claim [easily] if it takes us less time.}'' (P10)
\end{displayquote}

\subsubsection{Important but contextual dimensions.} 
Next, participants started considering secondary, \textit{important but contextual} dimensions. 
Those who initially looked at ``Verifiable'' claims 
tended to then consider ``Likely harmful'', but preferred to assess it alongside ``Interest to the public.'' If a harmful claim stood out, they used public interest as a benchmark to decide whether it warranted further investigation. Those who first prioritized ``Likely harmful'' also used ``Interest to the public" as a secondary gauge. Considering both together was thought to prevent misinformation amplification (P2, 13):
\begin{displayquote}
    ``\textit{If a claim has the potential to cause harm, but isn't very interesting to the public, then publishing a fact-check on it might just platform the claim and give it more fuel."} (P2) 
\end{displayquote}
However, both dimensions are very subjective and contextual. For example, participants pointed out that public interest might not be easily measured. 
Social media metrics could be used to understand how viral a claim is across platforms, then project the range of impact on public interest, 
but they also often relied on their intuition: 
\begin{displayquote}
    ``\textit{If we see a very harmful claim but not viral yet, I will beg to differ here... If that claim is reaching me on the WhatsApp helpline, that means it's viral in some sense. If I'm just typing keywords on Facebook or Twitter and I hardly find few posts on the pool, that means it's not viral on these platforms. If a claim is not viral anywhere, we are just discussing it internally in the newsroom; I would still take an editorial decision to fact-check it. If we are discussing it, it is also somewhere else being discussed.}'' (P10)
\end{displayquote}
Only a few participants (P3, 6, 14) mentioned that they would consider ``Likely false'' after first filtering ``Verifiable'' claims due to their organizational partnerships with social media platforms, which prioritize addressing false claims.

\subsubsection{Less important dimensions.} Participants occasionally checked claims that only partially satisfied the checkworthy dimensions in our study. 
These decisions tended to be driven by personal curiosity or journalistic intuition:
\begin{displayquote}
    ``\textit{We don't always fact-check things that are false or mostly false. Sometimes, we fact-check things that are half true or mostly true. We include those because sometimes we are curious: why is something true? I think that goes a bit like public interest; if an average person is curious about this topic, would they want to learn more about it? So we write those fact-checks to explain why something might be true and give more context.}'' (P14)
\end{displayquote}

\subsubsection{Comparing observational data with self-reporting.} 
By tracking participant slider weights used for each checkworthy dimension, we observed a pattern of slider weight usage consistent with the prioritization hierarchy posited above. First, for the \textit{prerequisite} dimensions, participants would assign weights that consistently remained higher than other dimensions. In addition, for \textit{important but contextual} dimensions, participants would change weights dynamically and frequently throughout multiple rounds of selecting claims. Finally, for \textit{less important} dimensions, participants would assign smaller weights or ignore these dimensions completely. 

To visualize our log data and the weighting patterns described above, we map each participant's actions onto a step-series diagram over the initial exploratory claim-finding task. Each diagram shows that participant slider weight changes for each of the four checkworthy dimensions natively supported, as well as a fifth checkworthy dimension once they created a custom LLM search filter for it. We show diagrams for all participants in Appendix \ref{appendix:weighting pattern} but include two illustrative examples below.  These examples highlight the hierarchical approach for claim prioritization we observed, and we accompany each diagram with explanations from each participant's retrospective think-aloud.

\begin{figure}[ht]
    \centering
    \includegraphics[width=1\textwidth]{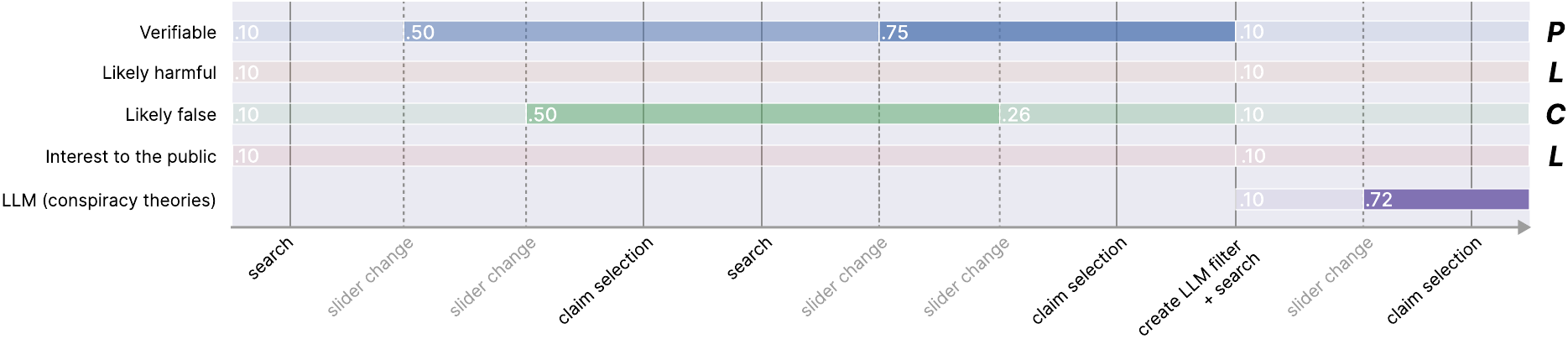}
    \caption{Slider weight changes by participant P7 during an exploratory search are mapped onto a step-series diagram. The X-axis shows key events during the task. The left Y-axis lists checkworthy dimensions: the four standard dimensions above a fifth LLM-customized dimension created by P7. For each dimension, horizontal bars show slider weight over time for that dimension, where bar opacity varies from transparent to opaque as slider weight varies in the range [0,1]. Slider weights are initialized to 0.10 and reset to this value when a customized LLM filter is created. The right Y-axis shows our manual analysis of the hierarchical importance of each dimension (Section \ref{subsec:finding-hierachy}), where \textit{\textbf{P}} denotes a \textit{prerequisite} checkworthy dimension, \textit{\textbf{C}} denotes \textit{important but contextual}, and \textit{\textbf{L}} denotes \textit{less important}.}
    \label{fig:Time-series-diagram P7}
\end{figure}

\textit{\textbf{Example 1}--Prerequisite dimension: ``Verifiable'' $\bullet$ Important but contextual dimension: ``Likely false'' $\bullet$  Less important dimensions: ``Likely harmful'' and ``Interest to the public''.} Figure \ref{fig:Time-series-diagram P7} presents participant P7's diagram. P7 first entered a search query, then adjusted the ``Verifiable'' and ``Likely false'' sliders to 0.5 each. After selecting his first claim, P7 then modified the search query and changed the slider values, increasing the ``Verifiable'' slider to 0.75 and reducing the ``Likely false'' slider to 0.26. 
In his retrospective think-aloud, he shared that he strongly preferred ``Verifiable'' in his searches: ``\textit{I didn't really see a reason to turn it down or use it less}.'' He viewed the other dimensions as more context-dependent, saying that ``\textit{The others, I felt were more situational. I only felt I needed to tweak them when I had done a search or when I was unsatisfied with the results}.'' This explains why he decreased the weight of "Likely false." 

\begin{figure}[ht]
    \centering
    \includegraphics[width=1\textwidth]{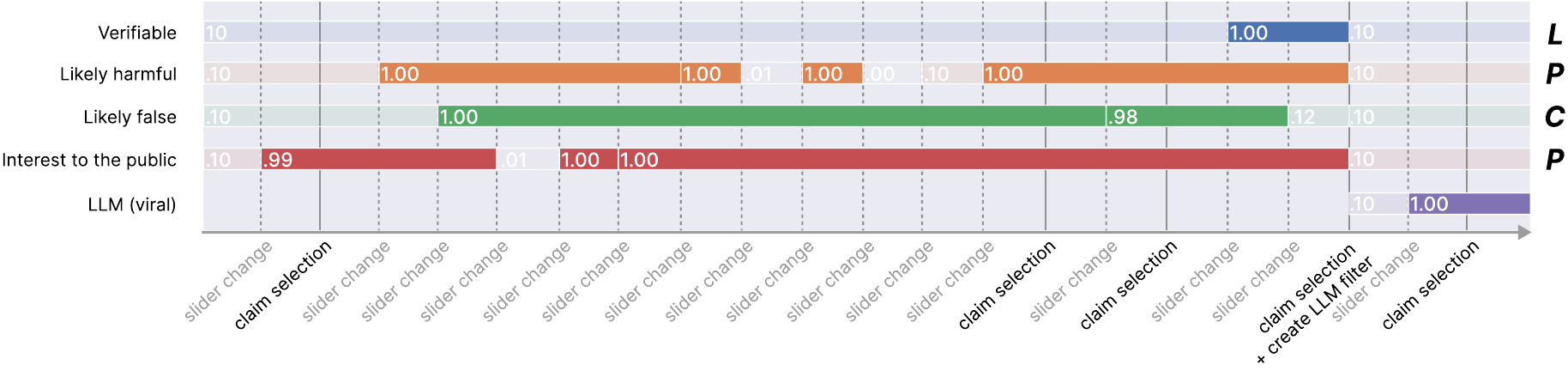}
    \caption{Participant P10's step-series diagram. See Figure \ref{fig:Time-series-diagram P7}'s caption for figure interpretation.}
    \label{fig:Time-series-diagram P10}
\end{figure}

\textit{\textbf{Example 2}--Prerequisite dimensions: ``Interest to the public'' and ``Likely harmful'' $\bullet$ Important but contextual dimension: ``Likely false'' $\bullet$ Less important dimension: ``Verifiable''.} Participant P10 had completely different priorities over the four checkworthy dimensions but exhibited a similar weighting pattern based on the three-level hierarchy (Figure \ref{fig:Time-series-diagram P10}). He first set the ``Interest to public'' slider to 0.99 and quickly selected his first claim. He then adjusted the ``Likely harmful'' and ``Likely false'' sliders to a maximum value of 1 (100\%). He briefly turned off the ``Interest to public'' slider but soon reactivated it and kept it to the maximum value consistently over time. He did the same for the ``Likely harmful'' slider. Following these adjustments, he selected the second and third times. Finally, he deactivated the ``Likely false'' slider, set the ``Verifiable'' slider to its maximum value of 1, then selected again.
In his retrospective think-aloud, P10 said that a claim with potential harm and relevance to public interest should take precedence over its verifiability: ``\textit{I can keep some [sliders] aside and look at something that will cause more harm than a verifiable claim.}'' As indicated in the step-series diagram, he kept the ``Likely harmful'' and ``Interest to the public'' sliders at their maximum weights when selecting claims. He occasionally used the ``Likely false'' slider 
and used the ``Verifiable'' slider only once.

We summarize all participant diagrams in Appendix \ref{appendix:weighting pattern}. Eight participant weighting patterns revealed a complete and clear three-level hierarchy (P2-4, P7, P10, and P14-16), while six participants only partially demonstrated a two-level hierarchy (P1, P5-6, P8, and P11-12). This might be due to limited interaction data (e.g., use frequency and changes in slider weights) or insufficient qualitative reflections from participants for us to validate each hierarchical level. With these participants, it seemed more difficult to distinguish between \textit{important but contextual} and \textit{less important} levels. The behavior of two participants (P9, P13) did not reflect any hierarchy. P9's weighting behavior was inconsistent with his reflections, and P13 did not use any checkworthy filter. 


\subsection{Targeted versus Abstract Prompting for Defining New Checkworthy Dimensions} \label{subsec:finding-prompting}

RQ2.1 explores how participants created customized LLM filters and the corresponding benefits and limitations. 
Customizable, user-generated filters enable users to filter for additional checkworthiness dimensions when keyword search and natively-supported checkworthy filters prove insufficient. When participants write LLM prompts in natural language, we can also explicitly understand what potential needs they have and how they express these needs. 

Prompts written by participants ranged from specific to abstract, including 1) \textit{targeted} prompts to retrieve claims with specific topics or particular types of claims and 2) \textit{abstract} prompts as benchmark relevance criteria for claim exploration. We categorize these prompts according to different user intents (Table \ref{tab:examples of customized dimensions}), which \replaced{helps }{help }us understand the reasons participants created them. These intents also emphasize both specific and general fact-checking information needs, given their familiarity with the fact-checking topics. 
We now proceed to describe how participants used LLMs to write targeted and abstract prompts. 

\begin{table}[ht]
\resizebox{\textwidth}{!}{%
\begin{tabular}{@{}>{\raggedright}p{3cm}>{\raggedright}p{2cm}>{\raggedright\arraybackslash}p{8.5cm}@{}} \toprule
User Intent  & Prompt name & Prompt text  \\ \midrule
\multirow{3}{3cm}{Search claims containing multiple queries or narratives} & VAERS  & I want to identify claims that mention the Vaccine Adverse Event Reporting System, COVID-19 vaccine-related deaths, and COVID-19 vaccine-related adverse events and reactions.  \\
& Covid and vaccine deaths & Identify claims that appear to cite different figures regarding COVID, including COVID deaths, deaths associated with COVID vaccines, and COVID cases. \\
& Vaccine denying & Claims that are in denial of vaccine efficacy, where users cite bogus reports, data, or exaggerated personal experiences to claim that vaccines are either useless or cause more harm than good. \\ \midrule
\multirow{3}{3cm}{Filter claims based on different claim attributes} & Opinions & Claims aim to confuse the public about the efficacy of prevention measures. \\
& Statistics & Claims made about numbers or percentages. \\
& Quotes & Claims that quote famous personalities regarding the data related to COVID-19. \\ \midrule
\multirow{2}{3cm}{Filter claims by multidimensional relevance} & Likelihood to spread & I want to identify claims that are likely to have a far-reaching spread. Claims that are extreme and are likely to create fear and panic have the potential to reach a wider audience. \\
& Chaos  & The content is aimed to cause chaos among the population. \\
& Public interest & Claim that it is important for the public health and its implication. \\ \bottomrule
\end{tabular}
}
\vspace{0.2cm}
\caption{Prompt examples written by participants, including how they named their prompts in our interface and their actual input prompt texts. We organize these examples according to different user intents.}
\label{tab:examples of customized dimensions}
\end{table}


\subsubsection{Targeted prompts.} This type of prompts represents fact-checker precise information need for topics they are familiar with. For example, in Table \ref{tab:examples of customized dimensions}, the prompt named ``VAERS'', is written by P13. The participant said:
\begin{displayquote}
    ``\textit{I've done a lot of reporting on this [VAERS] and I'm familiar with the language, that's like a huge subsection of COVID-19 claims... prompt like this helps me locate them}.'' (P13)
\end{displayquote}  
Participants explained that when using LLM filters, they combined multiple topically relevant keywords within a long-context window. In contrast, when using topical search, they typically entered each keyword separately. For example, the prompt description of `VAERS' consists of three phrases (see Table \ref{tab:examples of customized dimensions}). While these phrases are closely related, each conveys subtle differences in meaning. Participants noted that including multiple relevant keywords in the LLM prompt helped them save time. P14 directly compared it with keyword search, saying \textit{``I think the difference would be one less thing I would have to put in the search bar.''} 

Additionally, participants noted that using LLMs helped better refine searches with more precise natural language. This precision might not be easy to achieve with traditional keyword or semantic searches. For example, P15 mentioned that he could write LLM prompts to explicitly filter death-related COVID claims made only by anti-vaxxers, thereby excluding other death-related claims that often appeared together in traditional keyword or semantic searches:
\begin{displayquote}
    ``\textit{The main problem of keyword search is that [it] often does not bring out best results. If you'll give a prompt search. For instance, it could be a prompt that I want death claims that are made by anti-vaxxers, who think that vaccines cause harm, not death claims made by COVID. So those clear-cut narrow prompts bring out the exact result I want, just anti-vaxxers claiming that happened. If you search it as a term, like a keyword, both results will come and there won't be any difference.}'' (P15)
\end{displayquote}
However, this also requires participants to write sufficient details in the prompt to ensure LLMs can retrieve targeted results accurately. Many participants (P4, 9, 10) noted that there is a learning curve to writing effective prompts \cite{giray2023prompt}, perhaps akin to the effort required to write effective queries in the early days of search engines.  
As search engines increasingly exploit LLM technology to better interpret user queries \citep{google_genAI_2024-wl}, the added value of having a separate mechanism for writing LLM prompts as custom filters may correspondingly diminish. See further discussion in Section \ref{sec:discussion}. 

\subsubsection{Abstract prompts.} 
Unlike targeted prompts, where fact-checkers can explicitly specify the claims they want to investigate, abstract prompts are used to filter claims when fact-checkers are unsure which claims are worth checking. This reflects an exploratory fact-checking stage when they are less familiar with specific news topics or events at the beginning. Participants noted that writing these prompts aligned with strategies they typically used in faceted filtering and browsing, enabling them to efficiently extract relevant information based on specific attributes or relevant factors.

For example, in Table \ref{tab:examples of customized dimensions}, the prompt named ``likelihood to spread'', is written by P3. This participant noted that one key linguistic attribute of misinformation is the use of highly ``loaded language''. Featuring this type of language in the prompt helps identify misinformation that might spread widely: ``\textit{I add a facet that says likelihood to spread, basically the main criteria would be loaded language, usually used to evoke chaos, fear, and harm for the public.}'' (P3)

Additionally, P1 created the prompt ``chaos'' to identify claims that broadly hint at \replaced{potentially }{potential }harmful outcomes. This participant also created another prompt, ``confusion,'' aimed at uncovering ``\textit{claims aimed at misleading the public about the effectiveness of prevention measures.}''

Participants said these abstract prompts could be very useful for exploratory search, especially when \added{they are }unfamiliar with specific news topics. 
Their LLM goal in such cases was to explore what potential claims LLMs could offer them based on high-level relevance factors, including the semantics or outcomes suggested by the claims, as 
a starting point for pursuing more specific, targeted claims:
\begin{displayquote}
    ``\textit{I think that the LLM could be more useful if you've had less experience or you want to have a broader idea than a specific query to [let it] find things for you.}'' (P12) 
    ``\textit{For more general things like news that emerged recently, if I don't know what kind of information I'm searching for, giving a pointer instruction to the results could be very helpful.}'' (P15)
\end{displayquote}

\subsubsection{Comparing written prompts with existing checkworthy factors}
Since targeted prompts reflect fact-checker specific information needs, while abstract prompts align more closely with multi-dimensional relevance judgments, we compare participant abstract prompts to the checkworthy factors outlined in Table \ref{tab:multi-dimensional checkworthiness}. We found that some prompt descriptions written by participants align closely with existing checkworthy literature, particularly in their focus on harmfulness, public impact, and statistical claims, which correspond to the ``Harmful,'' ``Public interest,'' and ``Checkable'' factors, respectively described in Table \ref{tab:multi-dimensional checkworthiness}. Additionally, some prompts that describe the chaos and confusion caused by misinformation offer alternative interpretations of harmfulness.

However, a notable distinction in the written prompts is that participants would intentionally integrate the fact-checking topic, such as COVID-19, into the description of these checkworthy factors. For example, P9 created a customized facet called ``Vaccine causes harm'' (similar to the existing harmful filter) but elaborated \added{on}: ``\textit{Claims that vaccines are dangerous, often based on unreliable information, frequently include narratives that a specific group intends to cause harm or lead to death.}'' This highlights that while checkworthiness can be defined as several general relevant factors useful for exploratory claim searches, fact-checkers tend to contextualize these factors based on their familiarity with the topic, making them more topically relevant. Unlike traditional preset filters, which depend on curated datasets to train predictive models and may be less accurate when applied to out-of-distribution datasets, participants can create LLM filters that incorporate richer contextual information, including insights that might be underrepresented in the training data.

Additionally, we also found that certain factors from Table \ref{tab:multi-dimensional checkworthiness}, such as ``Already checked,'' ``Amplification,'' and ``Difficulty,'' were not explicitly written by participants in the prompts. Since this analysis is a post-hoc comparison, we were unable to ask fact-checkers whether these factors were overlooked intentionally or for other reasons. One hypothesis could be that for ``Already checked'' claims, fact-checkers \deleted{may }have already used the topical search or written targeted prompts to see if the dataset contains those already checked claims. In contrast, factors like ``Amplification'' and ``Difficulty'' might be perceived as too abstract. Fact-checkers might doubt the LLM ability to reliably capture these factors, underscoring the need for further investigation.

\subsection{Fact-checker Use Experience and Claim Selection Effectiveness} \label{subsec:finding-user experience}

We next report on participant overall use experience with the tool (RQ2.2), including 1) their usage behaviors in claim exploration (Section \ref{subsubsec:overall use behaviors}), 2) their effectiveness in selecting claims during search and filtering (Section \ref{subsubsec:effectiveness}), and 3) their subjective reflections (Section \ref{subsubsec:subjective reflections}). 


\subsubsection{Usage behaviors in claims exploration} \label{subsubsec:overall use behaviors}

Table \ref{tab:search behavior measures-uni versus multi} compares the mean (and standard deviation) of different behavioral measures between the unidimensional vs.\ multidimensional interfaces in the absence of any customized filters. Results show significantly greater use of checkworthy sliders with the multidimensional interface (\(M = 6.44\)) vs.\ the unidimensional one (\(M = 1.38\)). Regarding use of the main search box, we also see 
more queries (\(M_{uni} = 2.00\) vs.\ \(M_{multi} = 2.56\)) and query similarity slider changes (\(M_{uni} = 1.12\) vs.\ \(M_{multi} = 1.44\)), though these difference were not significant.  
As P6 noted, ``\textit{adding multidimensions helps you broaden your search a little bit}.''

\begin{table}[ht]
\begin{tabular}{@{}l>{\raggedleft}p{2.3cm}>{\raggedleft}p{2.3cm}>{\raggedleft\arraybackslash}p{2cm}@{}}
    \toprule
    Measures & Unidimensional \(M(SD)\) & Multidimensional \(M(SD)\) & Wilcoxon \linebreak \(W(p)\) \\
    \midrule
    \(\#\) Queries  & 2.00(2.48) & 2.56(3.41) & 10.00(0.25) \\
    \(\#\) Checkworthy slider changes   & 1.38(0.96) & 6.44(5.84) & \textbf{0.00(0.00)} \\
    \(\#\) Query similarity slider changes  & 1.12(2.47) & 1.44(2.63) & 1.5(0.19) \\
    \(\#\) Selected claims  & 5.25(5.56) & 5.44(3.52) & 43.50(0.57) \\
    \(\#\) Final claims found checkworthy  & 1.44(1.03) & 2.06(0.77) & \textbf{18.0(0.04)} \\
    Conversion rate   & 0.30(0.21) & 0.36(0.19) & 31.50(0.55) \\
    \bottomrule
\end{tabular}
\vspace{0.2cm}
\caption{Comparing unidimensional vs.\ multidimensional interfaces in the absence of any customized LLM  filters. Mean (M) and standard deviation (SD) of different behavioral measures (Section \ref{subsubsec:data collection and analysis}) are bolded if statistically significant for the Wilcoxon signed-rank test at \textit{p} < 0.05. Results show that the multidimensional interface generates significantly more interactions with the checkworthy dimension sliders (i.e., weight changes) and ultimately yields more checkworthy claims being selected.}
\label{tab:search behavior measures-uni versus multi}
\end{table}

\begin{table}[ht]
\resizebox{\textwidth}{!}{%
\begin{tabular}{@{}>{\raggedright}p{4.8cm}>{\raggedleft}p{1.4cm}>{\raggedleft}p{1.8cm}>{\raggedleft}p{1.5cm}|>{\raggedleft}p{1.4cm}>{\raggedleft}p{1.8cm}>{\raggedleft\arraybackslash}p{1.5cm}@{}}
    \toprule
    \multirow{2}{*}{Measures} & \multicolumn{3}{c}{\textbf{Unidimensional}} & \multicolumn{3}{c}{\textbf{Multidimensional}}  \\
    & Standard \(M(SD)\) & Standard + Customized \(M(SD)\) & Wilcoxon \(W(p)\) & Standard \(M(SD)\) & Standard + Customized \(M(SD)\) & Wilcoxon \(W(p)\) \\
    \midrule
    \(\#\) Queries  & 2.00(2.48) & 1.06(1.81) & 8.0(0.159) & 2.56(3.41) &  0.50(0.89) & \textbf{0.0(0.01)} \\
    \(\#\) Checkworthy slider changes   & 1.38(0.96) & 1.31(1.25) & 15.0(0.67)  & 6.44(5.84) &  3.69(3.74) & 23.0(0.12) \\
    \(\#\) Query similarity slider changes  & 1.12(2.47) & 0.56(1.03) & 6.0(0.68) & 1.44(2.63) &  0.38(0.72) & \textbf{2.5(0.04)} \\
    \(\#\) Selected claims  & 5.25(5.56) & 4.06(2.43) & 40.0(0.428) & 5.44(3.52)&  4.75(2.46) & 48.0(0.78)  \\
    \(\#\) Final claims found checkworthy  & 1.44(1.03) & 1.88(1.02) & 33.0(0.26) & 2.06(0.77) &  1.75(0.19) &  33.5(0.20) \\
    Conversion rate   & 0.30(0.21) &  0.45(0.28)  & 31.0(0.099) & 0.36(0.19) &  0.35(0.19) & 42.5(0.83)\\
    \bottomrule
\end{tabular}
}
\vspace{0.2cm}
\caption{Comparing behavioral measures (Section \ref{subsubsec:data collection and analysis}) with vs.\  without customized filters for unidimensional and multidimensional interfaces. Mean (M) and standard deviation (SD) results are bolded if statistically significant for the Wilcoxon signed-rank test at \textit{p} < 0.05. Results show that the number of queries and query similarity slider changes significantly decrease in the multidimensional interface with the customized LLM slider. A decrease was also seen with the unidimensional interface, but it was not significant.}

\label{tab:search behavior measures-standard vs custom}
\end{table}

Table \ref{tab:search behavior measures-standard vs custom} compares the mean (and standard deviation) of different behavioral measures with vs.\  without customized LLM filters for unidimensional and multidimensional interfaces. Once the customized filter was added, 
significantly fewer queries (\(M_{standard} = 2.56\) vs.\ \(M_{standard+customized} = 0.50\)) and query similarity slider changes (\(M_{standard} = 2.63\) vs.\ \(M_{standard+customized} = 0.72\)) were observed with the multidimensional interface. We also observed fewer checkworthy slider changes (\(M_{standard} = 6.44\) vs.\ \(M_{standard+customized} = 3.69\)) though this difference was not significant. Similar patterns were found when participants used the unidimensional interface, though no significant differences were observed. These results align with the qualitative responses (Section \ref{subsec:finding-prompting}). For example, P14 explained that, compared to keyword search, using LLM ``\textit{would be one less thing to put in the search bar}.'' 

\subsubsection{Success in finding checkworthy claims} \label{subsubsec:effectiveness}

In this section, we compare how successful participants were in finding checkworthy claims. We first compare unidimensional vs.\  multidimensional interfaces. We then compare success with or without using customized LLM filters. 


\textbf{Unidimensional vs. Multidimensional.} Table \ref{tab:search behavior measures-uni versus multi} shows that the multidimensional interface exerted a small but insignificant positive influence on claim selection vs.\ the unidimensional interface. All related measures showed a slight increase with the multidimensional interface, such as the number of selected claims (\(M_{uni} = 5.25\) vs.\ \(M_{multi} = 5.44\)), final checkworthy claims (\(M_{uni} = 1.44\) vs.\ \(M_{multi} = 2.06\)), and the conversion rate (\(M_{uni} = 0.30\) vs.\ \(M_{multi} = 0.36\)).

Table \ref{tab:claim analysis} compares the number of unique claims found cumulatively across all 16 participants in (first | second) task stages. While participants found more unique claims in the first exploratory stage using the unidimensional interface (\(T_{uni} = 63\) vs.\ \(T_{multi} = 43\)), this reversed in the second stage when participants narrowed down to the three claims they each found most checkworthy (\(T_{multi} = 23\) vs.\ \(T_{uni} = 19\)). Thus, while the unidimensional interface might generate more potentially checkable claims, the multidimensional interface ultimately yielded the most claims to be checked. This difference may be due to participants spending more time refining queries and reviewing claim results since they do not need extra time to explore the multi-faceted filters when using the unidimensional interface. Although we did not enforce the time limit and remind participants during the experiment, this could still lead to identifying more checkable claims.

\begin{table}[ht]
\begin{tabular}{@{}>{\raggedright}p{4cm}>{\raggedright}p{2.2cm}>{\raggedright}p{2.2cm}>{\raggedright\arraybackslash}p{3cm}@{}}
    \toprule
   Conditions & Unidimensional & Multidimensional & Total unique claims \\ \midrule
    Standard & 63 | 19 & 43 | 23 & 171 | 56  \\
    Standard + Customized & 49 | 26 & 50 | 19 & 141 | 58 \\
    Total unique claims & 87 | 36 & 72 | 31 & - \\                                  \bottomrule
\end{tabular}
\vspace{0.2cm}
\caption{Comparing the number of unique claims found cumulatively across all 16 participants in (first | second) task stages using unidimensional vs.\ multidimensional interfaces, with vs.\ without customized search filters. 
While participants find as many interesting claims as possible in the first stage, this set is narrowed in the second stage to three claims they each deem most checkworthy.}
\label{tab:claim analysis}
\end{table}

%

To address potential inefficiency in selecting claims using multidimensional checkworthiness in the exploratory phase, future work might initially prioritize AI-recommended claims that meet different dimensions to reduce the effort required for claim exploration (see Section \ref{subsec:finding-hierachy}). Once fact-checkers have quickly investigated these claims, they could then use the multi-faceted sliders to triage claims involving trade-offs between dimensions.


\textbf{With vs.\ without LLM customized filters.} While differences in claim selection with vs.\ without LLM use were not significant (at least with 16 participants), we discuss small differences observed that could be further investigated with more participants. In particular, success with customized filters to select claims appeared varied depending on unidimensional or multidimensional interface conditions. As shown in Table \ref{tab:search behavior measures-standard vs custom}, when participants used the unidimensional interface, adding customized dimensions led to a decrease in the number of selected claims (\(M_{standard} = 5.25\) vs.\ \(M_{standard+customized} = 4.06\)). However, there was also an increase in the number of final checkworthy claims (\(M_{standard} = 1.44\) vs.\ \(M_{standard+customized} = 1.88\)). There were also fewer unique claims selected in the first stage (\(T_{standard} = 63\) vs.\ \(T_{standard+customized} = 49\)) but more checkworthy claims selected in the second stage (\(T_{standard} = 19\) vs.\ \(T_{standard+customized} = 26\)). When participants used the unidimensional interface, customized filters might have helped them find more checkworthy claims despite finding fewer checkable claims in the exploratory phase. However, participant claim selections show a different pattern with the multidimensional interface. Adding customized dimensions reduced both the number of selected claims (\(M_{standard} = 5.44\) vs.\ \(M_{standard+customized} = 4.75\)) and final checkworthy claims (\(M_{standard} = 2.06\) vs.\ \(M_{standard+customized} = 1.75\)). Total unique claims increased in the first stage (\(T_{standard} = 43\) vs.\ \(T_{standard+customized} = 50\)) but decreased in the second stage (\(T_{standard} = 23\) vs.\ \(T_{standard+customized} = 19\)). 

Comparing these results with those from the unidimensional interface, adding customized dimensions seems to help broaden the scope of claim exploration with unidimensional ranking. However, customized filters might not be as effective as the four checkworthy dimensions implemented by our pre-trained classifiers. For example, many participants (P4, 9, 10) mentioned the difficulty of prompt writing: ``\textit{I don't know how to correctly word my idea in the prompt}'' (P4). It is widely known that successful prompt engineering with LLMs involves a learning curve \citep{giray2023prompt}. 

\subsubsection{Subjective reflections} \label{subsubsec:subjective reflections}

We analyzed participant self-reported metrics for their exploratory search experience between the unidimensional and multidimensional interface (Table \ref{tab:exploratory claim search}). 
The multidimensional interface scored significantly higher in several aspects: ``Understand topic scope'' (\(M_{uni} = 3.69\), \(M_{multi} = 4.44\)), ``Search specific topic'' (\(M_{uni} = 3.44\), \(M_{multi} = 4.56\)), ``Lookup many claims'' (\(M_{uni} = 3.44\), \(M_{multi} = 4.25\)), ``Investigate multiple criteria'' (\(M_{uni} = 3.38\), \(M_{multi} = 4.62\)), and ``Operationalize multiple criteria'' (\(M_{uni} = 3.69\), \(M_{multi} = 4.12\)). Overall, participants considered the multidimensional interface more useful in exploratory claim searches (despite finding more unique claims with the unidimensional interface). The multidimensional interface also received a high score regarding the perceived usefulness and ease of use to support claim prioritization (see details in Appendix \ref{appendix:post-hoc test}).

\begin{table}[ht]
\begin{tabular}{@{}>{\raggedright}p{5cm}>{\raggedright}p{2.5cm}>{\raggedright}p{2.5cm}>{\raggedleft\arraybackslash}p{2cm}@{}} \toprule
Measures & Unidimensional \textit{M(SD)} | \textit{Median} & Multidimensional \textit{M(SD)} | \textit{Median} & Wilcoxon \textit{Z(p)} \\ \midrule
Overall claim satisfaction & 3.75(1.24) | 4  &  4.19(0.83) | 4 &  15.0(0.37) \\
Understand topic scope &  3.69(1.20) | 4 & 4.44(0.51) | 4 &  \textbf{2.0(0.04)} \\
Acquire new perspective  &  3.12(1.31) | 3.5  & 3.81(1.28) | 4 & 12.0(0.05) \\ 
Search specific topic & 3.44(1.15) | 4 & 4.56(0.51) | 5 & \textbf{4.0(0.01)}\\
Lookup many claims &  3.44(1.21) | 4 & 4.25(0.45) | 4  &  \textbf{4.0(0.01)} \\ 
Select best claims &  3.56(0.96) | 4 & 4.12(0.81) | 4 & 2.5(0.08) \\
Uncover unexpected claims   &  3.75(1.00) | 4 & 4.06(0.93) | 4 &  7.0(0.21)\\
Investigate multiple criteria   & 3.38(1.31) | 3.5 & 4.62(0.72) | 5 & \textbf{5.5(0.01)}\\
Operationalize multiple criteria &  3.31(1.30) | 4 & 4.75(0.58) | 5 & \textbf{0.0(0.00)} \\
Operationalize new criteria  &  3.69(1.01) | 4 & 4.12(1.02) | 4 & 12.0(0.19)\\ \bottomrule
\end{tabular}
\vspace{0.2cm}
\caption{Comparing the mean (M), standard deviation (SD), and median of different participant self-reported responses of claim exploration (Section \ref{subsubsec:data collection and analysis}) between unidimensional vs.\ multidimensional interfaces. Results are bolded if statistically significant for the Wilcoxon signed-rank test at \textit{p} < 0.05. Five measures show significantly increased scores, including ``Understand the topic scope,'' ``Search a specific topic,'' ``Lookup many claims,'' ``Investigate multiple criteria,'' and ``Operationalize multiple criteria.'' This suggests that participants preferred using the multidimensional interface to explore claims.}
\label{tab:exploratory claim search}
\end{table}

Our participants also shared very positive reflections after using the multidimensional interface. First, the different customized functions empowered them with more control over using personal knowledge and experience to prioritize claims. P4 mentioned that ``\textit{fact checkers around the world have very local knowledge, specific and unique, that doesn't necessarily apply everywhere else. Now we can kind of add in our own things and make things more relevant to us}.'' 

Some participants further mentioned that these control levels enhance their sense of transparency and trust in AI. P5 and P6 noted that the existing tool they use daily operates like a black box, where they don't know what factors contribute to claims requested by the tool. In contrast, the multidimensional interface is more transparent, breaking down what checkworthiness represents. Even if the results are not entirely accurate, users can modify, change them, or specify a new dimension. This approach complements imperfect AI with human knowledge and oversight. 

Some participants also thought the tool created a playful experience. 
P7 stated that ``\textit{It did differentiate how I use this from how I use a lot of other tools. I could just change a slider instead of changing my search. And that was quite fun.}''

\section{Discussion} \label{sec:discussion}

%
In this section, we reflect on our key findings and connect them with literature in fact-checking and IR to discuss broader research insights.

We begin by discussing the design implications of the fact-checker dynamic and hierarchical claim prioritization process (Section \ref{subsec:implications}). Specifically, we summarize findings that reveal differences in fact-checker claim triage compared to prior research, highlighting underexplored aspects of their workflows (Section \ref{subsubsec:summary}). Building on these insights, we then propose design suggestions for personalized and efficient tool supports based on fact-checker feedback (Section \ref{subsubsec:future design}). Furthermore, we reflect on the dynamic nature of hierarchical relevance in claim prioritization and discuss its broader implications for other user activities (Section \ref{subsubsec:hierarchy}). 

Next, we move beyond viewing claim prioritization solely through the lens of information seeking and retrieval to examine its impact on broader fact-checking stakeholders, as well as the fact-checking ecosystem (Section \ref{subsec:tensions}). Specifically, we examine the evolving practice of claim prioritization across different fact-checking entities (Section \ref{subsubsec:shift of focus}), fact-checker partnerships with social media platforms (Section \ref{subsubsec:partnership}), and the regional differences (Section \ref{subsubsec:region}). In these sections, we discuss how personalized and efficient tools can support the changing nature of fact-checking practices with the potential of addressing tensions and diverse fact-checking objectives.

Finally, we describe the study limitations identified during our RtD process. We hope this helps guide future researchers in exploring multidimensional checkworthiness and developing more advanced claim prioritization tools (Section \ref{subsec:limitation}).

\subsection{Fact-checker Dynamic and Hierarchical Claim Prioritization} \label{subsec:implications}

\subsubsection{Summary of fact-checker claim prioritization process} \label{subsubsec:summary} 
As discussed in the related work (Section \ref{subsec:checkworthiness}), several open questions remain in understanding fact-checker claim prioritization. For example, prior studies have yet to clarify the relative importance of different checkworthiness dimensions and potential gaps between fact-checker self-reported perspectives and their actual behaviors involved in claim search and selection. By employing an RtD approach, which allows us to discover nuanced user practice and design knowledge grounded in making and doing, we have identified insights that further extend findings from earlier research.

First, by integrating qualitative and quantitative data, we find that fact-checker perceptions of the relative importance of different checkworthy dimensions vary across their organizations and regions (as described in Section \ref{subsec:finding-multi-dimensional}). Thus, we build upon and extend the prior qualitative work \citep{Sehat2023-xa, Liu2023-ur, Procter2023-lu} by incorporating supporting quantitative evidence.

We also observed an important contrast with prior work. In particular, while fact-checkers might describe their claim prioritization as unstructured or less systematic (in the context of searching, browsing, and filtering claims across multi-dimensional checkworthiness), their interactions with our prototype actually \replaced{developed }{reflected }a hierarchical claim triage process to enhance work efficiency (in Section \ref{subsec:finding-hierachy}). Moreover, by examining different intents reflected from their written LLM prompts, we found that fact-checker familiarity with fact-checking topics significantly influenced the strategies they employed when using LLMs to create customized filters (described in Section \ref{subsec:finding-prompting}).

Moreover, by employing a within-subject experiment, we have explored some semi-articulated hypotheses (see Section \ref{subsec:finding-user experience}) that provide a foundation for refining hypotheses in future work. In particular, we found that using multi-dimensional filtering and ranking helped participants identify more checkworthy claims compared to a uni-dimensional version (see Section \ref{subsubsec:effectiveness}). According to our participants, this approach not only helped address their concerns about algorithm transparency but also empowered them with greater control and flexibility. This empowered agency enabled them to adapt to varying fact-checking priorities (reported in Section \ref{subsubsec:subjective reflections}). We found that, however, directly integrating LLMs into multi-dimensional filters might reduce user performance (reported in Section \ref{subsubsec:effectiveness}), requiring further design improvements. 

The design integration of a personalized weighting mechanism and LLM-customized facets into our prototype primarily aimed to deepen our understanding of fact-checker claim prioritization practices. In addition to achieving this goal, participants also provided actionable suggestions to refine future designs. Thus, by reflecting on their hierarchical claim prioritization strategies, we expand on existing design knowledge about how to develop advanced tools to support personalized and efficient claim triage in Section \ref{subsubsec:future design}.

\subsubsection{Design implications for personalized and efficient claim triage} \label{subsubsec:future design} In this section, we present design recommendations synthesized from participant feedback, focusing on streamlining user hierarchical claim prioritization, leveraging LLMs to match a progressive fact-checking journey, and improving its use efficiency and transparency.

\textbf{Streamlining user hierarchical claim prioritization.} The hierarchical approach to claim prioritization reflects an underlying systematic process. Although less overt, this inherent structure can inform tool designs to help participants streamline their claim selection workflows. For example, if a claim mutually satisfies multiple checkworthy factors along with the prerequisite or important dimensions, this should be clearly shown in the interface. 
For example, P3 was interested in the ``Likely false'' and ``Verifiable''. The tool should prioritize or separate claims that meet both dimensions. 
P7 also remarked that some claims consistently ranked at the top across even when different dimensions were favored and 
suggested the tool could ``\textit{display them at the very top end so it was easier for me to review}.'' Thus, beyond providing a personalized weighting mechanism for users to prioritize their prerequisite or important dimensions, future designs could also allow fact-checkers to directly preset these priorities. The display of ranked claims should then explicitly indicate how each claim aligns with these priorities.

On the other hand, sometimes, there may be a trade-off between competing checkworthiness dimensions. For example, P1 found that some claims are ``\textit{flagged as very harmful but less interesting to the public}.'' Similar to P3 and P7 above, he suggested making such trade-offs clearly apparent to allow him to validate whether AI predictions for those checkworthy dimensions matched human judgment. If there is a mismatch between human judgment and AI predictions, the weighting mechanism could become particularly valuable, enabling fact-checkers to reduce AI influence.

In prior work, \citet{Sehat2023-xa} propose a conceptual framework for claim prioritization, which introduces a new and structured approach. However, its effective implementation heavily relies on fact-checker training and education. We expect our findings and design recommendations will complement this educational approach by specifying the AI tool supports needed to streamline claim triage, especially where different checkworthiness dimensions are either satisfied or in competition with one another.

\textbf{Leveraging LLMs to match a progressive fact-checking journey.} 
As outlined in Section \ref{subsec:finding-prompting}, fact-checkers employ targeted and abstract prompts to define new checkworthy dimensions, reflecting diverse fact-checking needs influenced by their familiarity with the topics. This process illustrates the real-world progression of fact-checking news events -- from their initial emergence to maturity -- and highlights two specific use cases for LLMs. 

First, when news events emerge and fact-checkers are uncertain about what topics or dis/ misinformation narratives would be the central of fact-checking, they create general faceted filters (i.e., abstract prompts) to understand the semantics or potential harmful outcomes of problematic information. These criteria are frequently-used filters across different news events, as P14 explained ``\textit{[they are] pretty broad category often used to filter out [less important] claims.}'' As fact-checkers become more familiar with specific narratives, direct LLM-integrated search would be more helpful as it can retrieve specific claims based on precise narratives (i.e., targeted prompts).

\textbf{Improving use efficiency and transparency of LLMs.}
However, as reported in Section \ref{subsubsec:effectiveness}, when participants created LLM-customized facets in the multi-dimensional interface, their claim selection performance decreased. This decline in effectiveness is likely due to the quality of prompt writing or the inherent limitations of the LLM used (we employed GPT-3.5 during testing). Based on participant feedback, we propose design recommendations to mitigate this reduced performance to enhance the use efficiency and transparency of LLMs.

As search queries and/or LLM prompts become more complex, it becomes more important to explain how search results relate to different portions of an input query/prompt (i.e., algorithmic transparency). In IR, this is typically the domain of \textit{query-biased summarization} \citep{sarwar-query-biased-summarization} or search result \textit{snippets} that serve to explain how each result relates to a user query. With our simple LLM integration,    
%
P2 reflected that ``\textit{the result might not be directly explainable based on each of the things I wrote down [in the prompt].} ''
%
P7 similarly remarked, ``\textit{it would be nice to tell me why things were being arranged and returned in a certain way. For example, if I had looked up the spike protein in the prompt, I would have seen claims highlighting the spike protein}.'' While users today are adept at using standard search engines and refining their queries, LLM and prompt engineering remain relatively unfamiliar and appear more complex.

This feedback highlights the importance of designing future LLM interfaces to help users better understand two key aspects: 1) how LLMs interpret lengthy prompts and demonstrate their alignment with user needs and 2) how they retrieve, organize, or generate results to address these needs. Some existing technical research might help shed light on the interface design, such as work on decomposing complex claims to retrieve evidence for claim verification \citep{Chen2022-zk} and explaining how different parts of a prompt influence the LLM output \citep{Feng2023-xh}.
For example, if we could similarly decompose complex prompts into smaller portions and illustrate the salience of each portion on the LLM output, this could be very informative to users. This would not only enable users to refine their prompts more precisely but also help address potential transparency and trust issues with LLM-generated content. 

In addition, because LLMs exhibit stronger reasoning and association ability to comprehend user intentions \citep{Zhu2024-bn}, P1 mentioned that if the LLM did not find results that matched the keywords written in the prompt but inferred other items, it would be helpful to point them out from the retrieved claims, e.g., semantic leaps from query terms to related terms. Note that with abstract prompting, conveying such semantic leaps becomes more important because query/prompt terms may be quite general or vague.
P15 explained that ``\textit{when you put it in a more general way [abstract prompts], we need a pointer on why LLM brings the most effective results from the data}.'' In general, LLM interpretability remains a very active area of research today \cite{Zhao2024-ek}. In our task setting, future designs might first provide a summary from LLM explaining how it interprets these subjective and general checkworthy factors and refer to the claims that match them. Additionally, keywords, phrases, or narratives within claims should be highlighted as evidence to support the explanation provided by LLM.

\subsubsection{The dynamic nature of hierarchical relevance in user information-seeking} \label{subsubsec:hierarchy}

Our analysis of self-reported and observational data revealed a hierarchical process in which fact-checkers prioritize claims across three levels: \textit{prerequisite}, \textit{important but contextual}, and \textit{less important}, either consciously or unconsciously (Section \ref{subsec:finding-hierachy}). Although this might be new in the area of claim prioritization, hierarchical structures are fundamental to how people cognitively organize information \citep{Novick2001-sz}, prioritize human needs \citep{McLeod2007-kt}, and make better decisions \citep{Saaty2008-bz}. To improve user information-seeking, hierarchies are commonly used in information architecture to rank topic relevance \citep{Spink1998-rg} and facilitate user interactions during multi-faceted browsing \citep{Capra2007-ph}. Our findings further highlight the dynamic nature of how people assess multidimensional relevance within this hierarchy, particularly in the context of fact-checking and journalistic practices. 

The changing news environment influences how journalists dynamically evaluate what news is worth reporting. Our participants argued that assessing the relative importance of multidimensional checkworthiness --- the hierarchy we identified from user patterns --- would change with different news contexts because fact-checking coverage evolves as events unfold. P3 said, ``\textit{between these factors, such as likely harm, spread, and topicality, if you ask me on different days, which of those I think is more important, I could give you a different answer.}'' He further explained that ``\textit{for the last two months, we solely focus on checking claims around Israel and Gaza, [where] topic and public interest were more important [but] during the height of the COVID, the ability and likelihood to cause harm is the most prominent factor}.'' Checkworthy dimensions following the hierarchy identified from user patterns may thus change depending on world events, necessitating a flexible design mechanism for claim prioritization. More generally, priorities over different dimensions of checkworthiness (e.g., Table \ref{tab:multi-dimensional checkworthiness}) can be expected to naturally and dynamically vary over diverse contexts such as time, location, and organizational and individual preferences. Beyond fact-checking, such dynamism also reflects a broader phenomenon familiar with other information-seeking contexts that are more personal, subjective, and situational, and where the relative importance of particular search criteria changes and new criteria emerge \citep{Schamber1990-uv}.

\subsection{Impacts of Claim Prioritization Tool Supports on Fact-checking Stakeholders and its Ecosystems} \label{subsec:tensions}

%

\subsubsection{Claim prioritization as a shift in fact-checking focus} \label{subsubsec:shift of focus} 
Over the past decade, changes in the media landscape, e.g., the surge of online social media, and news consumption behavior, e.g., the trend of reading news via social media platforms, have significantly transformed journalistic fact-checking. This transformation is also evident in the shift of claim prioritization from focusing solely on checking political claims to addressing online misinformation by considering a broader range of checkworthy factors (described in Section \ref{subsec:fact-checking}). Thus, claim prioritization has become a central challenge for fact-checkers. In this section, by illustrating the historical changes in claim prioritization and its differences across various fact-checking entities, we discuss how our study insights better inform the development of advanced tools in order to adapt to this evolving practice.

Traditionally, fact-checking was an ad-hoc process embedded within news production prior to publication, where editors worked closely with authors to verify the accuracy of arguments \citep{Smith2003-fe}. Ideally, every claim in ready-to-publish news was considered important to fact-check as part of internal journalistic work. This helps maintain the credibility and reputation of the media outlet. As fact-checking has evolved into a post-hoc activity focused on assessing the accuracy of existing public statements, claim prioritization has become a distinct step. In this step, journalists start assessing the newsworthiness of various claims and only select some of them that are worth checking. While this step is recognized as important, it is not generally viewed as particularly challenging. For example, as highlighted in the ethnographic work of \citet{Graves2016-tc}, post-hoc fact-checking was largely carried out by traditional newsrooms and media outlets that primarily focused on political claims. Journalists in these organizations often focus on statements made by politicians, other journalists, and public figures. This emphasis remains evident, as noted by some participants (P5, 12) and other related work \citep{Mena2019-ws, Micallef2022-vp}.

However, due to the surge of online misinformation, post-hoc fact-checking has evolved into a more sophisticated digital practice, focusing on combating a variety of problematic information on social media (e.g., rumors, hoaxes, and propaganda) \citep{Westlund_undated-po}. This digital practice also involves a larger group of stakeholders \citep{Juneja2022-fx}, including organizations solely checking online claims (e.g., Snopes, Lead Stories) and tech companies (e.g., Meta, Meedan) that provide automated fact-checking tools to support this effort \citep{Arnold2020-ly, Das2023-fg, Dierickx2023-db}. Claim prioritization has evolved into a more intricate and nuanced task, requiring alignment with the diverse values of multiple stakeholders \citep{Sehat2023-xa, Liu2023-ur, Procter2023-lu}.

Claim prioritization tool has the potential to accelerate the collaborative effort to meet the fact-checking needs of different stakeholders. For example, by using the tool to prioritize claims based on organizational checkworthiness, fact-checkers across these organizations can enhance their capacity to address a wider range of claims. Additionally, the tool might help address existing ethical considerations, particularly when fact-checkers navigate competing objectives in selecting claims. According to our participants, these competing objectives often stem from their partnerships with social media platforms and regional differences. We discuss these in the following sections.

\subsubsection{Partnerships with social media platforms} \label{subsubsec:partnership} 
\citet{Ananny2018-pr} argues that social media platform objectives can compete with fact-checking values. For example, although virality is an important criterion for claim prioritization, viral claims flagged by social media platforms might not always be worth checking. Fact-checkers believe fake stories with high advertising revenue are often excluded from the fact-check lists requested by the platforms. Additionally, fact-checkers interviewed by \citet{Vinhas2023-hx} reported their duty in maintaining the partnership with the platforms was more quantity-focused, e.g., to meet a quota requirement of fact-checks rather than focusing on potentially fewer checks with greater impact. \citet{Belair-Gagnon2023-fr} also reported that social media platforms urge quick fact-checks with less standard and transparent fact-checking procedures (e.g., not to disclose fact-checking sources and methodology), in opposition to the principles and practices of many fact-checking organizations.  


Some of our participants echoed the aforementioned situations, noting that they often served as outsourced workers for social media platforms to check misinformation rather than verifying content that aligns with the news values upheld by their organizations. For example, P2 and P8 highlighted the importance of fact-checking opinionated claims, such as political propaganda. Although these claims may not garner significant public attention, they can spread false ideologies through repeated exposure. Some participants also felt it was exploitative to maintain this partnership with social media platforms. For example, P1 and P13 emphasized that although they are journalists, the pressure from social media platforms has turned them into content moderators, a role p13 described as ``\textit{getting to be ridiculous}.'' 


\subsubsection{Regional differences} \label{subsubsec:region} 
Regional differences between Western and non-Western countries also matter. 
As noted in Section \ref{subsec:finding-hierachy}, three participants (P8, 10, 11) from India directly mentioned ``Likely harmful'' as the \textit{prerequisite} criterion rather than ``Verifiable''. This challenges a standard assumption that fact-checkers only check claims they believe to be verifiable. As P8 explained, they also report on extremely harmful but opinionated claims to prevent further misinformation. This suggests that local news environments can greatly influence the underlying values, priorities, and practices relating to claim prioritization and fact-checking more broadly. This merits further investigation.

Additionally, regional differences result in different fact-checking operations and funding. \citet{Graves2016-fc} reported that most political fact-checking sites in North America and Western Europe are led by legacy newsrooms, joined by a handful of independent outlets. However, most organizations in Asia, Africa, and South America are based on NGOs and alternative media outlets. Our participants, who conducted fact-checks mostly on an NGO model, mentioned prioritizing claims requested by social media platforms due to financial incentives. Some mentioned that their fact-checking organizations strive to balance checking claims between what the platform requests and their interests. This indicates a ``news judgment trade-off'' for claim prioritization \citep{Belair-Gagnon2023-fr}.


Such tensions and differing objectives suggest a strong need for fact-checkers to customize claim prioritization in a flexible and efficient manner. Based on the design implications discussed in Section \ref{subsubsec:future design}, we believe that effective tooling could empower fact-checkers to better prioritize multidimensional checkworthiness according to the stakeholder needs (e.g., what their organization wants vs. what social media platforms want), helping them to reduce the time and effort required for claim exploration. 
If different claims found relate to competing priorities (e.g., financial considerations vs.\ organizational missions.), better tooling could help fact-checkers balance these competing priorities.

\subsection{Study Limitations} \label{subsec:limitation}


\subsubsection{Search box vs.\ custom filters} The advent of LLMs has created tremendous excitement about democratizing AI capabilities. Our study empowers 
fact-checkers to create zero-code, custom search filters alongside a standard search box with four checkworthiness filters. 
Search engines have performed best for keyword-oriented queries, and users thus traditionally assume and write keyword queries. However, search engine capabilities have progressed tremendously in understanding more verbose and complex queries \citep{Bendersky2008-zq, Gupta2015-fh}, and today's search engines increasingly incorporate the latest LLM capabilities to support more powerful query interpretation \citep{google_genAI_2024-wl}. 
However, our prototype's search box only implemented semantic search using SentenceBERT (Section \ref{subsubsec:implementation}).
Future work could explore the integration of LLMs directly into the search box to enhance query interpretation without additional custom filters.

\subsubsection{Included vs. omitted checkworthy dimensions} Just as users can consider different dimensions of relevance in information seeking, many dimensions of checkworthiness have been identified in prior work (Table \ref{tab:multi-dimensional checkworthiness}). We adopted the COVID-19 claim dataset developed by \citet{alam-etal-2021-fighting-covid}, which annotated seven dimensions of checkworthiness, though we used only four of these dimensions to simplify our study. 
This assumption compromised realism since fact-checkers may prioritize claims using other dimensions, such as urgency and susceptibility (Appendix \ref{sec:other_dims}). Future research should investigate a more comprehensive set of checkworthy dimensions to support fact-checkers better.


\subsubsection{Use of a historical claims dataset} We adopted a COVID-19 dataset due to its existing annotation
for multiple dimensions of checkworthiness, making it easy for us to train predictive models for 
Participants were familiar with COVID-19, providing a solid foundation for an exploratory search task.
However, participants (P6) noted the limitations of using historical data, as current news judgment differs from past claims.
A stronger study design would instead let participants search the web for claims related to current events based on their current knowledge. However, working with live data presents a variety of different challenges, such as a lack of ground truth annotations for checkworthiness, as well as the risk of low classifier performance in predicting checkworthy dimensions due to distribution shift between training data (e.g., training classifiers on the COVID-19 dataset) and live data on which predictions are performed. 

\subsubsection{Small dataset scale} While the COVID-19 dataset used \citet{alam-etal-2021-fighting-covid} contains 4,542 tweets, we used only around 500 tweets with participants to balance experiment order effects and 
data splitting for model training. In addition, whenever participants created a custom search filter, running \texttt{gpt-3.5-turbo} on even 500 tweets required a noticeable delay, and using a larger dataset would have exacerbated this delay even further. 
Participants (P3, 13) desired larger datasets for more realistic evaluations and to better assess new dimensions created by LLMs. Additionally, they were interested in claims related to their regions, of which our small dataset had limited coverage. Future research should investigate larger datasets to enhance study realism and findings.

\subsubsection{Few participants for statistical testing}
It was important to us for realism to conduct this study with professional fact-checkers rather than surrogate journalism students. 
Our 16 participants included 14 full-time professional journalists and 2 researchers with previous professional experience as fact-checkers (Table \ref{tab:participants}). However, recruiting professionals was challenging due to their busy schedules, resulting in a small sample size that was insufficient for rigorous statistical analysis. 
That said, our study adopted an RtD approach \citep{Zimmerman2007-gh}, and as reflected by \citet{Zimmerman2014-cs}, the lab-based RtD aims to explore ``semi-articulated hypotheses.'' Therefore, by reporting the evaluation results of this preliminary prototype, we hope to assist future researchers in constructing clearer hypotheses for larger-scale testing. 

\section{Conclusion} \label{sec:conclusion}

With so many potentially false claims circulating online, claim prioritization is key to intelligently allocating limited human resources for fact-checking. Our study perceives claim prioritization through the lens of IR: just as relevance is multidimensional, with many factors influencing which search results a user deems relevant, checkworthiness is also multi-faceted, subjective, and even personal, with many factors influencing how fact-checkers prioritize claims to check. 

Our study investigated both the multidimensional nature of checkworthiness and effective tool support to assist fact-checkers in claim prioritization. Methodologically, we pursued \textit{Research through Design} combined with mixed-method evaluation. Our key artifact is an AI-assisted claim prioritization prototype developed as a probe to explore how fact-checkers use multidimensional checkworthiness factors in claim prioritization, simultaneously probing fact-checker needs while also exploring the design space to meet those needs. 

Our study revealed three key findings: 1) a hierarchical process of searching and filtering claims; 2) targeted vs.\ abstract approaches to writing LLM prompts to create custom search filters for checkworthiness; and 3) the value of using multidimensional checkworthiness to triage claims. Overall, our work offers insights into both fact-checker work practices and the need for more tailored and efficient claim prioritization, with corresponding design implications.

\section*{Acknowledgments}
We thank the professional fact-checkers who participated in our study for both their valuable work and for making our research possible. This research was supported in part by the Micron Foundation, the Knight Foundation\footnote{\scriptsize{\url{https://knightfoundation.org/articles/connective-democracy-a-new-approach-for-building-bridges-in-american-society/}}} and by Good Systems\footnote{\url{http://goodsystems.utexas.edu/}}, a UT Austin Grand Challenge to develop responsible AI technologies. The statements made herein are solely the opinions of the authors and do not reflect the views of the sponsoring agencies.

\bibliographystyle{ACM-Reference-Format}
\bibliography{reference}


\begin{thebibliography}{96}


\ifx \showCODEN    \undefined \def \showCODEN     #1{\unskip}     \fi
\ifx \showDOI      \undefined \def \showDOI       #1{#1}\fi
\ifx \showISBNx    \undefined \def \showISBNx     #1{\unskip}     \fi
\ifx \showISBNxiii \undefined \def \showISBNxiii  #1{\unskip}     \fi
\ifx \showISSN     \undefined \def \showISSN      #1{\unskip}     \fi
\ifx \showLCCN     \undefined \def \showLCCN      #1{\unskip}     \fi
\ifx \shownote     \undefined \def \shownote      #1{#1}          \fi
\ifx \showarticletitle \undefined \def \showarticletitle #1{#1}   \fi
\ifx \showURL      \undefined \def \showURL       {\relax}        \fi
\providecommand\bibfield[2]{#2}
\providecommand\bibinfo[2]{#2}
\providecommand\natexlab[1]{#1}
\providecommand\showeprint[2][]{arXiv:#2}

\bibitem[goo(2024)]%
        {google_genAI_2024-wl}
 \bibinfo{year}{2024}\natexlab{}.
\newblock \bibinfo{title}{Generative {AI} search trends \& transformations}.
\newblock \bibinfo{howpublished}{\url{https://www.thinkwithgoogle.com/marketing-strategies/search/generative-ai-search-trends-and-transformations/}}.
\newblock
\newblock
\shownote{Accessed: 2024-7-1}.


\bibitem[Alam et~al\mbox{.}(2021)]%
        {alam-etal-2021-fighting-covid}
\bibfield{author}{\bibinfo{person}{Firoj Alam}, \bibinfo{person}{Shaden Shaar}, \bibinfo{person}{Fahim Dalvi}, \bibinfo{person}{Hassan Sajjad}, \bibinfo{person}{Alex Nikolov}, \bibinfo{person}{Hamdy Mubarak}, \bibinfo{person}{Giovanni Da~San~Martino}, \bibinfo{person}{Ahmed Abdelali}, \bibinfo{person}{Nadir Durrani}, \bibinfo{person}{Kareem Darwish}, \bibinfo{person}{Abdulaziz Al-Homaid}, \bibinfo{person}{Wajdi Zaghouani}, \bibinfo{person}{Tommaso Caselli}, \bibinfo{person}{Gijs Danoe}, \bibinfo{person}{Friso Stolk}, \bibinfo{person}{Britt Bruntink}, {and} \bibinfo{person}{Preslav Nakov}.} \bibinfo{year}{2021}\natexlab{}.
\newblock \showarticletitle{Fighting the {COVID}-19 Infodemic: Modeling the Perspective of Journalists, Fact-Checkers, Social Media Platforms, Policy Makers, and the Society}. In \bibinfo{booktitle}{\emph{Findings of the Association for Computational Linguistics: EMNLP 2021}}, \bibfield{editor}{\bibinfo{person}{Marie-Francine Moens}, \bibinfo{person}{Xuanjing Huang}, \bibinfo{person}{Lucia Specia}, {and} \bibinfo{person}{Scott Wen-tau Yih}} (Eds.). \bibinfo{publisher}{Association for Computational Linguistics}, \bibinfo{address}{Punta Cana, Dominican Republic}, \bibinfo{pages}{611--649}.
\newblock
\urldef\tempurl%
\url{https://doi.org/10.18653/v1/2021.findings-emnlp.56}
\showDOI{\tempurl}


\bibitem[Allein and Moens(2020)]%
        {Allein2020-jy}
\bibfield{author}{\bibinfo{person}{Liesbeth Allein} {and} \bibinfo{person}{Marie-Francine Moens}.} \bibinfo{year}{2020}\natexlab{}.
\newblock \showarticletitle{Checkworthiness in Automatic Claim Detection Models: Definitions and Analysis of Datasets}. In \bibinfo{booktitle}{\emph{Disinformation in Open Online Media}}. \bibinfo{publisher}{Springer International Publishing}, \bibinfo{pages}{1--17}.
\newblock
\urldef\tempurl%
\url{https://doi.org/10.1007/978-3-030-61841-4\_1}
\showDOI{\tempurl}


\bibitem[Ananny(2018)]%
        {Ananny2018-pr}
\bibfield{author}{\bibinfo{person}{Mike Ananny}.} \bibinfo{year}{2018}\natexlab{}.
\newblock \bibinfo{booktitle}{\emph{The partnership press: Lessons for platform-publisher collaborations as Facebook and news outlets team to fight misinformation}}.
\newblock \bibinfo{type}{{T}echnical {R}eport}.
\newblock
\urldef\tempurl%
\url{https://doi.org/10.7916/D85B1JG9}
\showDOI{\tempurl}


\bibitem[Arif(2018)]%
        {Arif2018-hn}
\bibfield{author}{\bibinfo{person}{Ahmer Arif}.} \bibinfo{year}{2018}\natexlab{}.
\newblock \showarticletitle{Designing to Support Reflection on Values \& Practices to Address Online Disinformation}. In \bibinfo{booktitle}{\emph{Companion of the 2018 {ACM} Conference on Computer Supported Cooperative Work and Social Computing}} (Jersey City, NJ, USA) \emph{(\bibinfo{series}{CSCW '18 Companion})}. \bibinfo{publisher}{Association for Computing Machinery}, \bibinfo{address}{New York, NY, USA}, \bibinfo{pages}{61--64}.
\newblock
\showISBNx{9781450360180}
\urldef\tempurl%
\url{https://doi.org/10.1145/3272973.3272974}
\showDOI{\tempurl}


\bibitem[Arnold(2020)]%
        {Arnold2020-ly}
\bibfield{author}{\bibinfo{person}{Phoebe Arnold}.} \bibinfo{year}{2020}\natexlab{}.
\newblock \bibinfo{booktitle}{\emph{The challenges of online fact checking}}.
\newblock \bibinfo{type}{{T}echnical {R}eport}. \bibinfo{institution}{Full Fact}.
\newblock


\bibitem[Babaei et~al\mbox{.}(2022)]%
        {Babaei2022-so}
\bibfield{author}{\bibinfo{person}{Mahmoudreza Babaei}, \bibinfo{person}{Juhi Kulshrestha}, \bibinfo{person}{Abhijnan Chakraborty}, \bibinfo{person}{Elissa~M Redmiles}, \bibinfo{person}{Meeyoung Cha}, {and} \bibinfo{person}{Krishna~P Gummadi}.} \bibinfo{year}{2022}\natexlab{}.
\newblock \showarticletitle{Analyzing Biases in Perception of Truth in News Stories and Their Implications for Fact Checking}.
\newblock \bibinfo{journal}{\emph{IEEE Transactions on Computational Social Systems}} \bibinfo{volume}{9}, \bibinfo{number}{3} (\bibinfo{date}{June} \bibinfo{year}{2022}), \bibinfo{pages}{839--850}.
\newblock
\showISSN{2329-924X, 2373-7476}
\urldef\tempurl%
\url{https://doi.org/10.1109/TCSS.2021.3096038}
\showDOI{\tempurl}


\bibitem[Barrón-Cedeño et~al\mbox{.}(2024)]%
        {Barron-Cedeno2024-gi}
\bibfield{author}{\bibinfo{person}{Alberto Barrón-Cedeño}, \bibinfo{person}{Firoj Alam}, \bibinfo{person}{Tanmoy Chakraborty}, \bibinfo{person}{Tamer Elsayed}, \bibinfo{person}{Preslav Nakov}, \bibinfo{person}{Piotr Przybyła}, \bibinfo{person}{Julia~Maria Struß}, \bibinfo{person}{Fatima Haouari}, \bibinfo{person}{Maram Hasanain}, \bibinfo{person}{Federico Ruggeri}, \bibinfo{person}{Xingyi Song}, {and} \bibinfo{person}{Reem Suwaileh}.} \bibinfo{year}{2024}\natexlab{}.
\newblock \showarticletitle{The {CLEF}-2024 {CheckThat}! Lab: Check-worthiness, subjectivity, persuasion, roles, authorities, and adversarial robustness}.
\newblock In \bibinfo{booktitle}{\emph{Lecture Notes in Computer Science}}, \bibfield{editor}{\bibinfo{person}{Nazli Goharian}, \bibinfo{person}{Nicola Tonellotto}, \bibinfo{person}{Yulan He}, \bibinfo{person}{Aldo Lipani}, \bibinfo{person}{Graham McDonald}, \bibinfo{person}{Craig Macdonald}, {and} \bibinfo{person}{Iadh Ounis}} (Eds.). \bibinfo{series}{Lecture notes in computer science}, Vol.~\bibinfo{volume}{14612}. \bibinfo{publisher}{Springer Nature Switzerland}, \bibinfo{address}{Cham}, \bibinfo{pages}{449--458}.
\newblock
\showISBNx{9783031560682,9783031560699}
\showISSN{1611-3349,0302-9743}
\urldef\tempurl%
\url{https://doi.org/10.1007/978-3-031-56069-9\_62}
\showDOI{\tempurl}


\bibitem[Barrón-Cedeño et~al\mbox{.}(2023)]%
        {Barron-Cedeno2023-ec}
\bibfield{author}{\bibinfo{person}{Alberto Barrón-Cedeño}, \bibinfo{person}{Firoj Alam}, \bibinfo{person}{Andrea Galassi}, \bibinfo{person}{Giovanni Da~San~Martino}, \bibinfo{person}{Preslav Nakov}, \bibinfo{person}{Tamer Elsayed}, \bibinfo{person}{Dilshod Azizov}, \bibinfo{person}{Tommaso Caselli}, \bibinfo{person}{Gullal~S Cheema}, \bibinfo{person}{Fatima Haouari}, \bibinfo{person}{Maram Hasanain}, \bibinfo{person}{Mucahid Kutlu}, \bibinfo{person}{Chengkai Li}, \bibinfo{person}{Federico Ruggeri}, \bibinfo{person}{Julia~Maria Struß}, {and} \bibinfo{person}{Wajdi Zaghouani}.} \bibinfo{year}{2023}\natexlab{}.
\newblock \showarticletitle{Overview of the {CLEF–2023} {CheckThat}! Lab on Checkworthiness, Subjectivity, Political Bias, Factuality, and Authority of News Articles and Their Source}.
\newblock In \bibinfo{booktitle}{\emph{Experimental IR Meets Multilinguality, Multimodality, and Interaction}}, \bibfield{editor}{\bibinfo{person}{Avi Arampatzis}, \bibinfo{person}{Evangelos Kanoulas}, \bibinfo{person}{Theodora Tsikrika}, \bibinfo{person}{Stefanos Vrochidis}, \bibinfo{person}{Anastasia Giachanou}, \bibinfo{person}{Dan Li}, \bibinfo{person}{Mohammad Aliannejadi}, \bibinfo{person}{Michalis Vlachos}, \bibinfo{person}{Guglielmo Faggioli}, {and} \bibinfo{person}{Nicola Ferro}} (Eds.). \bibinfo{series}{Lecture Notes in Computer Science}, Vol.~\bibinfo{volume}{14163}. \bibinfo{publisher}{Springer Nature Switzerland}, \bibinfo{address}{Cham}, \bibinfo{pages}{251--275}.
\newblock
\showISBNx{9783031424472}
\showISSN{0302-9743,1611-3349}
\urldef\tempurl%
\url{https://doi.org/10.1007/978-3-031-42448-9\_20}
\showDOI{\tempurl}


\bibitem[Beers et~al\mbox{.}(2020)]%
        {Beers2020-cw}
\bibfield{author}{\bibinfo{person}{A Beers}, \bibinfo{person}{M~M~C Haughey}, \bibinfo{person}{A Arif}, {and} \bibinfo{person}{K Starbird}.} \bibinfo{year}{2020}\natexlab{}.
\newblock \showarticletitle{Examining the digital toolsets of journalists reporting on disinformation}. \bibinfo{publisher}{cj2020.northeastern.edu}.
\newblock
\showISSN{1406-0914}


\bibitem[B{\'e}lair-Gagnon et~al\mbox{.}(2023)]%
        {Belair-Gagnon2023-fr}
\bibfield{author}{\bibinfo{person}{Val{\'e}rie B{\'e}lair-Gagnon}, \bibinfo{person}{Rebekah Larsen}, \bibinfo{person}{Lucas Graves}, {and} \bibinfo{person}{Oscar Westlund}.} \bibinfo{year}{2023}\natexlab{}.
\newblock \showarticletitle{Knowledge Work in Platform {Fact-Checking} Partnerships}.
\newblock \bibinfo{journal}{\emph{International Journal of Communication Systems}} \bibinfo{volume}{17}, \bibinfo{number}{0} (\bibinfo{date}{29~Jan.} \bibinfo{year}{2023}), \bibinfo{pages}{21}.
\newblock
\showISSN{1074-5351}


\bibitem[Bendersky and Croft(2008)]%
        {Bendersky2008-zq}
\bibfield{author}{\bibinfo{person}{Michael Bendersky} {and} \bibinfo{person}{W~Bruce Croft}.} \bibinfo{year}{2008}\natexlab{}.
\newblock \showarticletitle{Discovering key concepts in verbose queries}. In \bibinfo{booktitle}{\emph{Proceedings of the 31st annual international {ACM} {SIGIR} conference on Research and development in information retrieval}} (Singapore, Singapore) \emph{(\bibinfo{series}{SIGIR '08})}. \bibinfo{publisher}{Association for Computing Machinery}, \bibinfo{address}{New York, NY, USA}, \bibinfo{pages}{491--498}.
\newblock
\showISBNx{9781605581644}
\urldef\tempurl%
\url{https://doi.org/10.1145/1390334.1390419}
\showDOI{\tempurl}


\bibitem[Bowers(2012)]%
        {Bowers2012-iq}
\bibfield{author}{\bibinfo{person}{John Bowers}.} \bibinfo{year}{2012}\natexlab{}.
\newblock \showarticletitle{The logic of annotated portfolios: communicating the value of 'research through design'}. In \bibinfo{booktitle}{\emph{Proceedings of the Designing Interactive Systems Conference}} \emph{(\bibinfo{series}{DIS '12})}. \bibinfo{publisher}{Association for Computing Machinery}, \bibinfo{address}{New York, NY, USA}, \bibinfo{pages}{68--77}.
\newblock
\showISBNx{9781450312103}
\urldef\tempurl%
\url{https://doi.org/10.1145/2317956.2317968}
\showDOI{\tempurl}


\bibitem[Cai et~al\mbox{.}(2019)]%
        {Cai2019-fz}
\bibfield{author}{\bibinfo{person}{Carrie~J Cai}, \bibinfo{person}{Emily Reif}, \bibinfo{person}{Narayan Hegde}, \bibinfo{person}{Jason Hipp}, \bibinfo{person}{Been Kim}, \bibinfo{person}{Daniel Smilkov}, \bibinfo{person}{Martin Wattenberg}, \bibinfo{person}{Fernanda Viegas}, \bibinfo{person}{Greg~S Corrado}, \bibinfo{person}{Martin~C Stumpe}, {and} \bibinfo{person}{Michael Terry}.} \bibinfo{year}{2019}\natexlab{}.
\newblock \showarticletitle{Human-centered tools for coping with imperfect algorithms during medical decision-making}. In \bibinfo{booktitle}{\emph{Proceedings of the 2019 {CHI} Conference on Human Factors in Computing Systems}} (Glasgow, Scotland Uk) \emph{(\bibinfo{series}{CHI '19}, \bibinfo{number}{Paper 4})}. \bibinfo{publisher}{Association for Computing Machinery}, \bibinfo{address}{New York, NY, USA}, \bibinfo{pages}{1--14}.
\newblock
\showISBNx{9781450359702}
\urldef\tempurl%
\url{https://doi.org/10.1145/3290605.3300234}
\showDOI{\tempurl}


\bibitem[Capra et~al\mbox{.}(2007)]%
        {Capra2007-ph}
\bibfield{author}{\bibinfo{person}{Robert Capra}, \bibinfo{person}{Gary Marchionini}, \bibinfo{person}{Jung~Sun Oh}, \bibinfo{person}{Fred Stutzman}, {and} \bibinfo{person}{Yan Zhang}.} \bibinfo{year}{2007}\natexlab{}.
\newblock \showarticletitle{Effects of structure and interaction style on distinct search tasks}. In \bibinfo{booktitle}{\emph{Proceedings of the 7th {ACM/IEEE-CS} joint conference on Digital libraries}} (Vancouver, BC, Canada) \emph{(\bibinfo{series}{JCDL '07})}. \bibinfo{publisher}{Association for Computing Machinery}, \bibinfo{address}{New York, NY, USA}, \bibinfo{pages}{442--451}.
\newblock
\showISBNx{9781595936448}
\urldef\tempurl%
\url{https://doi.org/10.1145/1255175.1255267}
\showDOI{\tempurl}


\bibitem[Chen et~al\mbox{.}(2022)]%
        {Chen2022-zk}
\bibfield{author}{\bibinfo{person}{Jifan Chen}, \bibinfo{person}{Aniruddh Sriram}, \bibinfo{person}{Eunsol Choi}, {and} \bibinfo{person}{Greg Durrett}.} \bibinfo{year}{2022}\natexlab{}.
\newblock \showarticletitle{Generating Literal and Implied Subquestions to Fact-check Complex Claims}. In \bibinfo{booktitle}{\emph{Proceedings of the 2022 Conference on Empirical Methods in Natural Language Processing}}, \bibfield{editor}{\bibinfo{person}{Yoav Goldberg}, \bibinfo{person}{Zornitsa Kozareva}, {and} \bibinfo{person}{Yue Zhang}} (Eds.). \bibinfo{publisher}{Association for Computational Linguistics}, \bibinfo{address}{Abu Dhabi, United Arab Emirates}, \bibinfo{pages}{3495--3516}.
\newblock
\urldef\tempurl%
\url{https://doi.org/10.18653/v1/2022.emnlp-main.229}
\showDOI{\tempurl}


\bibitem[{\c C}{\"o}mlek{\c c}i(2022)]%
        {Comlekci2022-ch}
\bibfield{author}{\bibinfo{person}{Mehmet~Fatih {\c C}{\"o}mlek{\c c}i}.} \bibinfo{year}{2022}\natexlab{}.
\newblock \showarticletitle{Why Do {Fact-Checking} Organizations Go Beyond {Fact-Checking}? A Leap Toward Media and Information Literacy Education}.
\newblock \bibinfo{journal}{\emph{International Journal of Communication Systems}} \bibinfo{volume}{16}, \bibinfo{number}{0} (\bibinfo{date}{25~Sept.} \bibinfo{year}{2022}), \bibinfo{pages}{21}.
\newblock
\showISSN{1074-5351}


\bibitem[Das et~al\mbox{.}(2023)]%
        {Das2023-fg}
\bibfield{author}{\bibinfo{person}{Anubrata Das}, \bibinfo{person}{Houjiang Liu}, \bibinfo{person}{Venelin Kovatchev}, {and} \bibinfo{person}{Matthew Lease}.} \bibinfo{year}{2023}\natexlab{}.
\newblock \showarticletitle{The state of human-centered {NLP} technology for fact-checking}.
\newblock \bibinfo{journal}{\emph{Information processing \& management}} \bibinfo{volume}{60}, \bibinfo{number}{2} (\bibinfo{date}{1~March} \bibinfo{year}{2023}), \bibinfo{pages}{103219}.
\newblock
\showISSN{0306-4573, 1873-5371}
\urldef\tempurl%
\url{https://doi.org/10.1016/j.ipm.2022.103219}
\showDOI{\tempurl}


\bibitem[{Design Council}(2015)]%
        {Design_Council2015-wu}
\bibfield{author}{\bibinfo{person}{{Design Council}}.} \bibinfo{year}{2015}\natexlab{}.
\newblock \bibinfo{title}{Framework for Innovation: Design Council's evolved Double Diamond}.
\newblock
\newblock


\bibitem[Dierickx et~al\mbox{.}(2023)]%
        {Dierickx2023-db}
\bibfield{author}{\bibinfo{person}{Laurence Dierickx}, \bibinfo{person}{Carl-Gustav Lindén}, {and} \bibinfo{person}{Andreas~Lothe Opdahl}.} \bibinfo{year}{2023}\natexlab{}.
\newblock \showarticletitle{Automated fact-checking to support professional practices: systematic literature review and meta-analysis}.
\newblock \bibinfo{journal}{\emph{International Journal of Communication}} \bibinfo{volume}{17}, \bibinfo{number}{0} (\bibinfo{date}{15~Aug.} \bibinfo{year}{2023}), \bibinfo{pages}{21--21}.
\newblock
\showISSN{1932-8036,1932-8036}


\bibitem[Feng et~al\mbox{.}(2023)]%
        {Feng2023-xh}
\bibfield{author}{\bibinfo{person}{Zijian Feng}, \bibinfo{person}{Hanzhang Zhou}, \bibinfo{person}{Zixiao Zhu}, \bibinfo{person}{Junlang Qian}, {and} \bibinfo{person}{Kezhi Mao}.} \bibinfo{year}{2023}\natexlab{}.
\newblock \showarticletitle{Unveiling and Manipulating Prompt Influence in Large Language Models}. In \bibinfo{booktitle}{\emph{The Twelfth International Conference on Learning Representations}}.
\newblock


\bibitem[Frankel and Racine(2010)]%
        {Frankel2010-sw}
\bibfield{author}{\bibinfo{person}{Lois Frankel} {and} \bibinfo{person}{Martin Racine}.} \bibinfo{year}{2010}\natexlab{}.
\newblock \showarticletitle{The Complex Field of Research: for Design, through Design, and about Design}. In \bibinfo{booktitle}{\emph{{DRS} Biennial Conference Series}}.
\newblock


\bibitem[Frayling(1993)]%
        {Frayling1993-sp}
\bibfield{author}{\bibinfo{person}{Christopher Frayling}.} \bibinfo{year}{1993}\natexlab{}.
\newblock \showarticletitle{Research in art and design}.
\newblock \bibinfo{journal}{\emph{Royal College of Art research papers}}  \bibinfo{volume}{1} (\bibinfo{year}{1993}), \bibinfo{pages}{1--5}.
\newblock


\bibitem[Gaver(2012)]%
        {Gaver2012-yw}
\bibfield{author}{\bibinfo{person}{William Gaver}.} \bibinfo{year}{2012}\natexlab{}.
\newblock \showarticletitle{What should we expect from research through design?}. In \bibinfo{booktitle}{\emph{Proceedings of the {SIGCHI} Conference on Human Factors in Computing Systems}} (Austin, Texas, USA) \emph{(\bibinfo{series}{CHI '12})}. \bibinfo{publisher}{Association for Computing Machinery}, \bibinfo{address}{New York, New York, USA}, \bibinfo{pages}{937--946}.
\newblock
\showISBNx{9781450310154}
\urldef\tempurl%
\url{https://doi.org/10.1145/2207676.2208538}
\showDOI{\tempurl}


\bibitem[Giray(2023)]%
        {giray2023prompt}
\bibfield{author}{\bibinfo{person}{Louie Giray}.} \bibinfo{year}{2023}\natexlab{}.
\newblock \showarticletitle{Prompt engineering with ChatGPT: a guide for academic writers}.
\newblock \bibinfo{journal}{\emph{Annals of biomedical engineering}} \bibinfo{volume}{51}, \bibinfo{number}{12} (\bibinfo{year}{2023}), \bibinfo{pages}{2629--2633}.
\newblock


\bibitem[Graves(2016)]%
        {Graves2016-tc}
\bibfield{author}{\bibinfo{person}{Lucas Graves}.} \bibinfo{year}{2016}\natexlab{}.
\newblock \bibinfo{booktitle}{\emph{Deciding What's True: The Rise of Political {Fact-Checking} in American Journalism}}.
\newblock \bibinfo{publisher}{Columbia University Press}.
\newblock
\showISBNx{9780231542227}


\bibitem[Graves(2017)]%
        {Graves2017-rb}
\bibfield{author}{\bibinfo{person}{Lucas Graves}.} \bibinfo{year}{2017}\natexlab{}.
\newblock \showarticletitle{Anatomy of a Fact Check: Objective Practice and the Contested Epistemology of Fact Checking}.
\newblock \bibinfo{journal}{\emph{Communication, Culture and Critique}} \bibinfo{volume}{10}, \bibinfo{number}{3} (\bibinfo{date}{1~Sept.} \bibinfo{year}{2017}), \bibinfo{pages}{518--537}.
\newblock
\showISSN{1753-9137}
\urldef\tempurl%
\url{https://doi.org/10.1111/cccr.12163}
\showDOI{\tempurl}


\bibitem[Graves(2018)]%
        {Graves2018-ne}
\bibfield{author}{\bibinfo{person}{Lucas Graves}.} \bibinfo{year}{2018}\natexlab{}.
\newblock \bibinfo{booktitle}{\emph{Understanding the Promise and Limits of Automated {Fact-Checking}}}.
\newblock \bibinfo{type}{{T}echnical {R}eport}. \bibinfo{institution}{Reuters Institute}. \bibinfo{pages}{1--7} pages.
\newblock
\showISBNx{9780874216561}
\showISSN{0717-6163}


\bibitem[Graves and Cherubini(2016)]%
        {Graves2016-fc}
\bibfield{author}{\bibinfo{person}{L Graves} {and} \bibinfo{person}{F Cherubini}.} \bibinfo{year}{2016}\natexlab{}.
\newblock \showarticletitle{The Rise of {Fact-Checking} Sites in Europe}.
\newblock In \bibinfo{booktitle}{\emph{Digital News Project Report}}. \bibinfo{publisher}{Reuters Institute for the Study of Journalism}.
\newblock
\showISBNx{9781907384257}


\bibitem[Graves and Mantzarlis(2020)]%
        {Graves2020-gw}
\bibfield{author}{\bibinfo{person}{Lucas Graves} {and} \bibinfo{person}{Alexios Mantzarlis}.} \bibinfo{year}{2020}\natexlab{}.
\newblock \showarticletitle{Amid political spin and online misinformation, fact checking adapts}.
\newblock \bibinfo{journal}{\emph{The Political quarterly}} \bibinfo{volume}{91}, \bibinfo{number}{3} (\bibinfo{date}{July} \bibinfo{year}{2020}), \bibinfo{pages}{585--591}.
\newblock
\showISSN{0032-3179, 1467-923X}
\urldef\tempurl%
\url{https://doi.org/10.1111/1467-923x.12896}
\showDOI{\tempurl}


\bibitem[Guo et~al\mbox{.}(2023)]%
        {Guo2023-iv}
\bibfield{author}{\bibinfo{person}{Mengtian Guo}, \bibinfo{person}{Zhilan Zhou}, \bibinfo{person}{David Gotz}, {and} \bibinfo{person}{Yue Wang}.} \bibinfo{year}{2023}\natexlab{}.
\newblock \showarticletitle{{GRAFS}: Graphical Faceted Search System to Support Conceptual Understanding in Exploratory Search}.
\newblock \bibinfo{journal}{\emph{ACM Trans. Interact. Intell. Syst.}} \bibinfo{volume}{13}, \bibinfo{number}{2} (\bibinfo{date}{5~May} \bibinfo{year}{2023}), \bibinfo{pages}{1--36}.
\newblock
\showISSN{2160-6455}
\urldef\tempurl%
\url{https://doi.org/10.1145/3588319}
\showDOI{\tempurl}


\bibitem[Guo et~al\mbox{.}(2022)]%
        {Guo2022-km}
\bibfield{author}{\bibinfo{person}{Zhijiang Guo}, \bibinfo{person}{Michael Schlichtkrull}, {and} \bibinfo{person}{Andreas Vlachos}.} \bibinfo{year}{2022}\natexlab{}.
\newblock \showarticletitle{{A Survey on Automated {Fact-Checking}}}.
\newblock \bibinfo{journal}{\emph{Transactions of the Association for Computational Linguistics}}  \bibinfo{volume}{10} (\bibinfo{year}{2022}), \bibinfo{pages}{178--206}.
\newblock
\showISSN{2307-387X}
\urldef\tempurl%
\url{https://doi.org/10.1162/tacl\_a\_00454}
\showDOI{\tempurl}


\bibitem[Gupta and Bendersky(2015)]%
        {Gupta2015-fh}
\bibfield{author}{\bibinfo{person}{Manish Gupta} {and} \bibinfo{person}{Michael Bendersky}.} \bibinfo{year}{2015}\natexlab{}.
\newblock \showarticletitle{Information Retrieval with Verbose Queries}. In \bibinfo{booktitle}{\emph{Proceedings of the 38th International {ACM} {SIGIR} Conference on Research and Development in Information Retrieval}} (Santiago, Chile) \emph{(\bibinfo{series}{SIGIR '15})}. \bibinfo{publisher}{Association for Computing Machinery}, \bibinfo{address}{New York, NY, USA}, \bibinfo{pages}{1121--1124}.
\newblock
\showISBNx{9781450336215}
\urldef\tempurl%
\url{https://doi.org/10.1145/2766462.2767877}
\showDOI{\tempurl}


\bibitem[Hameleers and van~der Meer(2020)]%
        {Hameleers2020-sp}
\bibfield{author}{\bibinfo{person}{Michael Hameleers} {and} \bibinfo{person}{Toni G L~A van~der Meer}.} \bibinfo{year}{2020}\natexlab{}.
\newblock \showarticletitle{Misinformation and Polarization in a {High-Choice} Media Environment: How Effective Are Political {Fact-Checkers}?}
\newblock \bibinfo{journal}{\emph{Communication research}} \bibinfo{volume}{47}, \bibinfo{number}{2} (\bibinfo{date}{1~March} \bibinfo{year}{2020}), \bibinfo{pages}{227--250}.
\newblock
\showISSN{0093-6502, 1552-3810}
\urldef\tempurl%
\url{https://doi.org/10.1177/0093650218819671}
\showDOI{\tempurl}


\bibitem[Hassan et~al\mbox{.}(2017)]%
        {Hassan2017-fn}
\bibfield{author}{\bibinfo{person}{Naeemul Hassan}, \bibinfo{person}{Fatma Arslan}, \bibinfo{person}{Chengkai Li}, {and} \bibinfo{person}{Mark Tremayne}.} \bibinfo{year}{2017}\natexlab{}.
\newblock \showarticletitle{Toward automated fact-checking: Detecting check-worthy factual claims by claimbuster}.
\newblock \bibinfo{journal}{\emph{Proceedings of the ACM SIGKDD International Conference on Knowledge Discovery and Data Mining}}  \bibinfo{volume}{Part F1296} (\bibinfo{year}{2017}), \bibinfo{pages}{1803--1812}.
\newblock
\urldef\tempurl%
\url{https://doi.org/10.1145/3097983.3098131}
\showDOI{\tempurl}


\bibitem[Horvitz(1999)]%
        {Horvitz1999-wp}
\bibfield{author}{\bibinfo{person}{Eric Horvitz}.} \bibinfo{year}{1999}\natexlab{}.
\newblock \showarticletitle{Principles of mixed-initiative user interfaces}. In \bibinfo{booktitle}{\emph{Proceedings of the {SIGCHI} conference on Human Factors in Computing Systems}} (Pittsburgh, Pennsylvania, USA) \emph{(\bibinfo{series}{CHI '99})}. \bibinfo{publisher}{Association for Computing Machinery}, \bibinfo{address}{New York, NY, USA}, \bibinfo{pages}{159--166}.
\newblock
\showISBNx{9780201485592}
\urldef\tempurl%
\url{https://doi.org/10.1145/302979.303030}
\showDOI{\tempurl}


\bibitem[Hvannberg et~al\mbox{.}(2006)]%
        {Hvannberg2006-nb}
\bibfield{author}{\bibinfo{person}{Ebba~Thora Hvannberg}, \bibinfo{person}{Effie Lai-Chong Law}, {and} \bibinfo{person}{Marta~Krist{\'\i}n L{\'e}rusd{\'o}ttir}.} \bibinfo{year}{2006}\natexlab{}.
\newblock \showarticletitle{Heuristic evaluation: Comparing ways of finding and reporting usability problems}.
\newblock \bibinfo{journal}{\emph{Interacting with computers}} \bibinfo{volume}{19}, \bibinfo{number}{2} (\bibinfo{date}{1~Dec.} \bibinfo{year}{2006}), \bibinfo{pages}{225--240}.
\newblock
\showISSN{0953-5438}
\urldef\tempurl%
\url{https://doi.org/10.1016/j.intcom.2006.10.001}
\showDOI{\tempurl}


\bibitem[Jack(2017)]%
        {Jack2017-cn}
\bibfield{author}{\bibinfo{person}{C Jack}.} \bibinfo{year}{2017}\natexlab{}.
\newblock \showarticletitle{Lexicon of lies: terms for problematic information}.
\newblock \bibinfo{journal}{\emph{Data \& Society}} \bibinfo{volume}{3}, \bibinfo{number}{22} (\bibinfo{date}{9~Aug.} \bibinfo{year}{2017}), \bibinfo{pages}{1094--1096}.
\newblock


\bibitem[Jasim et~al\mbox{.}(2022)]%
        {Jasim2022-rm}
\bibfield{author}{\bibinfo{person}{Mahmood Jasim}, \bibinfo{person}{Christopher Collins}, \bibinfo{person}{Ali Sarvghad}, {and} \bibinfo{person}{Narges Mahyar}.} \bibinfo{year}{2022}\natexlab{}.
\newblock \showarticletitle{Supporting Serendipitous Discovery and Balanced Analysis of Online Product Reviews with {Interaction-Driven} Metrics and {Bias-Mitigating} Suggestions}. In \bibinfo{booktitle}{\emph{Proceedings of the 2022 {CHI} Conference on Human Factors in Computing Systems}} (New Orleans, LA, USA) \emph{(\bibinfo{series}{CHI '22}, \bibinfo{number}{Article 9})}. \bibinfo{publisher}{Association for Computing Machinery}, \bibinfo{address}{New York, NY, USA}, \bibinfo{pages}{1--24}.
\newblock
\showISBNx{9781450391573}
\urldef\tempurl%
\url{https://doi.org/10.1145/3491102.3517649}
\showDOI{\tempurl}


\bibitem[Jiang et~al\mbox{.}(2014)]%
        {jiang2014searching}
\bibfield{author}{\bibinfo{person}{Jiepu Jiang}, \bibinfo{person}{Daqing He}, {and} \bibinfo{person}{James Allan}.} \bibinfo{year}{2014}\natexlab{}.
\newblock \showarticletitle{Searching, browsing, and clicking in a search session: changes in user behavior by task and over time}. In \bibinfo{booktitle}{\emph{Proceedings of the 37th international ACM SIGIR conference on Research \& development in information retrieval}}. \bibinfo{pages}{607--616}.
\newblock


\bibitem[Jiang et~al\mbox{.}(2017)]%
        {Jiang2017-fi}
\bibfield{author}{\bibinfo{person}{Jiepu Jiang}, \bibinfo{person}{Daqing He}, {and} \bibinfo{person}{James Allan}.} \bibinfo{year}{2017}\natexlab{}.
\newblock \showarticletitle{Comparing In Situ and Multidimensional Relevance Judgments}. In \bibinfo{booktitle}{\emph{Proceedings of the 40th International {ACM} {SIGIR} Conference on Research and Development in Information Retrieval}} (Shinjuku, Tokyo, Japan) \emph{(\bibinfo{series}{SIGIR '17})}. \bibinfo{publisher}{Association for Computing Machinery}, \bibinfo{address}{New York, NY, USA}, \bibinfo{pages}{405--414}.
\newblock
\showISBNx{9781450350228}
\urldef\tempurl%
\url{https://doi.org/10.1145/3077136.3080840}
\showDOI{\tempurl}


\bibitem[Juneja and Mitra(2022)]%
        {Juneja2022-fx}
\bibfield{author}{\bibinfo{person}{Prerna Juneja} {and} \bibinfo{person}{Tanushree Mitra}.} \bibinfo{year}{2022}\natexlab{}.
\newblock \showarticletitle{Human and technological infrastructures of fact-checking}.
\newblock \bibinfo{journal}{\emph{Proceedings of the ACM on human-computer interaction}} \bibinfo{volume}{6}, \bibinfo{number}{CSCW2} (\bibinfo{date}{7~Nov.} \bibinfo{year}{2022}), \bibinfo{pages}{1--36}.
\newblock
\showISSN{2573-0142}
\urldef\tempurl%
\url{https://doi.org/10.1145/3555143}
\showDOI{\tempurl}


\bibitem[Kern et~al\mbox{.}(2018)]%
        {Kern2018-vy}
\bibfield{author}{\bibinfo{person}{Dagmar Kern}, \bibinfo{person}{Wilko van Hoek}, {and} \bibinfo{person}{Daniel Hienert}.} \bibinfo{year}{2018}\natexlab{}.
\newblock \showarticletitle{Evaluation of a search interface for preference-based ranking: measuring user satisfaction and system performance}. In \bibinfo{booktitle}{\emph{Proceedings of the 10th Nordic Conference on {Human-Computer} Interaction}} (Oslo, Norway) \emph{(\bibinfo{series}{NordiCHI '18})}. \bibinfo{publisher}{Association for Computing Machinery}, \bibinfo{address}{New York, NY, USA}, \bibinfo{pages}{184--194}.
\newblock
\showISBNx{9781450364379}
\urldef\tempurl%
\url{https://doi.org/10.1145/3240167.3240170}
\showDOI{\tempurl}


\bibitem[Konstantinovskiy et~al\mbox{.}(2021)]%
        {Konstantinovskiy2021-ga}
\bibfield{author}{\bibinfo{person}{Lev Konstantinovskiy}, \bibinfo{person}{Oliver Price}, \bibinfo{person}{Mevan Babakar}, {and} \bibinfo{person}{Arkaitz Zubiaga}.} \bibinfo{year}{2021}\natexlab{}.
\newblock \showarticletitle{Towards Automated Factchecking: Developing an Annotation Schema and Benchmark for Consistent Automated Claim Detection}.
\newblock \bibinfo{journal}{\emph{Digital Threats}} \bibinfo{volume}{2}, \bibinfo{number}{2} (\bibinfo{date}{15~April} \bibinfo{year}{2021}), \bibinfo{pages}{1--16}.
\newblock
\showISSN{2576-5337}
\urldef\tempurl%
\url{https://doi.org/10.1145/3412869}
\showDOI{\tempurl}


\bibitem[Krieg et~al\mbox{.}(2022)]%
        {Krieg2022-wl}
\bibfield{author}{\bibinfo{person}{Klara Krieg}, \bibinfo{person}{Emilia Parada-Cabaleiro}, \bibinfo{person}{Markus Schedl}, {and} \bibinfo{person}{Navid Rekabsaz}.} \bibinfo{year}{2022}\natexlab{}.
\newblock \showarticletitle{Do Perceived Gender Biases in Retrieval Results Affect Relevance Judgements?}. In \bibinfo{booktitle}{\emph{Advances in Bias and Fairness in Information Retrieval}}. \bibinfo{publisher}{Springer International Publishing}, \bibinfo{pages}{104--116}.
\newblock
\urldef\tempurl%
\url{https://doi.org/10.1007/978-3-031-09316-6\_10}
\showDOI{\tempurl}


\bibitem[Liu et~al\mbox{.}(2024)]%
        {Liu2023-ur}
\bibfield{author}{\bibinfo{person}{Houjiang Liu}, \bibinfo{person}{Anubrata Das}, \bibinfo{person}{Alexander Boltz}, \bibinfo{person}{Didi Zhou}, \bibinfo{person}{Daisy Pinaroc}, \bibinfo{person}{Matthew Lease}, {and} \bibinfo{person}{Min~Kyung Lee}.} \bibinfo{year}{2024}\natexlab{}.
\newblock \showarticletitle{Human-centered {NLP} Fact-checking: Co-Designing with Fact-checkers using Matchmaking for {AI}}.
\newblock \bibinfo{journal}{\emph{Proceedings of the ACM on human-computer interaction}} \bibinfo{volume}{8}, \bibinfo{number}{CSCW2} (\bibinfo{date}{7~Nov.} \bibinfo{year}{2024}), \bibinfo{pages}{1--44}.
\newblock
\showISSN{2573-0142}
\urldef\tempurl%
\url{https://doi.org/10.1145/3686962}
\showDOI{\tempurl}


\bibitem[L{\o}vlie et~al\mbox{.}(2023)]%
        {Lovlie2023-df}
\bibfield{author}{\bibinfo{person}{Anders~Sundnes L{\o}vlie}, \bibinfo{person}{Astrid Waagstein}, {and} \bibinfo{person}{Peter Hyldg{\aa}rd}.} \bibinfo{year}{2023}\natexlab{}.
\newblock \showarticletitle{``How Trustworthy Is This Research?'' Designing a Tool to Help Readers Understand Evidence and Uncertainty in Science Journalism}.
\newblock \bibinfo{journal}{\emph{Digital Journalism}} \bibinfo{volume}{11}, \bibinfo{number}{3} (\bibinfo{date}{16~March} \bibinfo{year}{2023}), \bibinfo{pages}{431--464}.
\newblock
\showISSN{2167-0811}
\urldef\tempurl%
\url{https://doi.org/10.1080/21670811.2023.2193344}
\showDOI{\tempurl}


\bibitem[Majithia et~al\mbox{.}(2019)]%
        {Majithia2019-eu}
\bibfield{author}{\bibinfo{person}{Sarthak Majithia}, \bibinfo{person}{Fatma Arslan}, \bibinfo{person}{Sumeet Lubal}, \bibinfo{person}{Damian Jimenez}, \bibinfo{person}{Priyank Arora}, \bibinfo{person}{Josue Caraballo}, {and} \bibinfo{person}{Chengkai Li}.} \bibinfo{year}{2019}\natexlab{}.
\newblock \showarticletitle{{ClaimPortal}: Integrated monitoring, searching, checking, and analytics of factual claims on twitter}. In \bibinfo{booktitle}{\emph{Proceedings of the 57th Annual Meeting of the Association for Computational Linguistics: System Demonstrations}} (Florence, Italy). \bibinfo{publisher}{Association for Computational Linguistics}, \bibinfo{address}{Stroudsburg, PA, USA}, \bibinfo{pages}{153--158}.
\newblock
\urldef\tempurl%
\url{https://doi.org/10.18653/v1/p19-3026}
\showDOI{\tempurl}


\bibitem[Marchionini(2006)]%
        {Marchionini2006-wk}
\bibfield{author}{\bibinfo{person}{Gary Marchionini}.} \bibinfo{year}{2006}\natexlab{}.
\newblock \showarticletitle{Exploratory search: from finding to understanding}.
\newblock \bibinfo{journal}{\emph{Commun. ACM}} \bibinfo{volume}{49}, \bibinfo{number}{4} (\bibinfo{date}{1~April} \bibinfo{year}{2006}), \bibinfo{pages}{41--46}.
\newblock
\showISSN{0001-0782}
\urldef\tempurl%
\url{https://doi.org/10.1145/1121949.1121979}
\showDOI{\tempurl}


\bibitem[McLeod(2007)]%
        {McLeod2007-kt}
\bibfield{author}{\bibinfo{person}{Saul McLeod}.} \bibinfo{year}{2007}\natexlab{}.
\newblock \showarticletitle{Maslow's hierarchy of needs}.
\newblock \bibinfo{journal}{\emph{Simply psychology}} \bibinfo{volume}{1}, \bibinfo{number}{1-18} (\bibinfo{year}{2007}).
\newblock


\bibitem[Mena(2019)]%
        {Mena2019-ws}
\bibfield{author}{\bibinfo{person}{Paul Mena}.} \bibinfo{year}{2019}\natexlab{}.
\newblock \showarticletitle{Principles and boundaries of fact-checking: Journalists’ perceptions}.
\newblock \bibinfo{journal}{\emph{Journalism practice}} \bibinfo{volume}{13}, \bibinfo{number}{6} (\bibinfo{date}{3~July} \bibinfo{year}{2019}), \bibinfo{pages}{657--672}.
\newblock
\showISSN{1751-2786,1751-2794}
\urldef\tempurl%
\url{https://doi.org/10.1080/17512786.2018.1547655}
\showDOI{\tempurl}


\bibitem[Micallef et~al\mbox{.}(2022)]%
        {Micallef2022-vp}
\bibfield{author}{\bibinfo{person}{Nicholas Micallef}, \bibinfo{person}{Vivienne Armacost}, \bibinfo{person}{Nasir Memon}, {and} \bibinfo{person}{Sameer Patil}.} \bibinfo{year}{2022}\natexlab{}.
\newblock \showarticletitle{True or False: Studying the Work Practices of Professional {Fact-Checkers}}.
\newblock \bibinfo{journal}{\emph{Proceedings of the ACM on Human-Computer Interaction}} \bibinfo{volume}{6}, \bibinfo{number}{CSCW1} (\bibinfo{date}{7~April} \bibinfo{year}{2022}), \bibinfo{pages}{1--44}.
\newblock
\urldef\tempurl%
\url{https://doi.org/10.1145/3512974}
\showDOI{\tempurl}


\bibitem[Min et~al\mbox{.}(2023)]%
        {Min2023-yy}
\bibfield{author}{\bibinfo{person}{Bonan Min}, \bibinfo{person}{Hayley Ross}, \bibinfo{person}{Elior Sulem}, \bibinfo{person}{Amir Pouran~Ben Veyseh}, \bibinfo{person}{Thien~Huu Nguyen}, \bibinfo{person}{Oscar Sainz}, \bibinfo{person}{Eneko Agirre}, \bibinfo{person}{Ilana Heintz}, {and} \bibinfo{person}{Dan Roth}.} \bibinfo{year}{2023}\natexlab{}.
\newblock \showarticletitle{Recent Advances in Natural Language Processing via Large Pre-trained Language Models: A Survey}.
\newblock \bibinfo{journal}{\emph{ACM Comput. Surv.}} \bibinfo{volume}{56}, \bibinfo{number}{2} (\bibinfo{date}{14~Sept.} \bibinfo{year}{2023}), \bibinfo{pages}{1--40}.
\newblock
\showISSN{0360-0300}
\urldef\tempurl%
\url{https://doi.org/10.1145/3605943}
\showDOI{\tempurl}


\bibitem[Nakov et~al\mbox{.}(2021)]%
        {Nakov2021-sq}
\bibfield{author}{\bibinfo{person}{Preslav Nakov}, \bibinfo{person}{David Corney}, \bibinfo{person}{Maram Hasanain}, \bibinfo{person}{Firoj Alam}, \bibinfo{person}{Tamer Elsayed}, \bibinfo{person}{Alberto Barr{\'o}n-Cede{\~n}o}, \bibinfo{person}{Paolo Papotti}, \bibinfo{person}{Shaden Shaar}, {and} \bibinfo{person}{Giovanni Da~San~Martino}.} \bibinfo{year}{2021}\natexlab{}.
\newblock \showarticletitle{Automated {Fact-Checking} for Assisting Human {Fact-Checkers}}. In \bibinfo{booktitle}{\emph{Proceedings of the Thirtieth International Joint Conference on Artificial Intelligence, {IJCAI-21}}}, \bibfield{editor}{\bibinfo{person}{Zhi-Hua Zhou}} (Ed.). \bibinfo{publisher}{International Joint Conferences on Artificial Intelligence Organization}, \bibinfo{pages}{4551--4558}.
\newblock
\urldef\tempurl%
\url{https://doi.org/10.24963/ijcai.2021/619}
\showDOI{\tempurl}


\bibitem[Neumann et~al\mbox{.}(2022)]%
        {Neumann2022-ip}
\bibfield{author}{\bibinfo{person}{Terrence Neumann}, \bibinfo{person}{Maria De-Arteaga}, {and} \bibinfo{person}{Sina Fazelpour}.} \bibinfo{year}{2022}\natexlab{}.
\newblock \showarticletitle{Justice in Misinformation Detection Systems: An Analysis of Algorithms, Stakeholders, and Potential Harms}. In \bibinfo{booktitle}{\emph{Proceedings of the 2022 {ACM} Conference on Fairness, Accountability, and Transparency}} (Seoul, Republic of Korea) \emph{(\bibinfo{series}{FAccT '22})}. \bibinfo{publisher}{Association for Computing Machinery}, \bibinfo{address}{New York, NY, USA}, \bibinfo{pages}{1504--1515}.
\newblock
\showISBNx{9781450393522}
\urldef\tempurl%
\url{https://doi.org/10.1145/3531146.3533205}
\showDOI{\tempurl}


\bibitem[Novick and Hurley(2001)]%
        {Novick2001-sz}
\bibfield{author}{\bibinfo{person}{Laura~R Novick} {and} \bibinfo{person}{Sean~M Hurley}.} \bibinfo{year}{2001}\natexlab{}.
\newblock \showarticletitle{To Matrix, Network, or Hierarchy: That Is the Question}.
\newblock \bibinfo{journal}{\emph{Cognitive psychology}} \bibinfo{volume}{42}, \bibinfo{number}{2} (\bibinfo{date}{1~March} \bibinfo{year}{2001}), \bibinfo{pages}{158--216}.
\newblock
\showISSN{0010-0285}
\urldef\tempurl%
\url{https://doi.org/10.1006/cogp.2000.0746}
\showDOI{\tempurl}


\bibitem[Oard(2013)]%
        {Oard2013-qu}
\bibfield{author}{\bibinfo{person}{Douglas~W Oard}.} \bibinfo{year}{2013}\natexlab{}.
\newblock \showarticletitle{Information retrieval for {E}-discovery}.
\newblock \bibinfo{journal}{\emph{Foundations and Trends® in Information Retrieval}} \bibinfo{volume}{7}, \bibinfo{number}{2-3} (\bibinfo{year}{2013}), \bibinfo{pages}{99--237}.
\newblock
\showISSN{1554-0669,1554-0677}
\urldef\tempurl%
\url{https://doi.org/10.1561/1500000025}
\showDOI{\tempurl}


\bibitem[Papenmeier et~al\mbox{.}(2023)]%
        {Papenmeier2023-cx}
\bibfield{author}{\bibinfo{person}{Andrea Papenmeier}, \bibinfo{person}{Daniel Hienert}, \bibinfo{person}{Firas Sabbah}, \bibinfo{person}{Norbert Fuhr}, {and} \bibinfo{person}{Dagmar Kern}.} \bibinfo{year}{2023}\natexlab{}.
\newblock \showarticletitle{{UNDR}: {User-Needs-Driven} Ranking of Products in {E-Commerce}}.
\newblock  (\bibinfo{date}{13~Feb.} \bibinfo{year}{2023}).
\newblock
\showeprint[arxiv]{2302.06398}~[cs.IR]


\bibitem[Paul and Baron(2006)]%
        {Paul2006-mg}
\bibfield{author}{\bibinfo{person}{George~L Paul} {and} \bibinfo{person}{Jason~R Baron}.} \bibinfo{year}{2006}\natexlab{}.
\newblock \showarticletitle{Information inflation: Can the legal system adapt? Annual survey}.
\newblock \bibinfo{journal}{\emph{Richmond Journal of Law and Technology}} \bibinfo{volume}{13}, \bibinfo{number}{3} (\bibinfo{year}{2006}), \bibinfo{pages}{1--42}.
\newblock
\showISSN{1091-7322}


\bibitem[Pedregosa et~al\mbox{.}(2011)]%
        {Pedregosa2011-tl}
\bibfield{author}{\bibinfo{person}{Fabian Pedregosa}, \bibinfo{person}{Ga{\"e}l Varoquaux}, \bibinfo{person}{Alexandre Gramfort}, \bibinfo{person}{Vincent Michel}, \bibinfo{person}{Bertrand Thirion}, \bibinfo{person}{Olivier Grisel}, \bibinfo{person}{Mathieu Blondel}, \bibinfo{person}{Peter Prettenhofer}, \bibinfo{person}{Ron Weiss}, \bibinfo{person}{Vincent Dubourg}, {and} \bibinfo{person}{{Others}}.} \bibinfo{year}{2011}\natexlab{}.
\newblock \showarticletitle{Scikit-learn: Machine learning in Python}.
\newblock \bibinfo{journal}{\emph{the Journal of machine Learning research}}  \bibinfo{volume}{12} (\bibinfo{year}{2011}), \bibinfo{pages}{2825--2830}.
\newblock


\bibitem[Pierce(2014)]%
        {Pierce2014-am}
\bibfield{author}{\bibinfo{person}{James Pierce}.} \bibinfo{year}{2014}\natexlab{}.
\newblock \showarticletitle{On the presentation and production of design research artifacts in {HCI}}. In \bibinfo{booktitle}{\emph{Proceedings of the 2014 conference on Designing interactive systems}}. \bibinfo{publisher}{ACM}, \bibinfo{address}{New York, NY, USA}.
\newblock
\showISBNx{9781450329026}
\urldef\tempurl%
\url{https://doi.org/10.1145/2598510.2598525}
\showDOI{\tempurl}


\bibitem[Prochner and Godin(2022)]%
        {Prochner2022-lg}
\bibfield{author}{\bibinfo{person}{Isabel Prochner} {and} \bibinfo{person}{Danny Godin}.} \bibinfo{year}{2022}\natexlab{}.
\newblock \showarticletitle{Quality in research through design projects: Recommendations for evaluation and enhancement}.
\newblock \bibinfo{journal}{\emph{Design Studies}}  \bibinfo{volume}{78} (\bibinfo{date}{1~Jan.} \bibinfo{year}{2022}), \bibinfo{pages}{101061}.
\newblock
\showISSN{0142-694X}
\urldef\tempurl%
\url{https://doi.org/10.1016/j.destud.2021.101061}
\showDOI{\tempurl}


\bibitem[Procter et~al\mbox{.}(2023)]%
        {Procter2023-lu}
\bibfield{author}{\bibinfo{person}{Rob Procter}, \bibinfo{person}{Miguel~Arana Catania}, \bibinfo{person}{Yulan He}, \bibinfo{person}{Maria Liakata}, \bibinfo{person}{Arkaitz Zubiaga}, \bibinfo{person}{Elena Kochkina}, {and} \bibinfo{person}{Runcong Zhao}.} \bibinfo{year}{2023}\natexlab{}.
\newblock \showarticletitle{Some Observations on {Fact-Checking} Work with Implications for Computational Support}. In \bibinfo{booktitle}{\emph{Workshop Proceedings of the 17th International {AAAI} Conference on Web and Social Media}} (Limassol, Cyprus). \bibinfo{publisher}{workshop-proceedings.icwsm.org}.
\newblock
\urldef\tempurl%
\url{https://doi.org/10.36190/2023.28}
\showDOI{\tempurl}


\bibitem[Reimers and Gurevych(2019)]%
        {Reimers2019-pc}
\bibfield{author}{\bibinfo{person}{Nils Reimers} {and} \bibinfo{person}{Iryna Gurevych}.} \bibinfo{year}{2019}\natexlab{}.
\newblock \showarticletitle{Sentence-{BERT}: Sentence embeddings using Siamese {BERT}-networks}. In \bibinfo{booktitle}{\emph{Proceedings of the 2019 Conference on Empirical Methods in Natural Language Processing and the 9th International Joint Conference on Natural Language Processing (EMNLP-IJCNLP)}}, \bibfield{editor}{\bibinfo{person}{Kentaro Inui}, \bibinfo{person}{Jing Jiang}, \bibinfo{person}{Vincent Ng}, {and} \bibinfo{person}{Xiaojun Wan}} (Eds.). \bibinfo{publisher}{Association for Computational Linguistics}, \bibinfo{address}{Stroudsburg, PA, USA}, \bibinfo{pages}{3982--3992}.
\newblock
\urldef\tempurl%
\url{https://doi.org/10.18653/v1/d19-1410}
\showDOI{\tempurl}


\bibitem[Saaty(2008)]%
        {Saaty2008-bz}
\bibfield{author}{\bibinfo{person}{Thomas~L Saaty}.} \bibinfo{year}{2008}\natexlab{}.
\newblock \showarticletitle{Decision making with the analytic hierarchy process}.
\newblock \bibinfo{journal}{\emph{International journal of services sciences}} \bibinfo{volume}{1}, \bibinfo{number}{1} (\bibinfo{year}{2008}), \bibinfo{pages}{83}.
\newblock
\showISSN{1753-1446, 1753-1454}
\urldef\tempurl%
\url{https://doi.org/10.1504/ijssci.2008.017590}
\showDOI{\tempurl}


\bibitem[Saltz et~al\mbox{.}(2021)]%
        {Saltz2021-yg}
\bibfield{author}{\bibinfo{person}{Emily Saltz}, \bibinfo{person}{Soubhik Barari}, \bibinfo{person}{Claire Leibowicz}, {and} \bibinfo{person}{Claire Wardle}.} \bibinfo{year}{2021}\natexlab{}.
\newblock \showarticletitle{Misinformation interventions are common, divisive, and poorly understood}.
\newblock \bibinfo{journal}{\emph{Harvard Kennedy School Misinformation Review}} (\bibinfo{date}{27~Oct.} \bibinfo{year}{2021}).
\newblock
\showISSN{2766-1652}
\urldef\tempurl%
\url{https://doi.org/10.37016/mr-2020-81}
\showDOI{\tempurl}


\bibitem[Saracevic(1975)]%
        {saracevic1975relevance}
\bibfield{author}{\bibinfo{person}{Tefko Saracevic}.} \bibinfo{year}{1975}\natexlab{}.
\newblock \showarticletitle{Relevance: A review of and a framework for the thinking on the notion in information science}.
\newblock \bibinfo{journal}{\emph{Journal of the American Society for information science}} \bibinfo{volume}{26}, \bibinfo{number}{6} (\bibinfo{year}{1975}), \bibinfo{pages}{321--343}.
\newblock


\bibitem[Saracevic(2007a)]%
        {saracevic2007relevanceII}
\bibfield{author}{\bibinfo{person}{Tefko Saracevic}.} \bibinfo{year}{2007}\natexlab{a}.
\newblock \showarticletitle{Relevance: A review of the literature and a framework for thinking on the notion in information science. Part II: Nature and manifestations of relevance}.
\newblock \bibinfo{journal}{\emph{Journal of the American society for information science and technology}} \bibinfo{volume}{58}, \bibinfo{number}{13} (\bibinfo{year}{2007}), \bibinfo{pages}{1915--1933}.
\newblock


\bibitem[Saracevic(2007b)]%
        {saracevic2007relevanceIII}
\bibfield{author}{\bibinfo{person}{Tefko Saracevic}.} \bibinfo{year}{2007}\natexlab{b}.
\newblock \showarticletitle{Relevance: A review of the literature and a framework for thinking on the notion in information science. Part III: Behavior and effects of relevance}.
\newblock \bibinfo{journal}{\emph{Journal of the American Society for information Science and Technology}} \bibinfo{volume}{58}, \bibinfo{number}{13} (\bibinfo{year}{2007}), \bibinfo{pages}{2126--2144}.
\newblock


\bibitem[Sarwar et~al\mbox{.}(2021)]%
        {sarwar-query-biased-summarization}
\bibfield{author}{\bibinfo{person}{Sheikh~Muhammad Sarwar}, \bibinfo{person}{Felipe Moraes}, \bibinfo{person}{Jiepu Jiang}, {and} \bibinfo{person}{James Allan}.} \bibinfo{year}{2021}\natexlab{}.
\newblock \showarticletitle{Utility of Missing Concepts in Query-biased Summarization}. In \bibinfo{booktitle}{\emph{Proceedings of the 44th International ACM SIGIR Conference on Research and Development in Information Retrieval}} (Virtual Event, Canada) \emph{(\bibinfo{series}{SIGIR '21})}. \bibinfo{publisher}{Association for Computing Machinery}, \bibinfo{address}{New York, NY, USA}, \bibinfo{pages}{2056–2060}.
\newblock
\showISBNx{9781450380379}
\urldef\tempurl%
\url{https://doi.org/10.1145/3404835.3463121}
\showDOI{\tempurl}


\bibitem[Schamber et~al\mbox{.}(1990)]%
        {Schamber1990-uv}
\bibfield{author}{\bibinfo{person}{Linda Schamber}, \bibinfo{person}{Michael~B Eisenberg}, {and} \bibinfo{person}{Michael~S Nilan}.} \bibinfo{year}{1990}\natexlab{}.
\newblock \showarticletitle{A re-examination of relevance: toward a dynamic, situational definition}.
\newblock \bibinfo{journal}{\emph{Information processing \& management}} \bibinfo{volume}{26}, \bibinfo{number}{6} (\bibinfo{date}{1~Jan.} \bibinfo{year}{1990}), \bibinfo{pages}{755--776}.
\newblock
\showISSN{0306-4573}
\urldef\tempurl%
\url{https://doi.org/10.1016/0306-4573(90)90050-C}
\showDOI{\tempurl}


\bibitem[Sehat et~al\mbox{.}(2024)]%
        {Sehat2023-xa}
\bibfield{author}{\bibinfo{person}{Connie~Moon Sehat}, \bibinfo{person}{Ryan Li}, \bibinfo{person}{Peipei Nie}, \bibinfo{person}{Tarunima Prabhakar}, {and} \bibinfo{person}{Amy~X Zhang}.} \bibinfo{year}{2024}\natexlab{}.
\newblock \showarticletitle{Misinformation as a Harm: Structured Approaches for {Fact-Checking} Prioritization}.
\newblock \bibinfo{journal}{\emph{Proc. ACM Hum.-Comput. Interact.}} \bibinfo{volume}{8}, \bibinfo{number}{CSCW1} (\bibinfo{date}{26~April} \bibinfo{year}{2024}), \bibinfo{pages}{1--36}.
\newblock
\urldef\tempurl%
\url{https://doi.org/10.1145/3641010}
\showDOI{\tempurl}


\bibitem[Shaar et~al\mbox{.}(2020)]%
        {Shaar2020-eb}
\bibfield{author}{\bibinfo{person}{Shaden Shaar}, \bibinfo{person}{Nikolay Babulkov}, \bibinfo{person}{Giovanni Da~San~Martino}, {and} \bibinfo{person}{Preslav Nakov}.} \bibinfo{year}{2020}\natexlab{}.
\newblock \showarticletitle{That is a Known Lie: Detecting Previously {Fact-Checked} Claims}. In \bibinfo{booktitle}{\emph{Proceedings of the 58th Annual Meeting of the Association for Computational Linguistics}}, \bibfield{editor}{\bibinfo{person}{Dan Jurafsky}, \bibinfo{person}{Joyce Chai}, \bibinfo{person}{Natalie Schluter}, {and} \bibinfo{person}{Joel Tetreault}} (Eds.). \bibinfo{publisher}{Association for Computational Linguistics}, \bibinfo{address}{Online}, \bibinfo{pages}{3607--3618}.
\newblock
\urldef\tempurl%
\url{https://doi.org/10.18653/v1/2020.acl-main.332}
\showDOI{\tempurl}


\bibitem[Singh et~al\mbox{.}(2021)]%
        {Singh2021-du}
\bibfield{author}{\bibinfo{person}{Prakhar Singh}, \bibinfo{person}{Anubrata Das}, \bibinfo{person}{Junyi~Jessy Li}, {and} \bibinfo{person}{Matthew Lease}.} \bibinfo{year}{2021}\natexlab{}.
\newblock \showarticletitle{The Case for Claim Difficulty Assessment in Automatic Fact Checking}.
\newblock  (\bibinfo{date}{20~Sept.} \bibinfo{year}{2021}).
\newblock
\showeprint[arxiv]{2109.09689}~[cs.CL]


\bibitem[Smith(2003)]%
        {Smith2003-fe}
\bibfield{author}{\bibinfo{person}{Sarah~Harrison Smith}.} \bibinfo{year}{2003}\natexlab{}.
\newblock \bibinfo{booktitle}{\emph{The fact checker's bible}}.
\newblock \bibinfo{publisher}{Anchor Books}.
\newblock
\showISBNx{9780385721066}


\bibitem[Spink et~al\mbox{.}(1998)]%
        {Spink1998-rg}
\bibfield{author}{\bibinfo{person}{Amanda Spink}, \bibinfo{person}{Howard Greisdorf}, {and} \bibinfo{person}{Judy Bateman}.} \bibinfo{year}{1998}\natexlab{}.
\newblock \showarticletitle{Examining different regions of relevance: From highly relevant to not relevant}. In \bibinfo{booktitle}{\emph{Proceedings of the {ASIST} Annual Meeting}}, Vol.~\bibinfo{volume}{35}. \bibinfo{pages}{3--12}.
\newblock


\bibitem[Swire-Thompson and Lazer(2020)]%
        {Swire-Thompson2020-vi}
\bibfield{author}{\bibinfo{person}{Briony Swire-Thompson} {and} \bibinfo{person}{David Lazer}.} \bibinfo{year}{2020}\natexlab{}.
\newblock \showarticletitle{Public Health and Online Misinformation: Challenges and Recommendations}.
\newblock \bibinfo{journal}{\emph{Annual review of public health}}  \bibinfo{volume}{41} (\bibinfo{date}{2~April} \bibinfo{year}{2020}), \bibinfo{pages}{433--451}.
\newblock
\showISSN{0163-7525, 1545-2093}
\urldef\tempurl%
\url{https://doi.org/10.1146/annurev-publhealth-040119-094127}
\showDOI{\tempurl}


\bibitem[Tamine and Chouquet(2017)]%
        {Tamine2017-bx}
\bibfield{author}{\bibinfo{person}{Lynda Tamine} {and} \bibinfo{person}{Cecile Chouquet}.} \bibinfo{year}{2017}\natexlab{}.
\newblock \showarticletitle{On the impact of domain expertise on query formulation, relevance assessment and retrieval performance in clinical settings}.
\newblock \bibinfo{journal}{\emph{Information processing \& management}} \bibinfo{volume}{53}, \bibinfo{number}{2} (\bibinfo{date}{1~March} \bibinfo{year}{2017}), \bibinfo{pages}{332--350}.
\newblock
\showISSN{0306-4573}
\urldef\tempurl%
\url{https://doi.org/10.1016/j.ipm.2016.11.004}
\showDOI{\tempurl}


\bibitem[Tang and Solomon(1998)]%
        {Tang1998-co}
\bibfield{author}{\bibinfo{person}{Rong Tang} {and} \bibinfo{person}{Paul Solomon}.} \bibinfo{year}{1998}\natexlab{}.
\newblock \showarticletitle{Toward an understanding of the dynamics of relevance judgment: An analysis of one person's search behavior}.
\newblock \bibinfo{journal}{\emph{Information processing \& management}} \bibinfo{volume}{34}, \bibinfo{number}{2} (\bibinfo{date}{1~March} \bibinfo{year}{1998}), \bibinfo{pages}{237--256}.
\newblock
\showISSN{0306-4573}
\urldef\tempurl%
\url{https://doi.org/10.1016/S0306-4573(97)00081-2}
\showDOI{\tempurl}


\bibitem[Venkatagiri et~al\mbox{.}(2023)]%
        {Venkatagiri2023-wu}
\bibfield{author}{\bibinfo{person}{Sukrit Venkatagiri}, \bibinfo{person}{Anirban Mukhopadhyay}, \bibinfo{person}{David Hicks}, \bibinfo{person}{Aaron Brantly}, {and} \bibinfo{person}{Kurt Luther}.} \bibinfo{year}{2023}\natexlab{}.
\newblock \showarticletitle{{CoSINT}: Designing a Collaborative Capture the Flag Competition to Investigate Misinformation}. In \bibinfo{booktitle}{\emph{Proceedings of the 2023 {ACM} Designing Interactive Systems Conference}} (Pittsburgh, PA, USA) \emph{(\bibinfo{series}{DIS '23})}. \bibinfo{publisher}{Association for Computing Machinery}, \bibinfo{address}{New York, NY, USA}, \bibinfo{pages}{2551--2572}.
\newblock
\showISBNx{9781450398930}
\urldef\tempurl%
\url{https://doi.org/10.1145/3563657.3595997}
\showDOI{\tempurl}


\bibitem[Vinhas and Bastos(2023)]%
        {Vinhas2023-hx}
\bibfield{author}{\bibinfo{person}{Ot{\'a}vio Vinhas} {and} \bibinfo{person}{Marco Bastos}.} \bibinfo{year}{2023}\natexlab{}.
\newblock \showarticletitle{The {WEIRD} governance of fact-checking and the politics of content moderation}.
\newblock \bibinfo{journal}{\emph{New Media \& Society}} (\bibinfo{date}{29~Nov.} \bibinfo{year}{2023}), \bibinfo{pages}{14614448231213942}.
\newblock
\showISSN{1461-4448}
\urldef\tempurl%
\url{https://doi.org/10.1177/14614448231213942}
\showDOI{\tempurl}


\bibitem[Wallace et~al\mbox{.}(2013)]%
        {Wallace2013-ho}
\bibfield{author}{\bibinfo{person}{Byron~C Wallace}, \bibinfo{person}{Issa~J Dahabreh}, \bibinfo{person}{Christopher~H Schmid}, \bibinfo{person}{Joseph Lau}, {and} \bibinfo{person}{Thomas~A Trikalinos}.} \bibinfo{year}{2013}\natexlab{}.
\newblock \showarticletitle{Modernizing the systematic review process to inform comparative effectiveness: tools and methods}.
\newblock \bibinfo{journal}{\emph{Journal of comparative effectiveness research}} \bibinfo{volume}{2}, \bibinfo{number}{3} (\bibinfo{date}{May} \bibinfo{year}{2013}), \bibinfo{pages}{273--282}.
\newblock
\showISSN{2042-6305,2042-6313}
\urldef\tempurl%
\url{https://doi.org/10.2217/cer.13.17}
\showDOI{\tempurl}


\bibitem[Wardle and Derakhshan(2017)]%
        {Wardle2017-fq}
\bibfield{author}{\bibinfo{person}{Claire Wardle} {and} \bibinfo{person}{Hossein Derakhshan}.} \bibinfo{year}{2017}\natexlab{}.
\newblock \bibinfo{booktitle}{\emph{Information disorder: Toward an interdisciplinary framework for research and policymaking}}. Vol.~\bibinfo{volume}{27}.
\newblock \bibinfo{publisher}{Council of Europe}.
\newblock


\bibitem[Westlund et~al\mbox{.}(2022)]%
        {Westlund_undated-po}
\bibfield{author}{\bibinfo{person}{Oscar Westlund}, \bibinfo{person}{Rebekah Larsen}, \bibinfo{person}{Lucas Graves}, \bibinfo{person}{Lasha Kavtaradze}, {and} \bibinfo{person}{Steen Steensen}.} \bibinfo{year}{2022}\natexlab{}.
\newblock \showarticletitle{Technologies and fact-checking. A sociotechnical mapping}.
\newblock \bibinfo{journal}{\emph{Disinformations Studies: Perspectives from An Emerging Field}} (\bibinfo{year}{2022}).
\newblock


\bibitem[Wilner et~al\mbox{.}(2023)]%
        {Wilner2023-wo}
\bibfield{author}{\bibinfo{person}{Tamar Wilner}, \bibinfo{person}{Kayo Mimizuka}, \bibinfo{person}{Ayesha Bhimdiwala}, \bibinfo{person}{Jason~C Young}, {and} \bibinfo{person}{Ahmer Arif}.} \bibinfo{year}{2023}\natexlab{}.
\newblock \showarticletitle{It's About Time: Attending to Temporality in Misinformation Interventions}. In \bibinfo{booktitle}{\emph{Proceedings of the 2023 {CHI} Conference on Human Factors in Computing Systems}} (Hamburg, Germany) \emph{(\bibinfo{series}{CHI '23}, \bibinfo{number}{Article 404})}. \bibinfo{publisher}{Association for Computing Machinery}, \bibinfo{address}{New York, NY, USA}, \bibinfo{pages}{1--19}.
\newblock
\showISBNx{9781450394215}
\urldef\tempurl%
\url{https://doi.org/10.1145/3544548.3581068}
\showDOI{\tempurl}


\bibitem[Wilson(1981)]%
        {Wilson1981-hc}
\bibfield{author}{\bibinfo{person}{T~D Wilson}.} \bibinfo{year}{1981}\natexlab{}.
\newblock \showarticletitle{On User Studies and Information Needs}.
\newblock \bibinfo{journal}{\emph{Journal of Documentation}} \bibinfo{volume}{37}, \bibinfo{number}{1} (\bibinfo{date}{1~Jan.} \bibinfo{year}{1981}), \bibinfo{pages}{3--15}.
\newblock
\showISSN{0022-0418}
\urldef\tempurl%
\url{https://doi.org/10.1108/eb026702}
\showDOI{\tempurl}


\bibitem[Xu and Chen(2006)]%
        {Xu2006-sj}
\bibfield{author}{\bibinfo{person}{Yunjie~(calvin) Xu} {and} \bibinfo{person}{Zhiwei Chen}.} \bibinfo{year}{2006}\natexlab{}.
\newblock \showarticletitle{Relevance judgment: What do information users consider beyond topicality?}
\newblock \bibinfo{journal}{\emph{Journal of the American Society for Information Science and Technology}} \bibinfo{volume}{57}, \bibinfo{number}{7} (\bibinfo{date}{May} \bibinfo{year}{2006}), \bibinfo{pages}{961--973}.
\newblock
\showISSN{1532-2882, 1532-2890}
\urldef\tempurl%
\url{https://doi.org/10.1002/asi.20361}
\showDOI{\tempurl}


\bibitem[Yee et~al\mbox{.}(2003)]%
        {Yee2003-cr}
\bibfield{author}{\bibinfo{person}{Ka-Ping Yee}, \bibinfo{person}{Kirsten Swearingen}, \bibinfo{person}{Kevin Li}, {and} \bibinfo{person}{Marti Hearst}.} \bibinfo{year}{2003}\natexlab{}.
\newblock \showarticletitle{Faceted metadata for image search and browsing}. In \bibinfo{booktitle}{\emph{Proceedings of the {SIGCHI} Conference on Human Factors in Computing Systems}} (Ft. Lauderdale, Florida, USA) \emph{(\bibinfo{series}{CHI '03})}. \bibinfo{publisher}{Association for Computing Machinery}, \bibinfo{address}{New York, NY, USA}, \bibinfo{pages}{401--408}.
\newblock
\showISBNx{9781581136302}
\urldef\tempurl%
\url{https://doi.org/10.1145/642611.642681}
\showDOI{\tempurl}


\bibitem[Zade et~al\mbox{.}(2023)]%
        {Zade2023-vu}
\bibfield{author}{\bibinfo{person}{Himanshu Zade}, \bibinfo{person}{Megan Woodruff}, \bibinfo{person}{Erika Johnson}, \bibinfo{person}{Mariah Stanley}, \bibinfo{person}{Zhennan Zhou}, \bibinfo{person}{Minh~Tu Huynh}, \bibinfo{person}{Alissa~Elizabeth Acheson}, \bibinfo{person}{Gary Hsieh}, {and} \bibinfo{person}{Kate Starbird}.} \bibinfo{year}{2023}\natexlab{}.
\newblock \showarticletitle{Tweet Trajectory and {AMPS-based} Contextual Cues can Help Users Identify Misinformation}.
\newblock \bibinfo{journal}{\emph{Proc. ACM Hum.-Comput. Interact.}} \bibinfo{volume}{7}, \bibinfo{number}{CSCW1} (\bibinfo{date}{16~April} \bibinfo{year}{2023}), \bibinfo{pages}{1--27}.
\newblock
\urldef\tempurl%
\url{https://doi.org/10.1145/3579536}
\showDOI{\tempurl}


\bibitem[Zhang et~al\mbox{.}(2014)]%
        {Zhang2014-yk}
\bibfield{author}{\bibinfo{person}{Yinglong Zhang}, \bibinfo{person}{Jin Zhang}, \bibinfo{person}{Matthew Lease}, {and} \bibinfo{person}{Jacek Gwizdka}.} \bibinfo{year}{2014}\natexlab{}.
\newblock \showarticletitle{Multidimensional relevance modeling via psychometrics and crowdsourcing}. In \bibinfo{booktitle}{\emph{Proceedings of the 37th international {ACM} {SIGIR} conference on Research \& development in information retrieval}} (Gold Coast, Queensland, Australia) \emph{(\bibinfo{series}{SIGIR '14})}. \bibinfo{publisher}{Association for Computing Machinery}, \bibinfo{address}{New York, NY, USA}, \bibinfo{pages}{435--444}.
\newblock
\showISBNx{9781450322577}
\urldef\tempurl%
\url{https://doi.org/10.1145/2600428.2609577}
\showDOI{\tempurl}


\bibitem[Zhao et~al\mbox{.}(2024)]%
        {Zhao2024-ek}
\bibfield{author}{\bibinfo{person}{Haiyan Zhao}, \bibinfo{person}{Hanjie Chen}, \bibinfo{person}{Fan Yang}, \bibinfo{person}{Ninghao Liu}, \bibinfo{person}{Huiqi Deng}, \bibinfo{person}{Hengyi Cai}, \bibinfo{person}{Shuaiqiang Wang}, \bibinfo{person}{Dawei Yin}, {and} \bibinfo{person}{Mengnan Du}.} \bibinfo{year}{2024}\natexlab{}.
\newblock \showarticletitle{Explainability for Large Language Models: A Survey}.
\newblock \bibinfo{journal}{\emph{ACM Trans. Intell. Syst. Technol.}} \bibinfo{volume}{15}, \bibinfo{number}{2} (\bibinfo{date}{22~Feb.} \bibinfo{year}{2024}), \bibinfo{pages}{1--38}.
\newblock
\showISSN{2157-6904}
\urldef\tempurl%
\url{https://doi.org/10.1145/3639372}
\showDOI{\tempurl}


\bibitem[Zhou and Zafarani(2020)]%
        {Zhou2020-mf}
\bibfield{author}{\bibinfo{person}{Xinyi Zhou} {and} \bibinfo{person}{Reza Zafarani}.} \bibinfo{year}{2020}\natexlab{}.
\newblock \showarticletitle{A Survey of Fake News: Fundamental Theories, Detection Methods, and Opportunities}.
\newblock \bibinfo{journal}{\emph{ACM Comput. Surv.}} \bibinfo{volume}{53}, \bibinfo{number}{5} (\bibinfo{date}{28~Sept.} \bibinfo{year}{2020}), \bibinfo{pages}{1--40}.
\newblock
\showISSN{0360-0300}
\urldef\tempurl%
\url{https://doi.org/10.1145/3395046}
\showDOI{\tempurl}


\bibitem[Zhu et~al\mbox{.}(2024)]%
        {Zhu2024-bn}
\bibfield{author}{\bibinfo{person}{Qihao Zhu}, \bibinfo{person}{Leah Chong}, \bibinfo{person}{Maria Yang}, {and} \bibinfo{person}{Jianxi Luo}.} \bibinfo{year}{2024}\natexlab{}.
\newblock \showarticletitle{Reading users' minds with {LLMs}: Mental inference for artificial empathy in design}.
\newblock \bibinfo{journal}{\emph{Journal of mechanical design (New York, N.Y.: 1990)}} \bibinfo{volume}{147}, \bibinfo{number}{6} (\bibinfo{date}{24~Dec.} \bibinfo{year}{2024}), \bibinfo{pages}{1--38}.
\newblock
\showISSN{1050-0472,1528-9001}
\urldef\tempurl%
\url{https://doi.org/10.1115/1.4067527}
\showDOI{\tempurl}


\bibitem[Zimmerman and Forlizzi(2014)]%
        {Zimmerman2014-cs}
\bibfield{author}{\bibinfo{person}{John Zimmerman} {and} \bibinfo{person}{Jodi Forlizzi}.} \bibinfo{year}{2014}\natexlab{}.
\newblock \showarticletitle{Research Through Design in {HCI}}.
\newblock In \bibinfo{booktitle}{\emph{Ways of Knowing in {HCI}}}, \bibfield{editor}{\bibinfo{person}{Judith~S Olson} {and} \bibinfo{person}{Wendy~A Kellogg}} (Eds.). \bibinfo{publisher}{Springer New York}, \bibinfo{address}{New York, NY}, \bibinfo{pages}{167--189}.
\newblock
\showISBNx{9781493903788}
\urldef\tempurl%
\url{https://doi.org/10.1007/978-1-4939-0378-8\_8}
\showDOI{\tempurl}


\bibitem[Zimmerman et~al\mbox{.}(2007)]%
        {Zimmerman2007-gh}
\bibfield{author}{\bibinfo{person}{John Zimmerman}, \bibinfo{person}{Jodi Forlizzi}, {and} \bibinfo{person}{Shelley Evenson}.} \bibinfo{year}{2007}\natexlab{}.
\newblock \showarticletitle{Research through design as a method for interaction design research in {HCI}}. In \bibinfo{booktitle}{\emph{Proceedings of the {SIGCHI} Conference on Human Factors in Computing Systems}} (San Jose, California, USA) \emph{(\bibinfo{series}{CHI '07})}. \bibinfo{publisher}{Association for Computing Machinery}, \bibinfo{address}{New York, NY, USA}, \bibinfo{pages}{493--502}.
\newblock
\showISBNx{9781595935939}
\urldef\tempurl%
\url{https://doi.org/10.1145/1240624.1240704}
\showDOI{\tempurl}


\bibitem[Zimmerman et~al\mbox{.}(2010)]%
        {Zimmerman2010-fb}
\bibfield{author}{\bibinfo{person}{John Zimmerman}, \bibinfo{person}{Erik Stolterman}, {and} \bibinfo{person}{Jodi Forlizzi}.} \bibinfo{year}{2010}\natexlab{}.
\newblock \showarticletitle{An analysis and critique of Research through Design: towards a formalization of a research approach}. In \bibinfo{booktitle}{\emph{Proceedings of the 8th {ACM} Conference on Designing Interactive Systems}} (Aarhus, Denmark) \emph{(\bibinfo{series}{DIS '10})}. \bibinfo{publisher}{Association for Computing Machinery}, \bibinfo{address}{New York, NY, USA}, \bibinfo{pages}{310--319}.
\newblock
\showISBNx{9781450301039}
\urldef\tempurl%
\url{https://doi.org/10.1145/1858171.1858228}
\showDOI{\tempurl}


\end{thebibliography}

\appendix

\section*{APPENDIX} 
\label{appendix:appendix}

\section{Claim prioritization task} \label{appendix:task description}
Imagine that at the beginning of your everyday fact-checking, you want to find a set of claims across diverse sub-topics relevant to COVID-19. Finally, you want to recommend 3 claim candidates to editors from the list of candidates you have already found. Specific steps are:
\begin{enumerate}
    \item In the "Select claims" page, please use different tool features to select a set of claims across diverse topics relevant to COVID-19 and save it to the "Your selection" page. This step can be done multiple times.
    \item In the "Create facet" page, please create a new criterion. You can adopt a criterion template and revise it based on your understanding of that criterion. Please do not use the default template.
    \item Return to the "Select claims" page; please conduct Step 1 again. This time, feel free to use tool features in combination with the new criterion.
    \item You can make multiple rounds of selections by repeating the previous steps.
    \item At the end, go to the "Your selection" page. Select three top claims as the final candidates to be checked.
\end{enumerate}

\section{Measurements for data collection} \label{appendix:measurement}
We organized the measurement of data collection across the three-phased RtD evaluation process. First, participants were asked to answer the pre-screening survey during the tool familiarization phase. Then, we conducted interviews with participants before and after the experimental study. User interaction measures were collected when participants completed the within-subjects study. A post-task questionnaire was delivered after participants used each interface. A post-system questionnaire was conducted at the end of the experimental study. 

\begin{itemize}
\item \textbf{Pre-screening survey}
\begin{itemize}
    \item \textit{5-point Likert-scale questions measured for each checkworthy dimension <X>:}
    \begin{itemize}
        \item \textit{Perceived importance}: ``<X> is an important factor resulting in a final fact-checked claim.''
        \item \textit{Ease of finding}: ``It is easy for me to identify <X> claims.''
        \item \textit{Criterion accuracy}: ``Claims that I finally checked are usually <X> as they first appeared.''
    \end{itemize}
    \item \textit{Open-ended questions:}
    \begin{itemize}
        \item \textit{Implicit ranking}: Considering these four criteria, how would you characterize their relative importance vs. one another? Please rank these criteria from the most to least important.
    \end{itemize}
\end{itemize}
\item \textbf{User interaction logs and interview questions}
\begin{itemize}
    \item \textit{Interview questions before performing the within-subjects study:}
    \begin{itemize}
        \item Why would you prioritize some criteria over others? 
        \item How would you triage claims from these four criteria when using the new tool?
    \end{itemize}
    \item \textit{User measurements collected during the within-subjects study}:
    \begin{itemize}
        \item \textit{\# Queries}: The number of queries submitted by the participant.
        \item \textit{\# Checkworthy slider changes}: The number of times the checkworthy slider(s) were changed.
        \item \textit{\# Customized slider changes}: The number of times the customized slider(s) were changed.
        \item \textit{\# Query similarity slider changes}: The number of times the query similarity slider was changed.
        \item \textit{\# Selected claims}: The number of interesting claims identified in the initial exploratory stage (with or without using the customized filters).
        \item \textit{\# Final claims found checkworthy}: Out of the three final claims selected, the number of these that were initially found with or without customized filters.
        \item \textit{Conversion rate}: the ratio \textit{\# Final claims found checkworthy / \# Selected claims}
    \end{itemize}
    \item \textit{Retrospective think-aloud after the task:}
    \begin{itemize}
        \item Please describe how you used the four criteria sliders to prioritize claims and why these claims caught your attention.
    \end{itemize}
    \item \textit{Interview questions after performing the task}: 
    \begin{itemize}
        \item What new difficulties have you found when using the tool? \item Did you find the tool to be effective to find claims that match the criteria you previously mentioned? 
        \item Which criterion is particularly effective to find claims? \item What are the benefits or limitations for you to prioritize claims when using the tool? 
        \item How did the customized filter work created by ChatGPT? \item What other possibilities would you want GenAI to help you prioritize claims?
    \end{itemize}
\end{itemize}
\item \textbf{Post-task questionnaires} 5-point Likert-scale questions
\begin{itemize}
        \item \textit{Claim Satisfaction}: I was satisfied with the claim candidates I found by using this tool.
        \item \textit{Learn}: Using this tool supports me to
        \begin{itemize}
            \item \textit{Understand topic scope}: understand the gist of the main claims topics and the scope of the claim collections.
            \item \textit{Acquire new perspective}: acquire new perspectives of checkworthiness.
        \end{itemize}
        \item \textit{Lookup}: Using this tool supports me to:
        \begin{itemize}
            \item \textit{Search specific topic}: search relevant claims with specific topics.
            \item \textit{Lookup many claims}: lookup as many relevant claims as possible.
        \end{itemize}
        \item \textit{Investigate}: Using this tool supports me to
        \begin{itemize}
            \item \textit{Select best claims}: efficiently select the best claim candidates.
            \item \textit{Uncover unexpected claims}: uncover unexpected claims
            \item \textit{Investigate multiple criteria}: investigate multiple aspects of checkworthiness.
            \item \textit{Operationalize multiple criteria}: operationalize multiple criteria of checkworthiness.
            \item \textit{Operationalize new criteria}: operationalize personal criteria to find claims other fact-checkers and journalists might miss or choose to ignore.
        \end{itemize}
\end{itemize}
\item \textbf{Post-system questionnaires} 5-point Likert-scale questions
\begin{itemize}
    \item \textit{Perceived Usefulness}:
        \begin{itemize}
            \item Using this tool in my job would enable me to accomplish tasks more quickly.
            \item Using this tool would improve my job performance.
            \item Using this tool in my job would increase my productivity.
            \item Using this tool would enhance my effectiveness on the job.
            \item Using this tool would make it easier to do my job.
            \item I would find this tool useful in my job.
        \end{itemize}
    \item \textit{Ease of Use}:
        \begin{itemize}
            \item Learning to operate the tool would be easy for me.
            \item I would find it easy to get this tool to do what I want it to do.
            \item My interaction with the tool would be clear and understandable.
            \item I would find this tool to be clear and understandable.
            \item It would be easy for me to become skillful at using this tool.
            \item I found this tool easy to use.
        \end{itemize}
\end{itemize}
\end{itemize}

\section{Prompts} \label{appendix:Prompt}

\begin{itemize}
    \item Prompt placeholders:
    \begin{itemize}
        \item INPUT: The claim used for assessment.
        \item NAME: The user enters the name of the new dimension.
        \item CONTEXT: The user describes the dimension in detail.
    \end{itemize}
    \item Prompts:
        \begin{itemize}
            \item \texttt{Based on the new [NAME] and [CONTEXT]. Identify whether the [INPUT] follows the [CONTEXT] and output yes or no.}
        \end{itemize}
    \item Output values:
        \begin{itemize}
            \item \{\texttt{"tokens", "top-logprobs"}\}
        \end{itemize}
\end{itemize}


\section{Low-fidelity wireframe} \label{appendix:low-fidenlity wireframe}

\begin{figure}[h]
    \centering
    \includegraphics[width=0.6\textwidth]{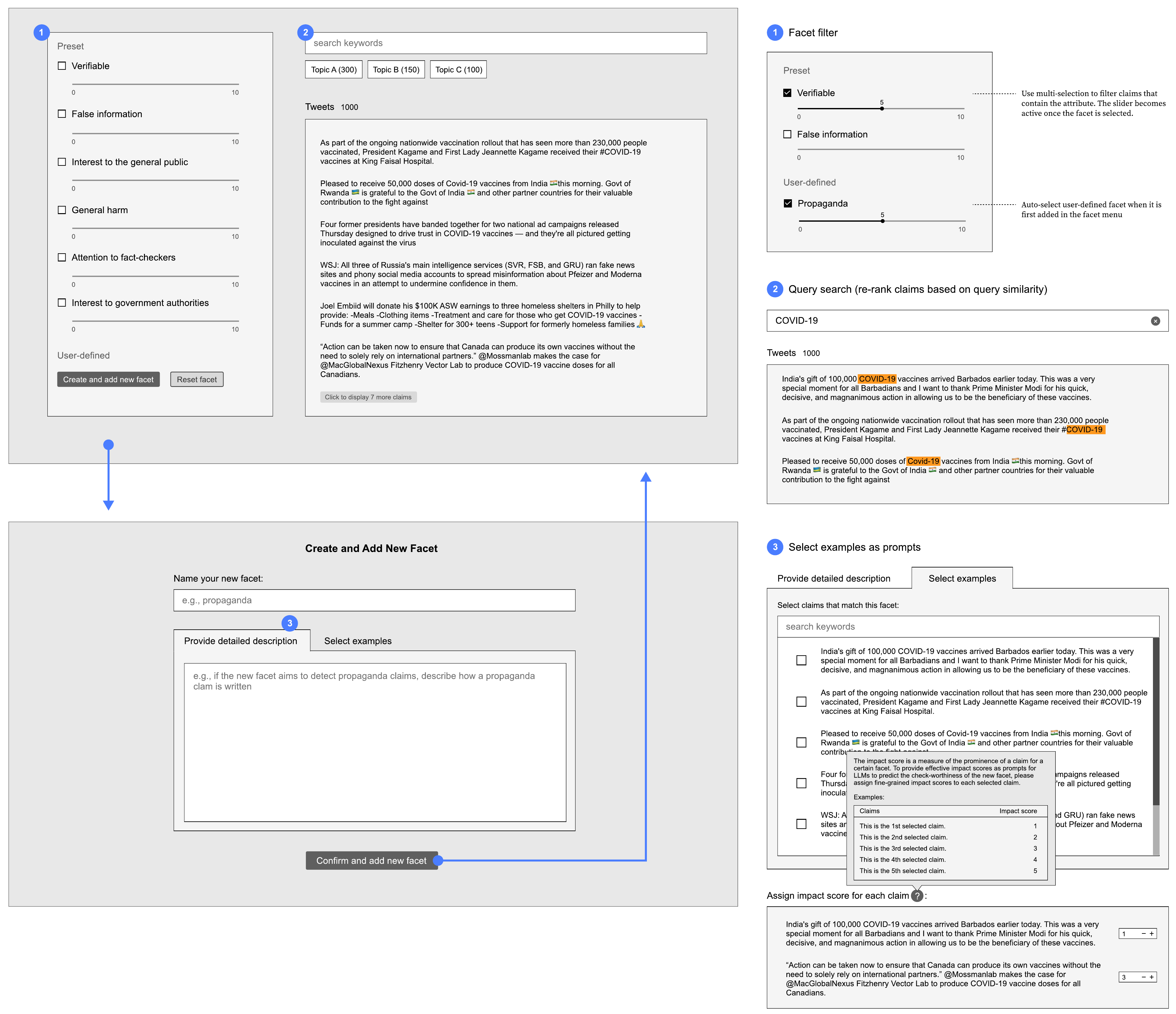}
    \caption{An example of an early low-fidelity wireframe and user workflow we created in Figma.}
    \Description{Low-fidelity wireframe}
    \label{fig:low-fidelity wireframe}
\end{figure}

\section{Preliminary findings} \label{sec:pilot_findings}

This section describes our preliminary findings from pilot tests to identify appropriate search tasks for fact-checkers. These tasks aim to help fact-checkers prioritize claims based on various dimensions of checkworthiness. 
It is well known in IR that user information needs often evolve over the course of a search session as they encounter and explore new information that expands their initial understanding of a topic \cite{jiang2014searching}. As this information needs to evolve, the criteria used to determine the search relevance adjust accordingly. A classic example demonstrated by \citet{Tang1998-co} is that users seek information driven by their broad interests and curiosity during the early stage of random browsing. Then, they apply more specific, case-based relevance criteria after better understanding their search objectives. These relevance factors are more related to knowledge construction and problem-solving. In the context of claim prioritization, this suggests that the relative importance that fact-checkers ascribe to different checkworthy dimensions used in their information-seeking may similarly evolve as they search. 

To identify an appropriate search scenario, we asked pilot study participants to conduct both exploratory and focused search tasks to filter and select claims with the tool. We found that in a focused search task (e.g., finding claims that mention the adverse effect of COVID-19 on marginalized populations), the search journey of pilot participants was very quick and precise, and they predominantly used certain facets, particularly ``Likely harmful'' in combination with keyword search. However, participants tended not to triage claims among multidimensional checkworthiness. In contrast, we observed that our pilot participants were more likely to use different facets and tool features in an exploratory search task (e.g., finding any claims that you found important to check). As a compromise, we hypothesized that many professional fact-checkers are already familiar with COVID-19 and related claims, so selecting this topic would provide a familiar foundation and starting point for an exploratory search task (as opposed to an exploratory task in which fact-checkers had no prior familiarity). 
We describe the task details in Section \ref{subsubsec:study procedure}.

We initially implemented an evaluative protocol combining formal usability tests and post-task interviews. 
After several rounds of pilot tests, we refined the protocol in two ways to better meet our two research goals (Section \ref{subsec:frame}). First, the protocol combines participant self-reported assessments, observed user behavior, and their post-hoc reflection to investigate how fact-checkers operationalize multidimensional checkworthiness before, during, and after using the tool. This allowed us to compare what participants said versus their actual actions. 

Second, we added another constrained claim selection task following the exploratory search. We created a separate page in our tool that displays claims participants selected during the exploratory search. Participants were asked to identify only the top three checkworthy claims from what they had selected.
We used this task to simulate a real-world scenario where fact-checkers pitch claims to editors \citep{Juneja2022-fx} (in this study, they were asked to provide comprehensive justifications of why these claims were selected). This enabled us to gain valuable insights into the fact-checker decision-making process of claim prioritization.

\section{Implicit ranking}
\label{sec:implicit_ranking}

Section \ref{subsubsec:data collection and analysis} mentioned that our pre-screening survey asked two related questions about the relative importance of different checkworthy dimensions. In this section, we compare how fact-checkers answer these two related questions. 

First, we asked fact-checkers to answer a 5-point Likert scale rating question about the \textit{Perceived importance} of each checkworthy dimension: ``This is an important factor resulting in the final fact-checked claim.'' 

We presented an analysis of the results of these answers across dimensions and participants in Section \ref{subsec:finding-multi-dimensional}, copied here for convenient access. As shown in Table \ref{tab:dimension-statistics}, ``Likely harmful'' had the mean average rating of (\(M = 4.81\)). The score decreased from ``Likely false'' (\(M_{false} = 4.63\)), ``Interest to the public,'' (\(M_{public-interest} = 4.50\)), to ``Verifiable'' (\(M_{verifiable} = 4.44\)). The median scores were the same for each dimension (\(Median = 5\)). ``Verifiable'' received the lowest average rating but had the largest standard deviation (\(SD = 1.21\)), indicating the greatest variation in opinions among our participants. No significant differences were found across these four dimensions (\(X^2 = 1.824\), \(p > 0.05\)). 

Complementing this rating question, we further asked participants to implicitly rank the relative importance of the four dimensions in an open-ended format: ``Considering these four criteria, how would you characterize their relative importance compared to one another? Please rank these criteria from most to least important.'' Because these answers were open-ended, participants often mentioned only a subset of the dimensions and used free-form language requiring manual analysis. Most participants identified several dimensions as ``most or equally important,'' while also indicating which dimensions they considered ``least important.'' We counted such responses and provided these counts in Table \ref{tab:dimension-responses}.

\begin{table}[ht]
  \begin{tabular}{ccclc}
    \toprule
    \multirow{2}{*}{Dimensions} & \multicolumn{4}{c}{\textbf{Implicit ranking}} \\
     &Least important&Ratio &Most or equally important & Ratio\\
    \midrule
    Verifiable & p9, 10, 16 & 3/16 & p2-7, 13-14 & 8/16 \\
    Likely false & - & 0/16 & p6 & 1/16 \\
    Likely harmful & p14 & 1/16 & p1-2, 4, 8-9, 11-12, 15 & 8/16\\
    Interest to the public & p5, 8, 13 & 3/16 & p9, 16 & 2/16 \\
  \bottomrule
\end{tabular}
\vspace{0.2cm}
\caption{Participant implicit ranking on the relative importance among four-dimensional checkworthiness. The results show that participants mostly agreed that ``Verifiable'' and ``Likely harmful'' were the most important or equally important dimensions.}
\label{tab:dimension-responses}
\end{table}

Table \ref{tab:dimension-responses} shows that eight participants identified ``Verifiable'' and ``Likely harmful'' as the most or equally important. This number was higher than the other two: only one participant rated ``Likely false'' and two considered ``Interest to the public'' as the most or equally important. Additionally, three participants rated ``Verifiable'' and ``Interest to the public'' as the least important. 

When we compare these implicit rankings to the \textit{Perceived importance} ratings, this further explains why ``Verifiable'' and ``Interest to the public'' showed a higher standard deviation in their importance ratings: for both dimensions, implicit rankings show that three participants thought these dimensions were among the least important of the four checkworthy dimensions. 

\section{Other important checkworthy dimensions}
\label{sec:other_dims}

Many dimensions of checkworthiness have been identified in prior work (Table \ref{tab:multi-dimensional checkworthiness}). We adopted the COVID-19 claim dataset developed by \citet{alam-etal-2021-fighting-covid}, which annotated seven dimensions of checkworthiness, though we used only four of these dimensions to simplify our study. To probe beyond these four dimensions, we also asked participants: ``If you found multiple claims that met all the criteria used in our study but couldn't check them all at once, how would you choose which claims to prioritize?'' To address such tie-breaking, participants began to invoke additional checkworthy dimensions beyond those in our study, such as urgency \citep{Sehat2023-xa}: 
\begin{displayquote}
    ``\textit{If the consequences of this are harm that will be caused immediately, the more immediate the harm that sometimes comes into play.}'' (P2) ``\textit{Potentially dangerous claims are considered more urgent. We often begin checking these claims even before determining that they have been widely shared. Only when a claim is clearly obscure or unlikely to be believed will a dangerous claim be dismissed after it has been established as being verifiable.}'' (P7)
\end{displayquote}
Susceptibility \citep{Babaei2022-so} was also invoked as another tie-breaking dimension beyond our study's scope. As \citet{Sehat2023-xa} note, susceptibility can serve as another indicator of harmfulness. P4 explained ``\textit{If a social media user believed the false claim, it could potentially result in a more harmful outcome.}''
Section \ref{subsec:limitation} further discusses our use of only four dimensions as a study limitation. 

\section{Additional statistical results} \label{appendix:post-hoc test}

\begin{table}[h]
\resizebox{0.85\textwidth}{!}{%
\begin{tabular}{@{}>{\raggedright}p{2cm}>{\raggedright}p{10cm}>{\raggedright}p{1.5cm}>{\raggedright\arraybackslash}p{1cm}@{}} \toprule
Dimensions & Agreement statement & \textit{M(SD)} & \textit{Median} \\ \midrule
Usefulness & Using this tool would enable me to accomplish tasks more quickly & 4.13(0.52) & 4   \\
& Using this tool would improve my job performance. &  4.07(0.59) & 4  \\
& Using this tool in my job would increase my productivity.  &  4.00(0.76) & 4  \\ 
& Using this tool would enhance my effectiveness on the job. & 4.07(0.59) & 4  \\
& Using this tool would make it easier to do my job. &  4.27(0.59) & 4  \\ 
& I would find this tool useful in my job. &  4.40(0.51) & 4 \\ \midrule
Ease of use & Learning to operate the tool would be easy for me.  &  4.38(0.89) & 5 \\ 
& I would find it easy to get this tool to do what I want it to do. & 4.06(0.77) & 4 \\
& My interaction with the tool would be clear and understandable. & 4.19(0.75) & 4 \\
& I would find this tool would be clear and understandable.  &  4.38(0.62) & 4 \\ 
& It would be easy for me to become skillful at using this tool. & 4.25(0.77) & 4 \\
& I found this tool easy to use. & 4.19(1.17) & 4 \\ \bottomrule
\end{tabular}
}
\vspace{0.1cm}
\caption{Descriptive statistics of mean (standard deviation) and median for participant self-reported responses of tool's perceived usefulness and ease of use.}
\end{table}

\begin{table}[h]
\resizebox{0.85\textwidth}{!}{%
\begin{tabular}{@{}l>{\raggedleft}p{1.8cm}>{\raggedleft}p{1.5cm}>{\raggedleft}p{2.2cm}>{\raggedleft}p{1.5cm}>{\raggedleft}p{2.5cm}>{\raggedleft\arraybackslash}p{1.5cm}@{}} \toprule
Measure & Likely false \textit{Mean} & Wilcoxon \linebreak \(W(p)\) & Likely harmful \linebreak \textit{Mean} & Wilcoxon \linebreak \(W(p)\) & Interest to public \linebreak \textit{Mean} & Wilcoxon \linebreak \(W(p)\) \\ \midrule
Verifiable  &  4.13 | 3.69 & 9.00(0.08) &  4.13 | 4.31 & 12.00(0.37) & 4.13 | 4.25 & 17.00(0.49) \\
Likely false  & - & - & 3.69 | 4.31 & \textbf{0.0(0.00)} & 3.69 | 4.25 & \textbf{10.50(0.03)} \\
Likely harmful & - & - & - & - & 4.31 | 4.25 & 14.00(0.50) \\ \bottomrule    
\end{tabular}
}
\vspace{0.1cm}
\caption{Comparing the mean of ``Ease of finding'' over four dimensions of checkworthiness. The mean values are presented as a pair (A | B) corresponding to dimensions over each row (A) and column (B). Results highlighted as bold are statistically significant for the pairwise Wilcoxon signed-rank test at \textit{p} < 0.05. }
\label{tab:post-hoc ease of findings}
\end{table}

\begin{table}[h]
\resizebox{0.85\textwidth}{!}{%
\begin{tabular}{@{}l>{\raggedleft}p{1.8cm}>{\raggedleft}p{1.5cm}>{\raggedleft}p{2.2cm}>{\raggedleft}p{1.5cm}>{\raggedleft}p{2.5cm}>{\raggedleft\arraybackslash}p{1.5cm}@{}} \toprule
Measure & Likely false \textit{Mean} & Wilcoxon \linebreak \(W(p)\) & Likely harmful \linebreak \textit{Mean} & Wilcoxon \linebreak \(W(p)\) & Interest to public \linebreak \textit{Mean} & Wilcoxon \linebreak \(W(p)\) \\ \midrule
Verifiable  &  0.77 | 0.73 & 24.00(0.24) &  0.77 | 0.62 & 19.00(0.06) & 0.77 | 0.39 & \textbf{7.00(0.00)} \\
Likely false  & - & - & 0.73 | 0.62 & 27.00(0.20) & 0.73 | 0.39 & \textbf{8.50(0.01)} \\
Likely harmful & - & - & - & - & 0.62 | 0.39 & \textbf{11.00(0.02)} \\ \bottomrule    
\end{tabular}
}
\vspace{0.1cm}
\caption{Comparing the mean of ``Overall weights'' over four dimensions of checkworthiness. The mean values are presented as a pair (A | B) corresponding to dimensions over each row (A) and column (B). Results highlighted as bold are statistically significant for the pairwise Wilcoxon signed-rank test at \textit{p} < 0.05.}
\label{tab:post-hoc overall weights}
\end{table}

\pagebreak

\section{Participant weighting patterns} \label{appendix:weighting pattern}

Eight participant weighting patterns reveal a complete three-level hierarchy (P2-4, P7, P10, and P14-16). Six participant weighting patterns reveal a two-level hierarchy (P1, P5-6, P8, and P11-12). 
Two participant weighting patterns do not reveal any hierarchy (P9, P13). See Figure \ref{fig:Time-series-diagram P7}'s caption for figure interpretation. 

\begin{center}
\begin{longtable}{c}
\label{tab:diagrams_longtable} \\
\endfirsthead
\endhead

\endfoot

\endlastfoot

\includegraphics[width=0.9\textwidth]{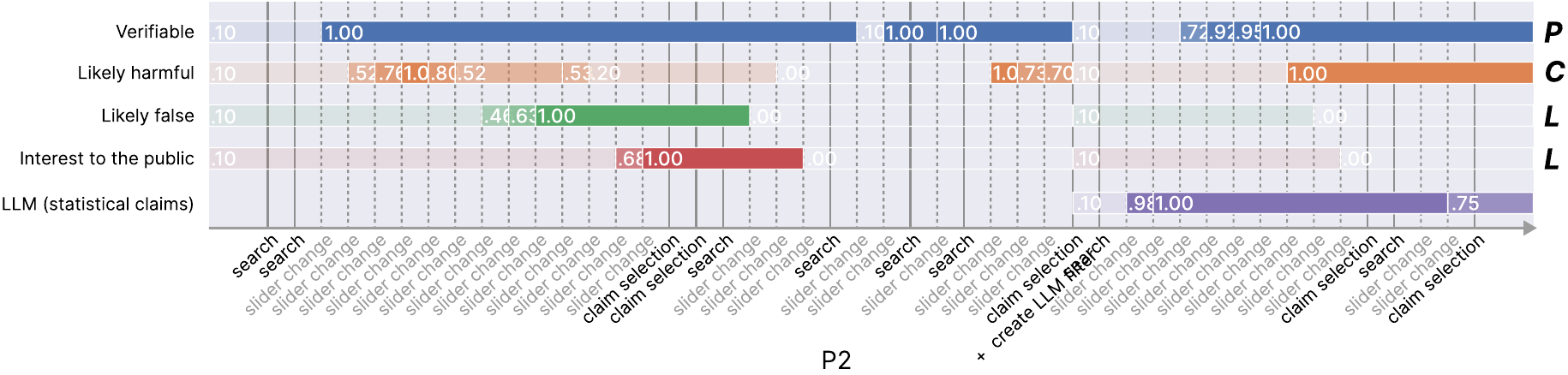} \\
\includegraphics[width=0.9\textwidth]{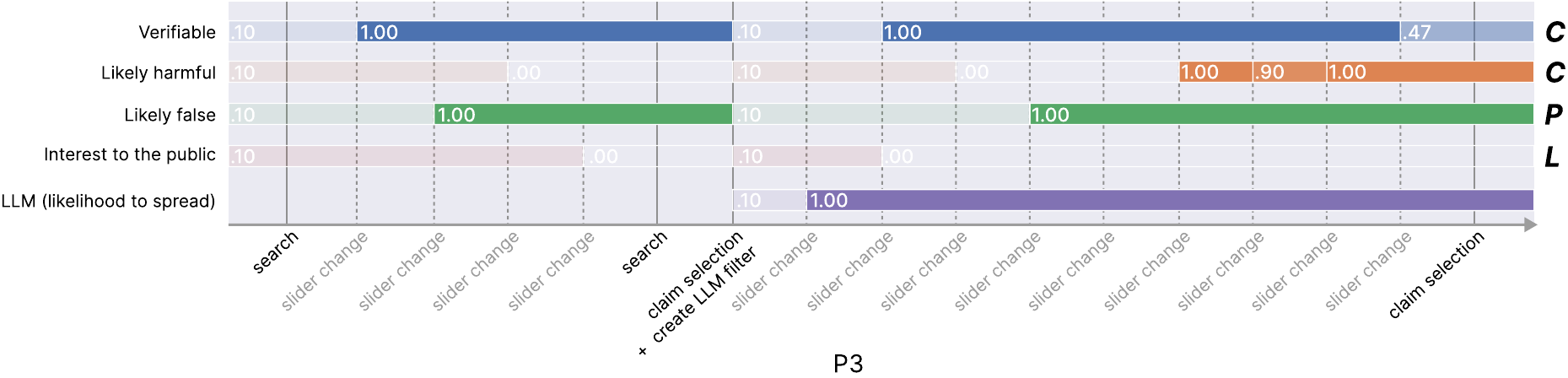} \\
\includegraphics[width=0.9\textwidth]{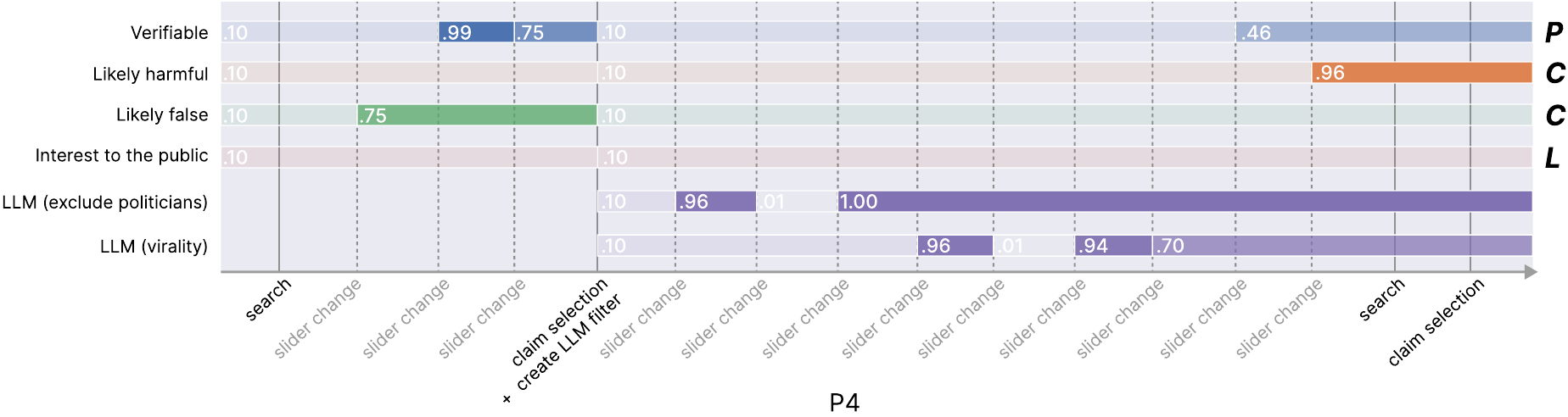} \\
\includegraphics[width=0.9\textwidth]{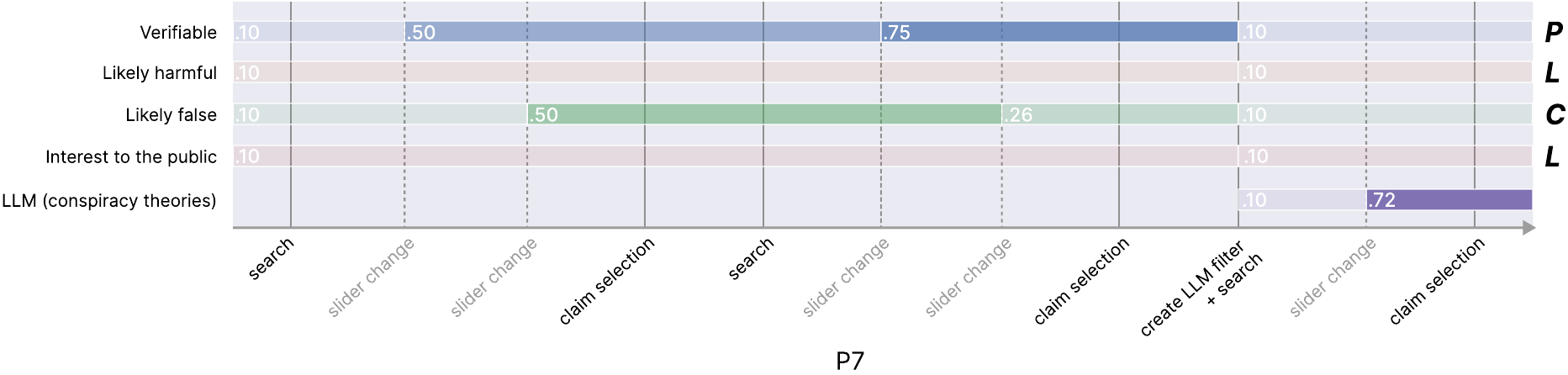} \\
\includegraphics[width=0.9\textwidth]{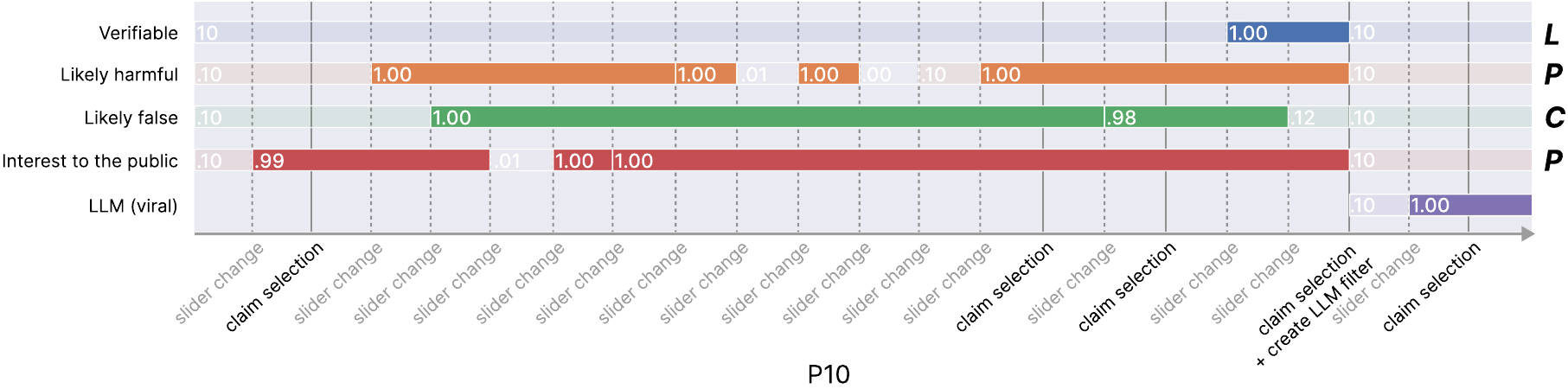} \\
\includegraphics[width=0.9\textwidth]{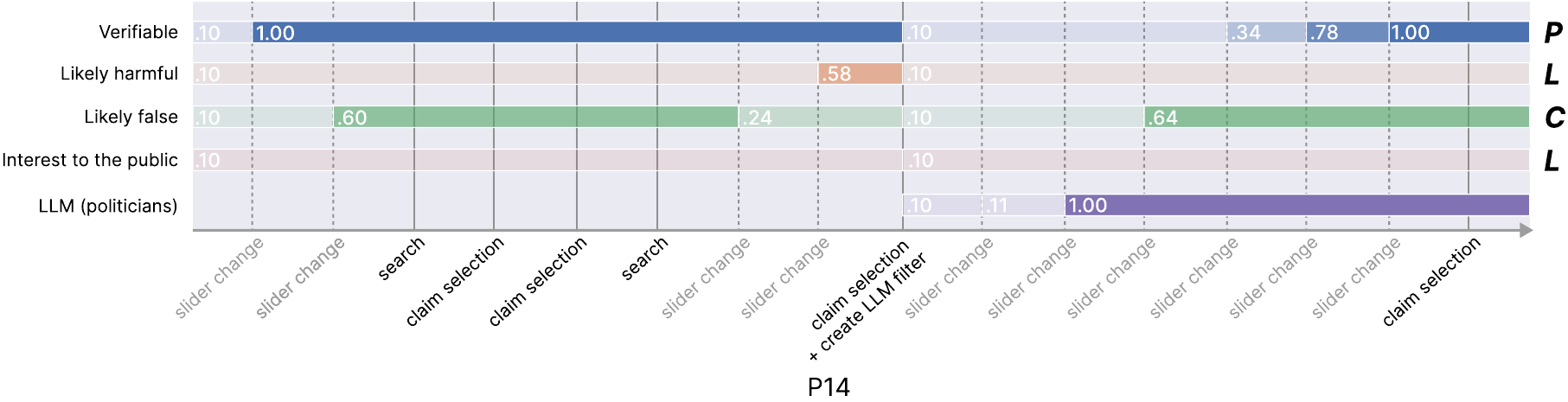} \\
\includegraphics[width=0.9\textwidth]{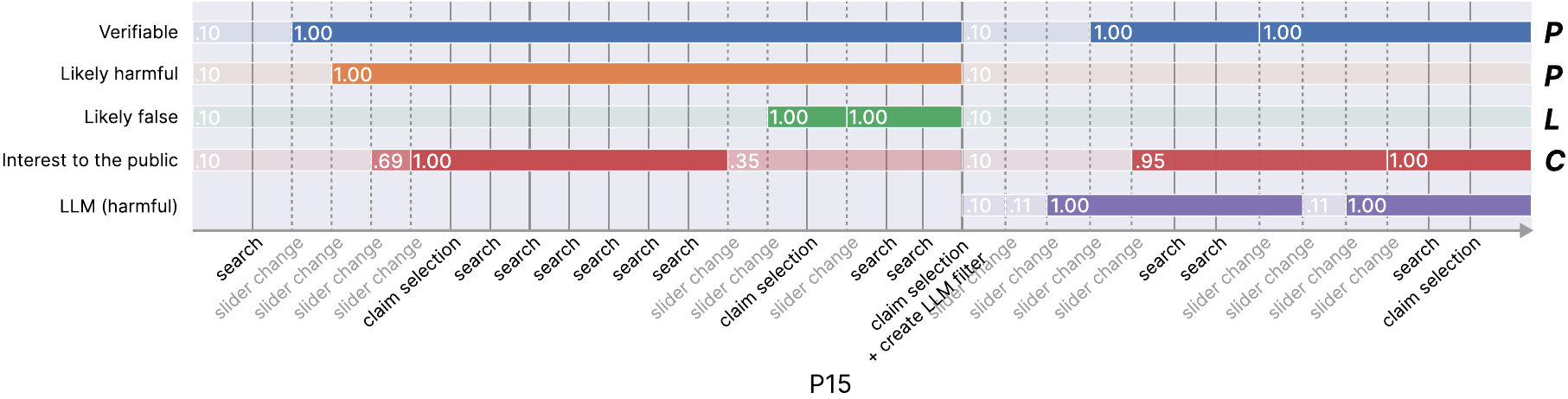} \\
\includegraphics[width=0.9\textwidth]{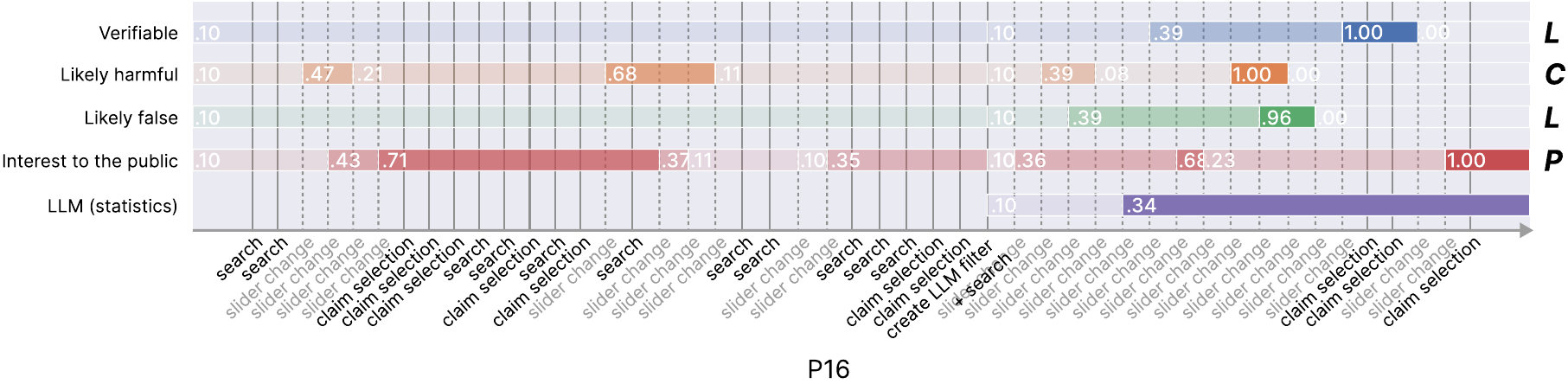} \\
\includegraphics[width=0.9\textwidth]{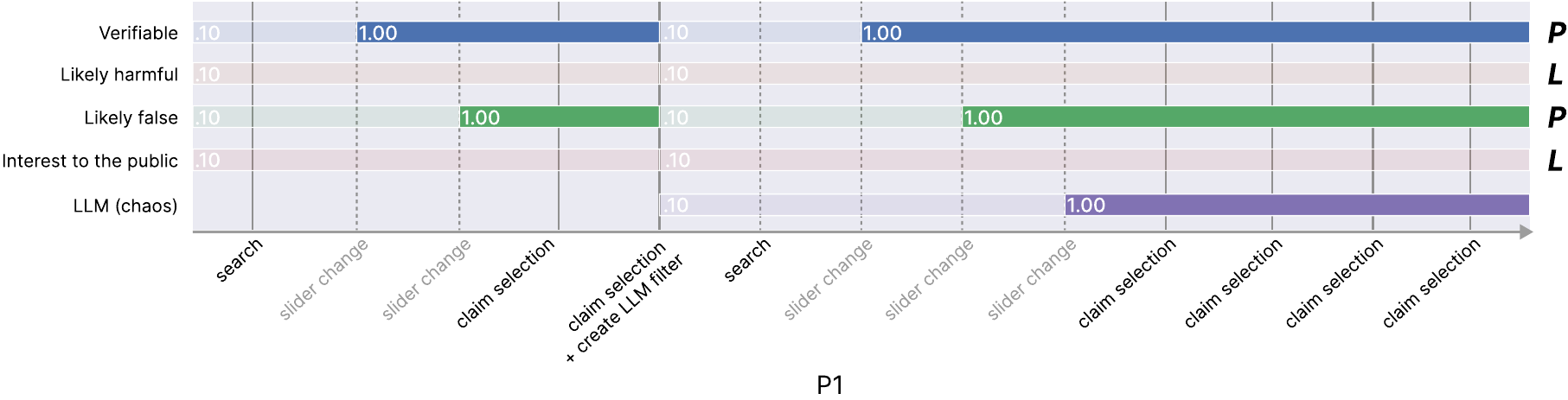} \\
\includegraphics[width=0.9\textwidth]{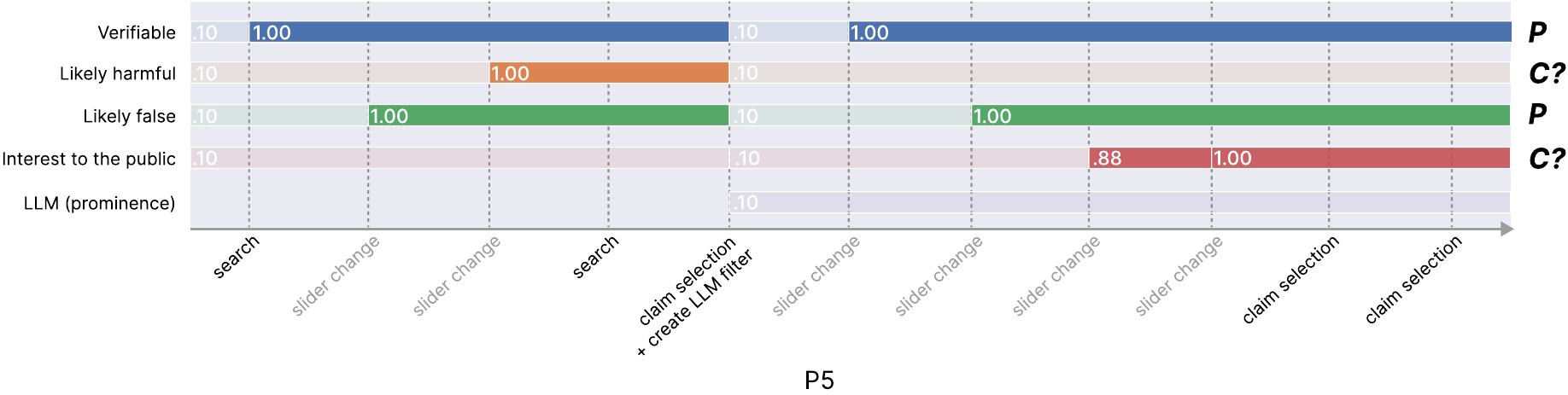} \\
\includegraphics[width=0.9\textwidth]{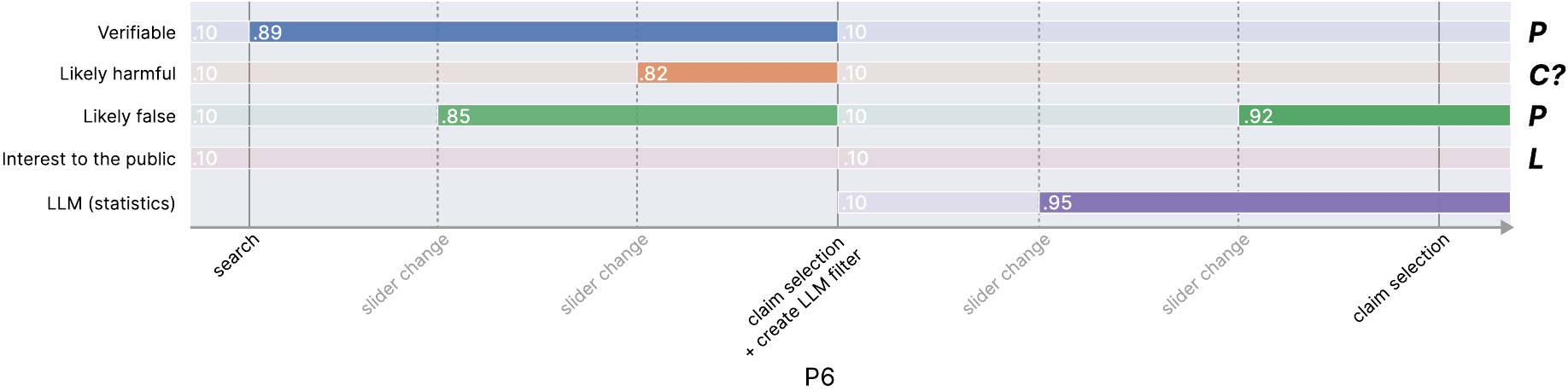} \\
\includegraphics[width=0.9\textwidth]{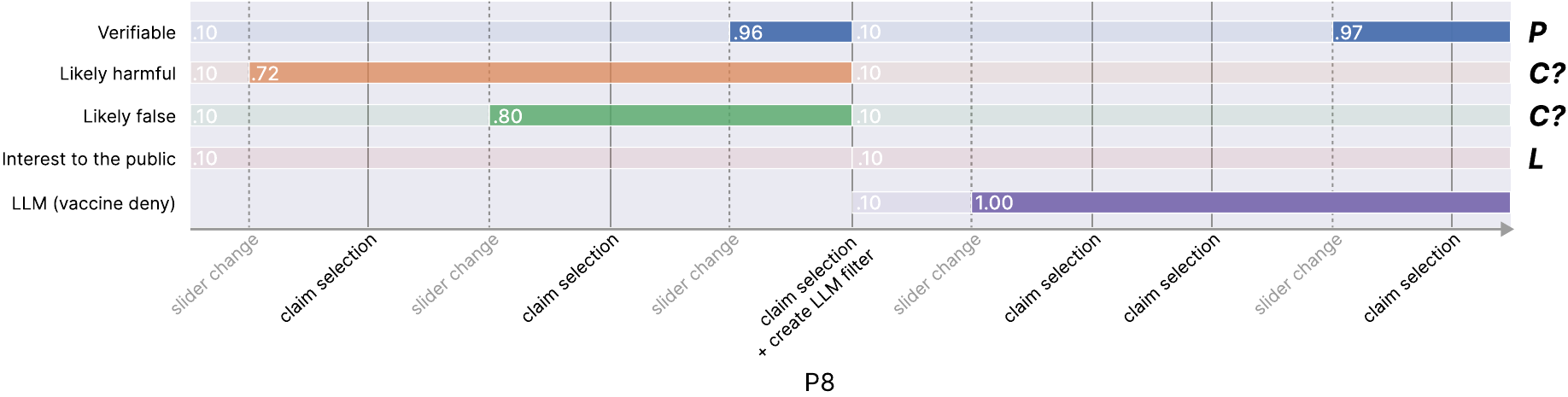} \\
\includegraphics[width=0.9\textwidth]{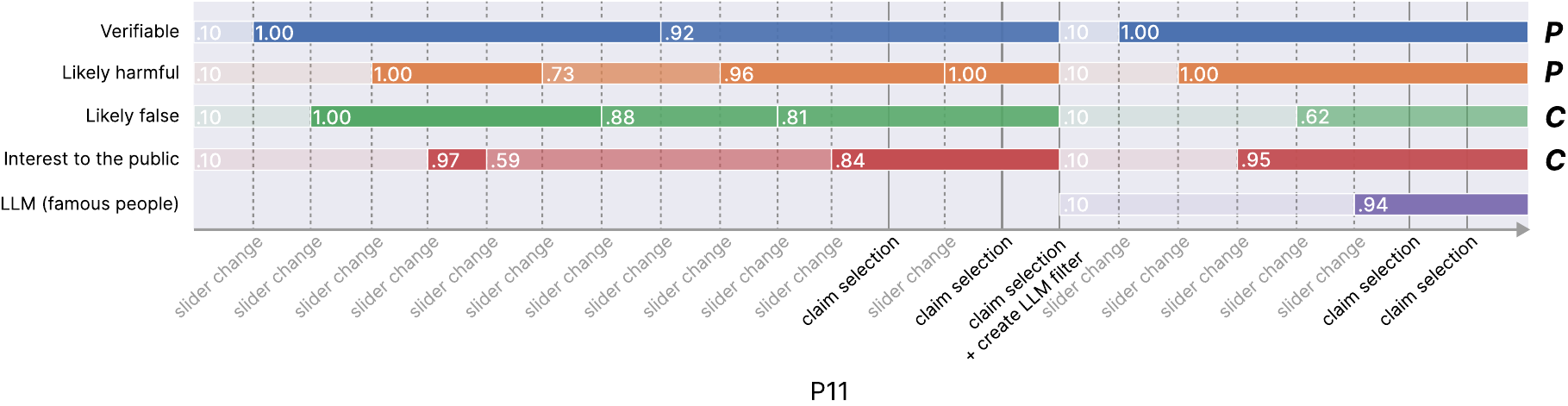} \\
\includegraphics[width=0.9\textwidth]{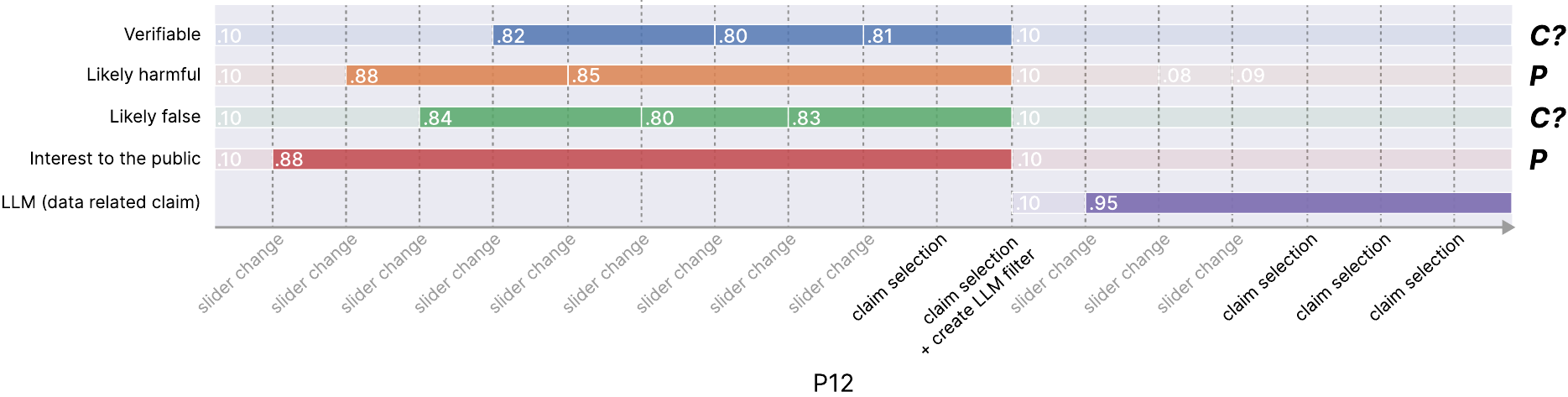} \\
\includegraphics[width=0.9\textwidth]{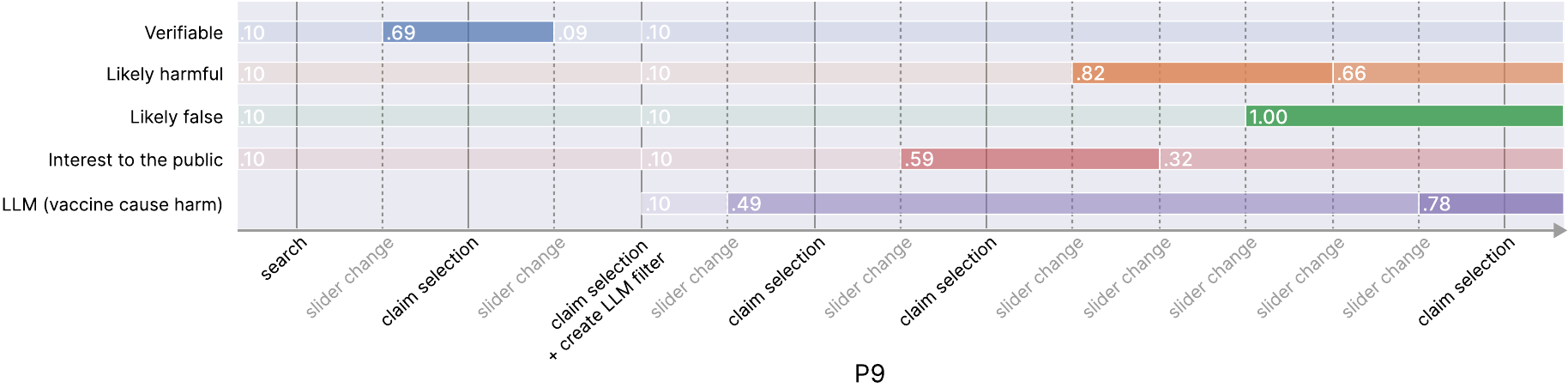} \\
\includegraphics[width=0.9\textwidth]{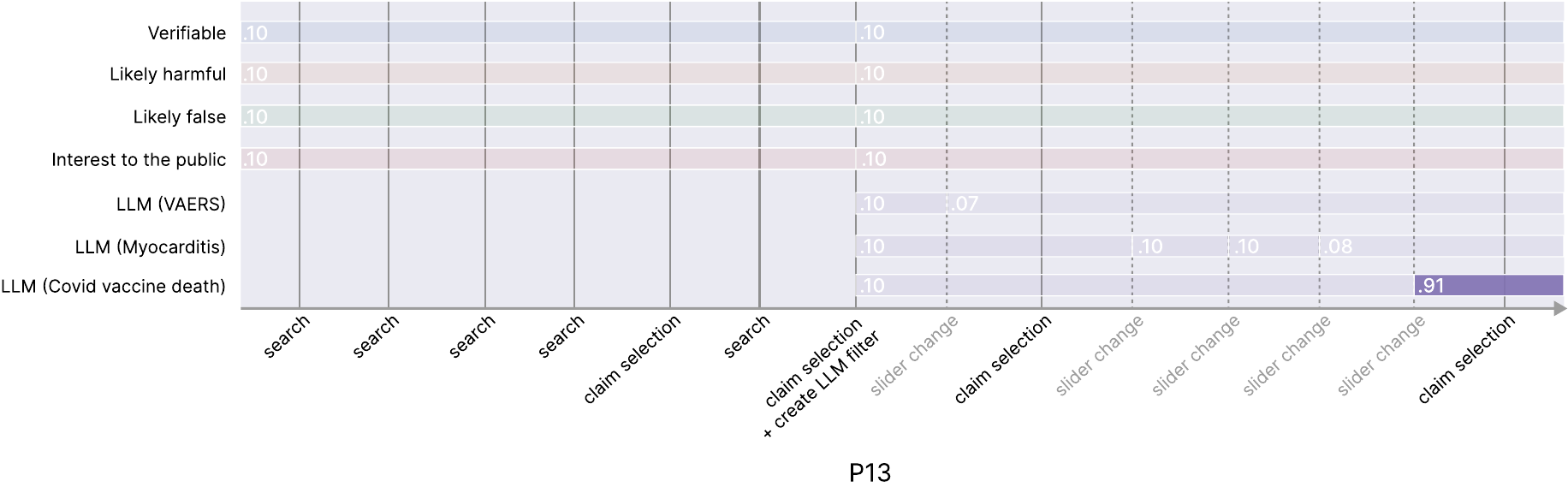} \\

\end{longtable}
\end{center}


\end{document}